\documentclass[10pt, a4paper]{article} 
\usepackage[square,numbers,sort&compress]{natbib}
\usepackage{booktabs}
\usepackage{bbm} 
\usepackage[shortlabels]{enumitem}
\usepackage{setspace}
\usepackage{subfloat}
\usepackage{mathtools}
\usepackage{amsmath,amsfonts,amssymb,amsthm,mathrsfs, bm}
\usepackage[utf8]{inputenc}
\usepackage[swedish, english]{babel}
\usepackage{csquotes}
\usepackage{subfig}
\usepackage{graphicx}
\usepackage{supertabular}                                             
\usepackage[table]{xcolor} 
\usepackage{textgreek} 
\usepackage{graphicx}
\usepackage{epstopdf}
\usepackage{microtype}
\usepackage{multirow}
\usepackage{xcolor}
\usepackage{authblk}
\usepackage{datetime2}	
\usepackage[margin=1in]{geometry} 
\usepackage{setspace}	
\theoremstyle{plain}
\newtheorem{theorem}{Theorem}[section]
\newtheorem{lemma}{Lemma}[section]

\newtheorem{assumption}{Assumption}[section]

\usepackage{amsmath,amsfonts,amssymb,amsthm,mathrsfs, bm}
\usepackage{stmaryrd}

\usepackage{graphicx}
\usepackage{tikz}
\usepackage[utf8]{inputenc}
\usepackage{pgfplots} 
\usepackage{pgfgantt}
\usepackage{pdflscape}
\pgfplotsset{compat=newest} 
\pgfplotsset{plot coordinates/math parser=false}

\usepackage{endnotes}
\usepackage{commath}
\usepackage{gensymb}
\usepackage{mathtools}
\usepackage{tikz} 
\usetikzlibrary{calc,quotes,angles}
\usepackage{tkz-euclide}
\usetikzlibrary{automata,positioning} 

\usepackage{xcolor}
\usepackage{tikz,tkz-euclide,siunitx}
\usepackage{etoolbox} 

\usetikzlibrary{quotes,arrows.meta,angles,calc,shadings,positioning,babel}

\newtheorem{remark}{Remark}[section]

\usepackage{endnotes}
\usepackage{commath}
\usepackage{gensymb}
\usepackage{tcolorbox}

\makeatletter
\patchcmd{\tkz@DrawLine}{\begingroup}{\begingroup\makeatletter}{}{}
\makeatother

\allowdisplaybreaks
\DeclareMathOperator{\interior}{int}

\makeatletter
\newcommand\makebig[2]{%
  \@xp\newcommand\@xp*\csname#1\endcsname{\bBigg@{#2}}%
  \@xp\newcommand\@xp*\csname#1l\endcsname{\@xp\mathopen\csname#1\endcsname}%
  \@xp\newcommand\@xp*\csname#1r\endcsname{\@xp\mathclose\csname#1\endcsname}%
}
\makeatother

\providecommand*{\ped}[1]{%
\ensuremath{_\textnormal{#1}}}
\providecommand*{\ap}[1]{%
\ensuremath{^\textnormal{#1}}}
\providecommand*{\eu}%
{\ensuremath{\mathrm{e}}}
\providecommand*{\im}%
{\ensuremath{\mathrm{i}}}
\providecommand*{\GammaF}%
{\ensuremath{\mathrm{\Gamma}}}
\providecommand*{\BetaF}%
{\ensuremath{\mathrm{\Beta}}}

\makebig{biggg} {3.0}
\makebig{Biggg} {3.5}
\makebig{bigggg}{4.0}
\makebig{Bigggg}{4.5}
\makebig{biggggg}{5.0}
\makebig{Biggggg}{5.5}
\usepackage{longtable}
\usepackage{xcolor}
\DeclareMathSymbol{\Gamma}{\mathalpha}{letters}{"00}
\DeclareMathSymbol{\Delta}{\mathalpha}{letters}{"01}
\DeclareMathSymbol{\Theta}{\mathalpha}{letters}{"02}
\DeclareMathSymbol{\Lambda}{\mathalpha}{letters}{"03}
\DeclareMathSymbol{\Xi}{\mathalpha}{letters}{"04}
\DeclareMathSymbol{\Pi}{\mathalpha}{letters}{"05}
\DeclareMathSymbol{\Sigma}{\mathalpha}{letters}{"06}
\DeclareMathSymbol{\Upsilon}{\mathalpha}{letters}{"07}
\DeclareMathSymbol{\Phi}{\mathalpha}{letters}{"08}
\DeclareMathSymbol{\Psi}{\mathalpha}{letters}{"09}
\DeclareMathSymbol{\Omega}{\mathalpha}{letters}{"0A}

\definecolor{matblue}{rgb}{0.0000,0.4470,0.7410}
\definecolor{matred}{rgb}{0.8500,0.3250,0.0980}
\definecolor{matyellow}{rgb}{0.9290,0.6940,0.1250}
\definecolor{matpurple}{rgb}{0.4940,0.1840,0.5560}
\definecolor{matgreen}{rgb}{0.4660,0.6740,0.1880}
\definecolor{matcyan}{rgb}{0.3010,0.7450,0.9330}
\definecolor{matmaroon}{rgb}{0.6350,0.0780,0.1840}

\usepackage{hyperref}
\hypersetup{
    colorlinks = true,
    linkbordercolor = {white},
            linkcolor=blue,
            bookmarks=true,
            filecolor=blue,
            urlcolor=blue,
            citecolor=blue
}

\newtcolorbox[auto counter]{modelbox}[2][]{%
  colback=white, colframe=black,
  fonttitle=\bfseries,
  title=Models~\thetcbcounter: #2,
  label=#1
}

\begin{document}

\title{The role of distributed friction dynamics on heave-induced motion of offshore drilling systems: modelling, analysis, and simulations\footnote{This document is the author version of the manuscript.}}
\date{}
\author[a,b,c]{Luigi Romano\thanks{Corresponding author. Email: luigi.romano@liu.se.}}
\author[b]{Ole Morten Aamo\thanks{Corresponding author. Email: ole.morten.aamo@ntnu.no.}}
\affil[a]{\footnotesize{Division of Vehicular Systems, Department of Electrical Engineering, Linköping University, SE-581 83 Linköping, Sweden}}
\affil[b]{\footnotesize{Department of Engineering Cybernetics, Norwegian University of Science and Technology, O. S. Bragstads plass 2, NO-7034, Trondheim, Norway}}
\affil[c]{\footnotesize{Control Systems Technology Group, Department of Mechanical Engineering, Eindhoven University of Technology, Groene Loper 1, 5612 AZ Eindhoven, the Netherlands}}

\maketitle

\begin{abstract}
This paper investigates the influence of friction dynamics on heave-induced disturbance propagation using a coupled hydraulic-structural drilling model formulated as a semilinear interconnection of \emph{ordinary} and hyperbolic \emph{partial differential equations} (ODE--PDEs), and capturing spatially distributed memory effects in the drillstring-wellbore interaction. First, local and global existence and uniqueness results are established, and sufficient conditions for boundedness and input-to-state stability are deduced via a small-gain analysis. Subsequently, extensive numerical investigations are carried out, including harmonic excitation studies, nonlinear distortion analysis, spatial amplification investigations, mechanism-isolation experiments, parameter sensitivity analysis, and broadband simulations under realistic offshore sea-state excitation. The simulation results indicate that, whilst the hydraulic response is governed primarily by wave propagation phenomena, compressibility, and boundary reflections, the structural dynamics display substantially greater sensitivity to distributed friction effects. Furthermore, drillstring-wellbore friction is shown to have a relatively limited influence on well-posedness but a potentially significant impact on stability margins. These findings highlight the importance of accounting for coupled hydraulic-structural dynamics and distributed friction mechanisms when analysing, monitoring, and controlling offshore drilling systems.
\end{abstract}
\section*{Keywords}
Drilling systems; distributed friction dynamics; nonlinear vibrations; distributed parameter systems; semilinear ODE--PDEs

\section{Introduction}\label{intro}

Offshore drilling operations are increasingly performed in deep-water environments, where the interaction between the floating rig, the drilling fluid, and the drillstring plays a fundamental role in determining operational safety, drilling efficiency, and well-control. Wave-induced rig heave generates oscillatory drillstring motion, which acts as a source of distributed hydraulic and structural disturbances throughout the well \cite{Petersen2008,Mitchell1988,Wagner1993,Pavlov2010}. These disturbances propagate through both the drilling fluid and the drillstring in the form of pressure, flow, velocity, and stress waves \cite{Samuel2003,Stecki1986a,Stecki1986b,Kaasa2012}, generating fluctuations that may deteriorate drilling performance, increase fatigue loading, accelerate equipment degradation, and, in severe cases, compromise well integrity and operational safety \cite{Aamo00,Aamo01}. In addition to hydraulic transients, drillstrings exhibit a broad range of nonlinear vibration phenomena, including axial bit bounce, torsional stick-slip oscillations, lateral whirl, and coupled vibration mechanisms. These phenomena are recognised as major contributors to drilling inefficiency, excessive tool wear, and drillstring failures \cite{Spanos2003}. 

Consequently, significant research efforts have been devoted to the modelling and analysis of drillstring dynamics, with particular attention to the mechanisms underlying self-excited vibrations. Seminal works \cite{Leine2002,Leine2005,Richard2004,Richard2007,Besselink2011} introduced nonlinear drillstring models capable of reproducing stick-slip oscillations, whirl interactions, and self-excited responses induced by bit-rock interaction and friction. Subsequently, reduced-order and experimentally validated formulations were proposed in \cite{Liao2011}, whilst the instability mechanisms responsible for spontaneous drilling vibrations were clarified in \cite{Kovalyshen2014,Depouhon2014}.

In parallel with these developments, increasing attention has been devoted to the hydraulic effects induced by heave motion in offshore \emph{managed-pressure drilling} (MPD) systems. Maintaining the downhole pressure within a narrow operational window is essential to prevent formation influx, wellbore collapse, and fracture of the surrounding formation. During drilling from floating platforms, however, the rigid connection between the drillstring and the rig causes the oscillatory heave motion to act as a piston, generating surge and swab pressures that propagate throughout the well. Since the characteristic time scale of offshore wave excitation is comparable with the travelling time of pressure waves in deep wells, distributed-parameter models are required to accurately describe these transport phenomena. In this context, infinite-dimensional formulations were introduced in \cite{Landet2013} to analyse and control heave-induced pressure fluctuations in MPD systems, demonstrating the necessity of accounting for distributed dynamics in pressure regulation problems. Feedback control for attenuating downhole pressure fluctuations were investigated further in \cite{Aamo03,Aamo04,Aamo05}. More recently, comprehensive hydraulic-structural models incorporating compressible mud dynamics, drillstring elasticity, distributed drag, realistic well geometries, and non-Newtonian Herschel--Bulkley rheology were developed in \cite{Strecker2017}, and utilised for feedback control in \cite{Aamo06}. The investigations presented in \cite{Landet2013,Strecker2017}, in particular, showed that pressure amplification depends strongly on operating conditions, mud rheology, well geometry, and excitation frequency. Furthermore, these studies highlighted two important features of offshore drilling systems: significant downhole pressure oscillations may remain largely invisible from surface measurements, and the drilling system behaves as a distributed waveguide exhibiting resonant amplification and substantial spatial variations of the response. 

In the context outlined above, the interaction between distributed friction dynamics and hydraulic-structural wave propagation remains, however, poorly understood. Existing heave-induced pressure models generally describe friction using static or quasi-static constitutive relationships that depend instantaneously on local velocities and flow rates \cite{Leine2002,Leine2005,Sheppard}. By contrast, the drillstring-dynamics literature consistently demonstrates that friction is inherently nonstationary and may strongly affect vibration amplitudes, damping characteristics, stress transmission, and stability margins. These observations suggest that friction may influence not only structural vibrations but also the generation, propagation, attenuation, and amplification of hydraulic disturbances in coupled drilling systems, despite having received comparatively little attention. 

Motivated by this gap, the present paper investigates the propagation and amplification of heave-induced disturbances using a distributed hydraulic-structural drilling model incorporating nonlinear friction-memory dynamics. The proposed formulation combines compressible fluid transport, drillstring elasticity, hydraulic drag, and a distributed FrBD friction model, resulting into an interconnection of \emph{ordinary} and hyperbolic \emph{partial differential equations} (PDEs). Unlike conventional static descriptions of friction, the proposed model captures the local dynamic evolution of stresses and strains, introducing distributed memory effects that interact directly with hydraulic and structural wave propagation mechanisms. Beyond the development of the mathematical model itself, the objective of the paper is to clarify the physical mechanisms governing disturbance transmission in offshore drilling systems. To this end, local and global well-posedness results are established for the resulting ODE--PDE interconnection and a sufficient input-to-state stability criterion is derived using small-gain arguments. Complementary to the theoretical analysis, extensive numerical studies are performed to investigate the gains from the rig velocity input to the hydraulic and structural variables of interest. In particular, the numerical simulations reported in this paper focus on harmonic frequency-response analyses, nonlinear distortion quantification, spatial amplification studies, mechanism-isolation experiments, parameter sensitivity analysis, and broadband offshore simulations. In this context, particular attention is devoted to identifying the respective roles of hydraulic transport, compressibility, drillstring elasticity, hydraulic drag, and friction-memory effects in shaping the pressure, flow, velocity, and stress responses. 

Summarising, the contributions of this paper are fourfold. First, a distributed hydraulic-structural drilling model incorporating dynamic friction-memory effects is developed within a unified semilinear ODE--PDE framework. Second, local and global well-posedness results are established for the resulting infinite-dimensional system. Third, a small-gain-based stability analysis is developed to quantify the influence of distributed friction on system stability. Fourth, extensive numerical investigations are used to characterise how hydraulic and frictional mechanisms affect disturbance amplification and propagation. The obtained results reveal that the drilling system behaves as a distributed waveguide exhibiting pronounced frequency selectivity and substantial downhole amplification. Furthermore, whilst hydraulic effects appear to primarily govern well-posedness and hydraulic amplification, distributed friction dynamics is found to influence stability margins and structural responses.

The remainder of this manuscript is organised as follows. Section~\ref{ect:Model} introduces the main equations of the model in terms of physical variables, clarifying their meaning and the underlying assumptions. The resulting ODE--PDE system is then reformulated in Sect.~\ref{sect:RiemETeho} in terms of Riemann variables, which facilitates the well-posedness and stability analyses. The results of an exhaustive campaign of numerical simulations are then discussed in Sect.~\ref{sect:Numerics}, where the roles of viscous and friction dynamics on the hydraulic and structural responses of the system are investigated. The main conclusions of the paper are then summarised in Sect.~\ref{sect:Conclusions}, where some recommendations for future research are also given. Finally, technical results are contained in Appendix~\ref{app:techn}, whilst Appendix~\ref{app:details} provides additional implementation details.

\subsection*{Notation}
In this paper, $\mathbb{R}$ denotes the set of real numbers; $\mathbb{R}_{>0}$ and $\mathbb{R}_{\geq 0}$ indicate the set of positive real numbers excluding and including zero, respectively. 
The set of $n\times m$ matrices with values in $\mathbb{F}$ ($\mathbb{F} = \mathbb{R}$, $\mathbb{R}_{>0}$, $\mathbb{R}_{\geq0}$) is denoted by $\mathbf{M}_{n\times m}(\mathbb{F})$ (abbreviated as $\mathbf{M}_{n}(\mathbb{F})$ whenever $m=n$). $\mathbf{Sym}_n(\mathbb{R})$ and $\mathbf{GL}_n(\mathbb{R})$ represent respectively the groups of symmetric and invertible matrices with values in $\mathbb{R}$; the identity matrix on $\mathbb{R}^n$ is indicated with $\mathbf{I}_n$. A positive-definite matrix is noted as $\mathbf{M}_n(\mathbb{R}) \ni \mathbf{Q} \succ \mathbf{0}$; a positive semidefinite one as $\mathbf{M}_n(\mathbb{R}) \ni \mathbf{Q} \succeq \mathbf{0}$. A $n$-dimensional vector whose entries are all equal to one is denoted by $\bm{1}_n$, that is, $\mathbb{R}^n \ni \bm{1}_n \triangleq [1\; 1\; \dots \; 1]^{\mathrm{T}}$.

For an element $\bm{f} = [f_1\; f_2\; \dots \; f_n]^{\mathrm{T}}$ of $\mathbb{R}^n$, the following equivalent norms are defined:
\begin{align}
\norm{\bm{f}}_{\ell^1(\mathbb{R}^n)} \triangleq \sum_{i=1}^n \abs{f_i}, && \norm{\bm{f}}_{\ell^2(\mathbb{R}^n)} \triangleq \sqrt{\sum_i^n \abs{f_i}^2},
\end{align}
where $\abs{f}$ indicates the absolute value of $f \in \mathbb{R}$. Matrix norms are simply denoted by $\norm{\cdot}$.
Given a domain $\Omega$ with closure $\overline{\Omega}$, $C^k(\overline{\Omega};\mathcal{Z})$ ($p, k \in \{1, 2, \dots, \infty\}$) denotes the spaces of $k$-times continuously differentiable functions on $\overline{\Omega}$ with values in a Banach space $\mathcal{Z}$ (for $T = \infty$, the interval $[0,T]$ is identified with $\mathbb{R}_{\geq 0}$). In particular, the following norms are defined on $C^0(\overline{\Omega};\mathbb{R}^n)$:
\begin{align}\label{eq:normsonC0}
\norm{\bm{f}(\cdot)}_{C^0(\overline{\Omega};\ell^1(\mathbb{R}^n))} \triangleq \sup_{\bm{z}\in\overline{\Omega}} \norm{\bm{f}(\bm{z})}_{\ell^1(\mathbb{R}^n)} , && \norm{\bm{f}(\cdot)}_{C^0(\overline{\Omega};\ell^2(\mathbb{R}^n))} \triangleq \sup_{\bm{z}\in\overline{\Omega}}\norm{\bm{f}(\bm{z})}_{\ell^2(\mathbb{R}^n)}.
\end{align}
$C^0(\overline{\Omega};\mathbb{R}^n)$ is a Banach space naturally equipped with norm $\norm{\bm{f}(\cdot)}_{C^0(\overline{\Omega};\mathbb{R}^n)}$, where $\norm{\bm{f}(\cdot)}_{C^0(\overline{\Omega};\mathbb{R}^n)}$ is any of the two equivalent norms in Eq.~\eqref{eq:normsonC0}.



\section{Model}\label{ect:Model}

\begin{figure}
\centering
\includegraphics[width=1\linewidth]{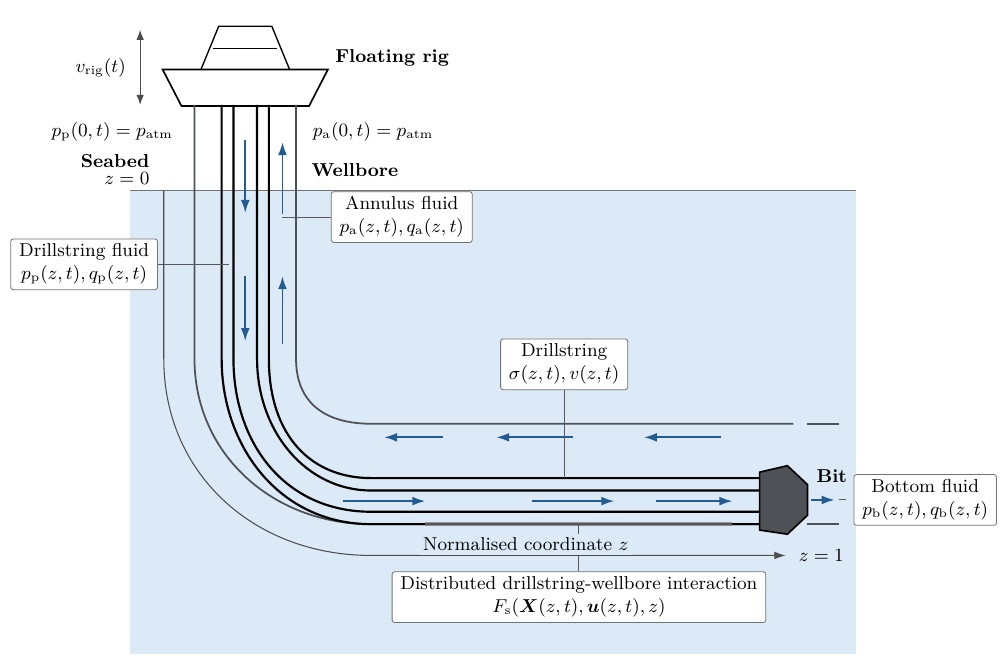} 
\caption{Schematic of the system with hydraulic drillstring variables $(p\ped{p}(z,t),q\ped{p}(z,t)) \in \mathbb{R}^2$, annulus variables $(p\ped{a}(z,t),q\ped{a}(z,t)) \in \mathbb{R}^2$, bottom fluid variables $(p\ped{b}(z,t),q\ped{b}(z,t)) \in \mathbb{R}^2$, and structural drillstring variables $(\sigma(z,t), v(z,t)) \in \mathbb{R}^2$.}
\label{fig:SchemDrill}
\end{figure}

This paper considers a nonlinear model comprising coupled one-dimensional transmission lines describing the dynamics of the mud pressure $p\ped{p}(z,t) \in \mathbb{R}$ and volumetric flow $q\ped{p}(z,t) \in \mathbb{R}$ inside the drillstring, the mud pressure $p\ped{a}(z,t) \in \mathbb{R}$ and volumetric flow $q\ped{a}(z,t) \in \mathbb{R}$ inside the annulus, and the mud pressure $p\ped{b}(z,t) \in \mathbb{R}$ and volumetric flow $q\ped{b}(z,t) \in \mathbb{R}$ below the drill bit, together with the drillstring deformation, represented by the axial stress $\sigma(x,t) \in \mathbb{R}$ and velocity $v(x,t) \in \mathbb{R}$. The coordinate $z \in [0,1]$ is measured from the rig to the bit. The governing equations -- formulated as systems of nonlinear, hyperbolic ODE--PDEs -- are detailed in Sects.~\ref{sect:pipeNDs} and~\ref{sect:ODE}. For ease of notation, the state variables are collected in the sequel into the vector $\mathbb{R}^8 \ni \bm{u}(z,t) \triangleq [p\ped{p}(z,t) \; q\ped{p}(z,t) \; p\ped{a}(z,t) \; q\ped{a}(z,t) \; p\ped{b}(z,t) \; q\ped{b}(z,t) \; \sigma(z,t) \; v(z,t)]^{\mathrm{T}}$. On the other hand, $ \bm{X}(z,t) \in \mathbb{R}^{n_X}$ denotes the vector of friction variables, whose meaning will be clarified in Sect.~\ref{sect:ODE}. A schematic illustration of the drillstring system is offered in Fig.~\ref{fig:SchemDrill}.

It is worth clarifying that the derivation of the following equations may be carried out analogously to \cite{Landet2013}, and is not repeated here for brevity.

\subsection{Coupled drillstring and fluid dynamics}\label{sect:pipeNDs}
The first two sets of PDEs describe the coupled evolution of the pressure and volumetric flow inside the drillstring and annulus, rescaled on a unit domain:
\begin{subequations}\label{eq:PDE1}
\begin{align}
\dpd{p\ped{p}(z,t)}{t} & = -\dfrac{\beta}{A\ped{p}(z)L\ped{s}}\dpd{q\ped{p}(z,t)}{z}, \\
\begin{split}
\dpd{q\ped{p}(z,t)}{t} & = -\dfrac{A\ped{p}(z)}{\rho L\ped{s}}\dpd{p\ped{p}(z,t)}{z}+f\ped{p}\bigl(\bm{X}(z,t),\bm{u}(z,t),z\bigr)
\end{split} \label{eq:qp}\\
\dpd{p\ped{a}(z,t)}{t} & = -\dfrac{\beta}{A\ped{a}(z)L\ped{s}}\dpd{q\ped{a}(z,t)}{z}, \\
\begin{split}
\dpd{q\ped{a}(z,t)}{t} & = -\dfrac{A\ped{a}(z)}{\rho L\ped{s}}\dpd{p\ped{a}(z,t)}{z} +f\ped{a}\bigl(\bm{X}(z,t),\bm{u}(z,t),z\bigr), \quad (z,t) \in (0,1)\times (0,T),
\end{split} \label{eq:qa}
\end{align}
\end{subequations}
with $f\ped{p}(\bm{X}(z,t),\bm{u}(z,t),z), f\ped{a}(\bm{X}(z,t),\bm{u}(z,t),z) \in \mathbb{R}$ defined as
\begin{subequations}\label{eqs:fpfa}
\begin{align}
\begin{split}
f\ped{p}\bigl(\bm{X}(z,t),\bm{u}(z,t),z\bigr) & \triangleq  A\ped{p}(z)\mathrm{g}\cos\theta\ped{p}(z) -\dfrac{A\ped{p}(z)}{\rho}D\ped{p}\Bigl(v\ped{rel,p}\bigl(q\ped{p}(z,t), v(z,t),z\bigr),z\Bigr), 
\end{split} \\
\begin{split}
f\ped{a}\bigl(\bm{X}(z,t),\bm{u}(z,t),z\bigr) & \triangleq  A\ped{a}(z)\mathrm{g}\cos\theta\ped{p}(z) \\
& \quad -\dfrac{A\ped{a}(z)}{\rho}\biggl[D\ped{a,i}\Bigl(v\ped{rel,a}\bigl(q\ped{a}(z,t), v(z,t),z\bigr),z\Bigr)+D\ped{a,o}\Bigl(v\ped{rel,a}\bigl(q\ped{a}(z,t), v(z,t),z\bigr),z\Bigr)\biggr].
\end{split}
\end{align}
\end{subequations}
In Eqs.~\eqref{eq:PDE1} and~\eqref{eqs:fpfa}, $L\ped{s} \in \mathbb{R}_{>0}$ indicates the drillstring length, $A\ped{p}, A\ped{a} \in C^1([0,1];\mathbb{R}_{>0})$ are the cross-sectional areas flow in the drillstring and annulus pipes, respectively, $\rho \in \mathbb{R}$ is the fluid density, $\mathrm{g}$ the acceleration of gravity, and $\theta\ped{p}(z) \in \mathbb{R}$ is the direction of the drillstring relative to gravity. Finally, the terms $D\ped{p}(v\ped{rel,p},z)$, $D\ped{a,i}(v\ped{rel,a},z)$, and $D\ped{a,o}(v\ped{rel,a},z)$ represent, respectively, the viscous drags exerted on the fluid from the internal string wall, on the fluid from the external drillstring wall, and on the fluid from the well wall. The velocities $v\ped{p}(q\ped{p},z)$ and $v\ped{a}(q\ped{a},z)$ in~\eqref{eq:qp} and~\eqref{eq:qa} are calculated as
\begin{subequations}
\begin{align}
v\ped{p}(q\ped{p},z) = \frac{q\ped{p}}{A\ped{p}(z)}, \\
v\ped{a}(q\ped{a},z) = \frac{q\ped{a}}{A\ped{a}(z)}.
\end{align}
\end{subequations}
A second set of nonlinear, hyperbolic PDEs describes the dynamics of the drillstring deformation on a unit domain, in terms of stresses and velocity: 
\begin{subequations}\label{eq:PDE2}
\begin{align}
\dpd{\sigma(z,t)}{t} & = \dfrac{E}{L\ped{s}}\dpd{v(z,t)}{z}, \\
\begin{split}
\dpd{v(z,t)}{t} & = \dfrac{1}{\rho\ped{s}L\ped{s}}\dpd{\sigma(z,t)}{z}+f\ped{s}\bigl(\bm{X}(z,t),\bm{u}(z,t),z\bigr), \quad (z,t) \in (0,1) \times (0,T),
\end{split}
\end{align}
\end{subequations}
with $f\ped{s}(\bm{X}(z,t),\bm{u}(z,t),z) \in \mathbb{R}$ defined as
\begin{align}\label{eq:fsPD}
\begin{split}
f\ped{s}\bigl(\bm{X}(z,t),\bm{u}(z,t),z\bigr) & \triangleq \dfrac{\sigma(z,t)}{\rho\ped{s}A\ped{s}(z) L\ped{s}}\dod{A\ped{s}(z)}{z}+\dfrac{F\ped{s}\bigl(\bm{X}(z,t),\bm{u}(z,t),z\bigr)}{\rho\ped{s}A\ped{s}(z)} +\mathrm{g}\cos\theta\ped{p}(z)\\
& \quad + \dfrac{1}{\rho\ped{s}A\ped{s}(z)L\ped{s}}\biggl( p\ped{a}(z,t)\dod{A\ped{i}(z)}{z}-p\ped{p}(z,t)\dod{A\ped{p}(z)}{z}\biggr) \\
 & \quad +\dfrac{A\ped{p}(z)D\ped{p}\Bigl(v\ped{rel,p}\bigl(q\ped{p}(z,t), v(z,t),z\bigr),z\Bigr)+A\ped{a}(z)D\ped{a,i}\Bigl(v\ped{rel,a}\bigl(q\ped{a}(z,t), v(z,t),z\bigr),z\Bigr)}{A\ped{s}(z)\rho\ped{s}},
\end{split}
\end{align}
where $E \in \mathbb{R}_{>0}$ is the Young's modulus, $\rho\ped{s} \in \mathbb{R}_{>0}$ the density of the string, $F\ped{s}(\bm{X},\bm{u},z)$ is the contact friction between the string and the well wall, $D\ped{p}(v\ped{rel,p},z)$ the viscous drag, and the internal and external areas of the drillstring pipe, $A\ped{i}, A\ped{p} \in C^1([0,1];\mathbb{R}_{>0})$, are computed as
\begin{align}
A\ped{i}(z) & = \dfrac{\pi}{4}d\ped{a,i}^2(z), && A\ped{p}(z) = \dfrac{\pi}{4}d\ped{p}^2(z),
\end{align}
being $d\ped{a,i}, d\ped{p} \in C^1([0,1];\mathbb{R}_{>0})$ the inner annulus and inner pipe diameters, respectively, and $C^1([0,1];\mathbb{R}_{>0}) \ni A\ped{s} = A\ped{i}-A\ped{p}$.
Finally, below the string, the equations governing the pressure and volumetric flow of the mud inside the annulus (bottom hole) are
\begin{subequations}\label{eq:PDE3}
\begin{align}
\dpd{p\ped{b}(z,t)}{t} & = -\dfrac{\beta}{A\ped{b}(z)L\ped{b}}\dpd{q\ped{b}(z,t)}{z}, \\
\begin{split}
\dpd{q\ped{b}(z,t)}{t} & = -\dfrac{A\ped{b}(z)}{\rho L\ped{b}}\dpd{p\ped{b}(z,t)}{z}+f\ped{b}\bigl(\bm{X}(z,t),\bm{u}(z,t),z\bigr), \quad (z,t) \in (0,1)\times (0,T),
\end{split} 
\end{align}
\end{subequations}
with $f\ped{b}(\bm{X}(z,t),\bm{u}(z,t),z) \in \mathbb{R}$ reading
\begin{align}\label{eq:fb}
\begin{split}
f\ped{b}\bigl(\bm{X}(z,t),\bm{u}(z,t),z\bigr) & \triangleq  A\ped{b}(z)\mathrm{g}\cos\theta\ped{b}(z)  -\dfrac{A\ped{b}(z)}{\rho}D\ped{b}\Bigl(v\ped{rel,b}\bigl(q\ped{b}(z,t), z\bigr),z\Bigr),
\end{split}
\end{align}
where $A\ped{b}\in C^1([0,1];\mathbb{R}_{>0})$ denotes the cross-sectional are of the pipe, $\theta\ped{p}(z) \in \mathbb{R}$ is the direction of the bit relative to gravity, and $D\ped{b}(v\ped{rel,b},z)$ the viscous drag. As previously, the velocity $v\ped{b}(q\ped{b},z) \in \mathbb{R}$ is computed as
\begin{align}
v\ped{b}(q\ped{b},z) = \dfrac{q\ped{b}}{A\ped{b}(z)}.
\end{align}
The viscous drags exerted on the fluid, appearing in Eqs.~\eqref{eqs:fpfa},~\eqref{eq:fsPD}, and~\eqref{eq:fb}, are modelled as
\begin{align}\label{eq;dragModel}
D_k(v_{\textnormal{rel},k},z) & = \dfrac{2\rho v_{\textnormal{rel},k}^2 f_k(v_{\textnormal{rel},k},z)}{d_{\textnormal{hyd},k}(z)}, \quad k = \textnormal{p},\textnormal{a},\textnormal{b},
\end{align}
with the hydraulic diameters given by
\begin{subequations}
\begin{align}
d\ped{hyd,p}(z) &= d\ped{p}(z), \\
d\ped{hyd,a}(z) &= d\ped{a,o}(z)-d\ped{a,i}(z), \\
d\ped{hyd,b}(z) &= d\ped{b}(z),
\end{align}
\end{subequations}
where $d\ped{a,o}(z), d\ped{b}(z) \in \mathbb{R}_{>0}$ denote outer annulus and bottom-hole (below the string) diameters, and $f\ped{p}( v\ped{rel,p},z)$, $f\ped{a}( v\ped{rel,a},z)$, $f\ped{b}( v\ped{b},z)$ are the Fanning friction factors.

In Eq.~\eqref{eq;dragModel}, $v\ped{rel,b}(q\ped{b},z) = v\ped{b}(q\ped{b},z)$, the relative velocity between the pipe and the fluid in the drill string reads
\begin{align}
v\ped{rel,p}(q\ped{p},v,z) = v\ped{p}(q\ped{p},z)-v, 
\end{align}
and the relative velocity between the fluid and the well wall for the annulus is approximated as
\begin{align}
v\ped{rel,a}(q\ped{a},v,z) \approx v\ped{a}(p\ped{a},z)-\Bigggl( \dfrac{1}{2\ln\Bigl(\frac{d\ped{a,o}(z)}{d\ped{a,i}(z)}\Bigr)}- \dfrac{d\ped{a,i}^2(z)}{d\ped{a,o}^2(z)-d\ped{a,i}^2(z)}\Bigggr)v.
\end{align}
Starting with Eq.~\eqref{eq;dragModel}, the drags exerted on the fluid in the annulus by the drill string and well wall may be calculated as
\begin{subequations}
\begin{align}
D\ped{a,i}(v\ped{rel,a},z) & = \dfrac{d\ped{a,i}(z)}{d\ped{a,o}(z) + d\ped{a,i}(z)}D\ped{a}(v\ped{rel,a},z), \\
D\ped{a,o}(v\ped{rel,a},z) & = \dfrac{d\ped{a,o}(z)}{d\ped{a,o}(z) + d\ped{a,i}(z)}D\ped{a}(v\ped{rel,a},z).
\end{align}
\end{subequations}

\subsection{Friction dynamics}\label{sect:ODE}
A last set of equations is needed to model the contact force between the drillstring and well wall.
Inspired by recent contributions \cite{FrBD1,FrBD2,FrBD}, this paper adopts a distributed \emph{friction with bristle dynamics} (FrBD) model with Generalised Kelvin-Voigt rheology\footnote{Alternatively, the formulation adopted in this paper can be recast into an equivalent Generalised Maxwell representation. Although the equivalence between linear rheological models is well established \cite{FrBD2}, the Kelvin-Voigt formulation is adopted here because of its simpler implementation.}.

Accordingly, the governing equations may be derived in the form of spatially varying ODEs as
\begin{subequations}\label{eq:ODEi}
\begin{align}
\begin{split}
\dpd{f(z,t)}{t} & =-\bar{k}_0(z)g\bigl(v(z,t),z\bigr)f(z,t)  - \bar{k}_0(z)v(z,t)\\
& \quad-\bar{k}_0(z)\sum_{i=1}^{n_1}\dpd{x_{i,1}(z,t)}{t} + \bar{k}_0(z)\sum_{i=1}^{n_2}\dpd{x_{i,2}(z,t)}{t}, 
\end{split}\label{eq:ODEf} \\
\dpd{x_{i,1}(z,t)}{t} & = -\dfrac{x_{i,1}(z,t)}{\tau_{i,1}(z)} + \dfrac{f(z,t)}{\bar{c}_{i,1}(z)}, \quad i \in \{1,\dots,n_1\}, \label{eq:ODEx1} \\
\dpd{x_{i,2}(z,t)}{t} & = -\dfrac{x_{i,2}(z,t)}{\tau_{i,2}(z)} - \dfrac{f(z,t)}{\bar{c}_{i,2}(z)}, \quad i \in \{1,\dots,n_2\},\; (z,t) \in (0,1)\times (0,T),\label{eq:ODEx2}
\end{align}
\end{subequations}
being $f(z,t) \in \mathbb{R}$ the normalised friction force, $x_{i,1}(z,t) \in \mathbb{R}$, $i \in \{1,\dots,n_1\}$, and $x_{i,2}(z,t) \in \mathbb{R}$, $i \in \{1,\dots,n_2\}$, internal states describing the deformation of the dissipative branches of the element, $\bar{k}_0 \in C^0([0,1];\mathbb{R}_{>0})$ the normalised micro-stiffnesses of the connection, $\tau_{i,1}, \bar{c}_{i,1} \in C^0([0,1];\mathbb{R}_{>0})$, $i \in \{1,\dots,n_1\}$, the time relaxation and normalised damping constants of the dissipative branches of the first element, $\tau_{i,2}, \bar{c}_{i,2} \in C^0([0,1];\mathbb{R}_{>0})$, $i \in \{1,\dots,n_2\}$, the time relaxation and normalised damping constants of the dissipative branches of the second element, and $g (\cdot,z)$ a nonlinear function of the relative velocity, which captures the frictional interaction between the contacting bodies. Often, $g(v,z)$ is postulated in the form
\begin{align}\label{eq:g_mu}
g(v,z) = \dfrac{\abs{v}_\varepsilon}{\mu(v,z)},
\end{align}
where $\mu : \mathbb{R}\times[0,1]\to [\mu\ped{min},\infty)$ denotes the friction coefficient, with $\mu\ped{min}\in \mathbb{R}_{>0}$, and the function $\abs{\cdot}_\varepsilon \in C^0(\mathbb{R};\mathbb{R}_{\geq 0})$, with $\varepsilon \in \mathbb{R}_{\geq0}$, is a regularisation of the absolute value $\abs{\cdot}$ for $\varepsilon\in \mathbb{R}_{>0}$, converging uniformly to $\abs{\cdot}$ in $C^0(\mathbb{R};\mathbb{R}_{\geq 0})$ for $\varepsilon \to 0$ (e.g., $\abs{v}_\varepsilon= \sqrt{v^2 +\varepsilon}$), and with $\abs{\cdot}_\varepsilon \in C^1(\mathbb{R};\mathbb{R}_{\geq 0})$ for $\varepsilon \in \mathbb{R}_{>0}$ \cite{FrBD}.

Equation~\eqref{eq:ODEi} describes the generation of frictional forces at the interface between contacting bodies as a dynamical process governed by the viscoelastic deformation of bristle-like elements. This conceptual framework is closely related to other phenomenological representations of friction, including the Dahl and LuGre models. However, although the FrBD formulation generalises these models, it is derived entirely from physical principles rather than being introduced empirically. Such a physical foundation is also reflected in its mathematical properties (the model is passive for any choice of parameters \cite{FrBD1,FrBD2,FrBD}).

It is worth noting that, although distributed, Eq.~\eqref{eq:ODEi} comprises $n_X = n_1+n_2+1$ nonlinear ODEs, making the interconnection with Eqs.~\eqref{eq:PDE1},~\eqref{eq:PDE2}, and~\eqref{eq:PDE3} a semilinear, hyperbolic ODE--PDE system. In particular, defining the vector $\mathbb{R}^{n_X}\ni \bm{X}(z,t) \triangleq [f(z,t)\; x_{1,1}(z,t)\; \dots \; x_{n_1,1}(z,t)\; x_{1,2}(z,t)\; \dots \; x_{n_2,2}(z,t)]^{\mathrm{T}}$, the contact friction between the string and the well wall showing in Eq.~\eqref{eq:fsPD} may be modelled as
\begin{align}\label{eq.FS}
F\ped{s}\bigl(\bm{X}(z,t), \bm{u}(z,t), z\bigr) =F\ped{n}\bigl(\bm{u}(z,t),z\bigr) f(z,t),
\end{align}
where the normal contact load per unit of length, $F\ped{n}\bigl(\bm{u}(z,t),z\bigr) \in \mathbb{R}_{\geq 0}$, is modelled as \cite{Strecker2017}
\begin{align}
F\ped{n}\bigl(\bm{u}(z,t),z\bigr) & = (\rho\ped{s}-\rho)\mathrm{g}A\ped{s}(z)\sin\theta\ped{p}(z) + \dfrac{A\ped{s}(z)\sigma\ped{e}\bigl(\bm{u}(z,t),z\bigr)}{L\ped{s}}\dod{\theta\ped{p}(z)}{z}, \label{eq:nomN} 
\end{align}
with the effective tension $A\ped{s}(z)\sigma\ped{e}(\bm{u}(z,t),z) \in \mathbb{R}$ reading \cite{Aarsnes2,Sheppard}
\begin{align}\label{eq:effstress}
A\ped{s}(z)\sigma\ped{e}\bigl(\bm{u}(z,t),z\bigr) & = A\ped{s}(z)\sigma(z,t) + A\ped{i}(z)p\ped{a}(z,t)-A\ped{p}(z)p\ped{p}(z,t).
\end{align}

\subsection{Boundary conditions}
Equations~\eqref{eq:PDE1},~\eqref{eq:PDE2}, and~\eqref{eq:PDE3} constitute a system of eight coupled hyperbolic PDEs, and require the specification of appropriate \emph{boundary conditions} (BCs). In particular, the following BCs are prescribed in this paper:
\begin{subequations}\label{eq:BCs}
\begin{align}
p\ped{a}(0,t) & = p\ped{ext}(t) = p\ped{atm}, \\
p\ped{p}(0,t) & = p\ped{ext}(t) = p\ped{atm}, \\
v(0,t) & = v\ped{rig}(t), \\
q\ped{b}(1,t) & = 0, \\
p\ped{b}(0,t) &= p\ped{a}(1,t), \\
\sigma(1,t) &= \dfrac{k\ped{bit}F\ped{bit}(t)-A\ped{i}(1)p\ped{a}(1,t)+A\ped{p}(1)p\ped{p}(1,t)}{A\ped{s}(1)}, \label{eq:bCsigma} \\
q\ped{b}(0,t) & = q\ped{a}(1,t) + q\ped{p}(1,t) + A\ped{s}(1)v(1,t), \quad t \in (0,T),
\end{align}
\end{subequations}
where $p\ped{atm}\in \mathbb{R}_{>0}$ indicates the atmospheric pressure, $p\ped{ext}(t) \in \mathbb{R}$ the applied external pressure, $v\ped{rig}(t) \in \mathbb{R}$ the velocity of the rig, and $F\ped{bit}(t) \in \mathbb{R}_{>0}$ is the bit force, regarded as exogenous inputs, and $k\ped{bit}\in \{0,1\}$ is a dummy variable introduced to distinguish between the open/closed pipe ($k\ped{bit} = 0$) and bit ($k\ped{bit} = 1$) configurations.

In addition to Eq.~\eqref{eq:BCs}, another BC is needed that depends on the system's configuration. Two different cases are considered in this paper:
\begin{itemize}
\item \textbf{Open pipe:} \begin{align} p\ped{a}(1,t) = p\ped{p}(1,t), \quad t \in (0,T),  \label{BC:open}\end{align}
\item \textbf{Closed pipe/bit:} \begin{align} q\ped{p}(1,t) = A\ped{p}(1)v(1,t) + k\ped{bit}C\ped{d}A\ped{bit}\sqrt{\dfrac{2\max\bigl\{\Delta p\bigl(p\ped{p}(1,t),p\ped{a}(1,t)\bigr), 0\bigr\}}{\rho}}, \quad t \in (0,T), \label{BC:closed}\end{align}
where
\begin{align}
\Delta p(p\ped{p},p\ped{a}) \triangleq p\ped{p}-p\ped{a},\label{eq:Deltap}
\end{align}
and $C\ped{d}, A\ped{bit} \in \mathbb{R}_{>0}$ are constant parameters.
\end{itemize}

Equations~\eqref{eq:BCs}-\eqref{eq:Deltap} conclude the presentation of the coupled hydraulic-structural model. Next, Sect.~\ref{sect:RiemETeho} recasts Eqs.~\eqref{eq:PDE1},~\eqref{eq:PDE2},~\eqref{eq:PDE3}, and~\eqref{eq:ODEi} into a form that is more amenable to mathematical analysis.

\section{System in Riemann variables and theoretical analysis}\label{sect:RiemETeho}
The formulation introduced in Sect.~\ref{ect:Model} provides a clear physical description of the hydraulic-structural system, but is not particularly suited for well-posedness and stability studies. To address this limitation, Sect.~\ref{sect:riemann} reformulates the PDEs~\eqref{eq:PDE1},~\eqref{eq:PDE2}, and~\eqref{eq:PDE3} in terms of Riemann variables, yielding a compact representation that is amenable to the rigorous analysis conducted in Sect.~\ref{sect:HeoNal}.

\subsection{ODE--PDE system in Riemann variables}\label{sect:riemann}
For what follows, it is convenient to introduce the vectors $\mathbb{R}^2 \ni \bm{u}_k(z,t) = [p_k(z,t)\; q_k(z,t)]^{\mathrm{T}}$, $k = \textnormal{p},\textnormal{a},\textnormal{b}$, and $\mathbb{R}^2 \ni \bm{u}\ped{s}(z,t) = [\sigma(z,t)\; v(z,t)]^{\mathrm{T}}$, so that $\mathbb{R}^8\ni \bm{u}(z,t) = [\bm{u}\ped{p}^{\mathrm{T}}(z,t)\;  \bm{u}\ped{a}^{\mathrm{T}}(z,t)\; \bm{u}\ped{b}^{\mathrm{T}}(z,t)\; \bm{u}\ped{s}^{\mathrm{T}}(z,t)]^{\mathrm{T}}$.

Accordingly, the Riemann variables $\mathbb{R}^2 \ni \bm{w}_k(z,t) = [w_k^+(z,t)\; w_k^-(z,t)]^{\mathrm{T}}$ for the PDE subsystems~\eqref{eq:PDE1},~\eqref{eq:PDE2}, and~\eqref{eq:PDE3} may be deduced as
\begin{subequations}\label{eq:RiemannS}
\begin{align}
\bm{w}_k(z,t) &= \begin{bmatrix} w_k^+(z,t)\\ w_k^-(z,t) \end{bmatrix} = \mathbf{R}_k(z)\bm{u}_k(z,t), \quad k = \textnormal{p},\textnormal{a},\textnormal{b}, \\
\bm{w}\ped{s}(z,t) &= \begin{bmatrix} w\ped{s}^+(z,t) \\ w\ped{s}^-(z,t) \end{bmatrix}= \mathbf{R}\ped{s}\bm{u}\ped{s}(z,t), 
\end{align}
\end{subequations}
where the spatially varying matrices $\mathbf{R}_k(z) \in \mathbf{GL}_2(\mathbb{R})$, $k = \textnormal{p}, \textnormal{a}, \textnormal{b}$, and $\mathbf{R}\ped{s} \in \mathbf{GL}_2(\mathbb{R})$ are given by
\begin{subequations}
\begin{align}
 \mathbf{R}_k(z) & = \begin{bmatrix} \dfrac{A_k(z)}{\sqrt{\beta \rho}} & 1 \\- \dfrac{A_k(z)}{\sqrt{\beta \rho}} & 1\end{bmatrix}, \quad k = \textnormal{p},\textnormal{a},\textnormal{b}, \\
\mathbf{R}\ped{s} &=  \begin{bmatrix}-\dfrac{1}{\sqrt{E \rho\ped{s}}} & 1 \\ \dfrac{1}{\sqrt{E \rho\ped{s}}} & 1\end{bmatrix}. \label{eq:matrixRs}
\end{align}
\end{subequations}
The inverse transformations of Eq.~\eqref{eq:RiemannS} are $\bm{u}_k(z,t) = \mathbf{L}_k(z)\bm{w}_k(z,t)$ for $k = \textnormal{p},\textnormal{a},\textnormal{b}$, and $\bm{u}\ped{s}(z,t) = \mathbf{L}\ped{s}\bm{w}\ped{s}(z,t)$, with
\begin{subequations}
\begin{align}
\mathbf{L}_k(z) & \triangleq \mathbf{R}_k^{-1}(z) = \dfrac{1}{2}\begin{bmatrix} \dfrac{\sqrt{\beta \rho}}{A_k(z)} & - \dfrac{\sqrt{\beta \rho}}{A_k(z)} \\ 1 & 1\end{bmatrix}, \quad  k = \textnormal{p},\textnormal{a},\textnormal{b}, \\
\mathbf{L}\ped{s} & \triangleq \mathbf{R}\ped{s}^{-1} =  \dfrac{1}{2} \begin{bmatrix} -\sqrt{E \rho\ped{s}} & \sqrt{E \rho\ped{s}} \\ 1 & 1 \end{bmatrix}. \label{eq:invSigmaV}
\end{align}
\end{subequations}
For ease of notation, it is also useful to assemble the matrices
\begin{subequations}
\begin{align}
\mathbf{R}(z) & \triangleq \begin{bmatrix}1 & 0 & 0 & 0 & 0 & 0 & 0 & 0 \\ 0 & 0 & 1 & 0 & 0 & 0 & 0 & 0 \\ 0 & 0 & 0 & 0 & 1 & 0 & 0 & 0  \\ 0 & 0 & 0 & 0 & 0 & 0 & 1 & 0 \\0 & 1 & 0 & 0 & 0 & 0 & 0 & 0\\0 & 0 & 0 & 1 & 0 & 0 & 0 & 0\\0 & 0 & 0 & 0 & 0 & 1 & 0 & 0 \\0 & 0 & 0 & 0 & 0 & 0 & 0 & 1 \end{bmatrix}\begin{bmatrix} \mathbf{R}\ped{p}(z) & \mathbf{0} & \mathbf{0} & \mathbf{0} \\
\mathbf{0} &  \mathbf{R}\ped{a}(z) & \mathbf{0} & \mathbf{0} \\
\mathbf{0} & \mathbf{0} & \mathbf{R}\ped{b}(z) & \mathbf{0} \\
 \mathbf{0} & \mathbf{0} & \mathbf{0} & \mathbf{R}\ped{s}\end{bmatrix}, \\
\mathbf{L}(z) & \triangleq \begin{bmatrix} \mathbf{L}\ped{p}(z) & \mathbf{0} & \mathbf{0} & \mathbf{0} \\
\mathbf{0} &  \mathbf{L}\ped{a}(z) & \mathbf{0} & \mathbf{0} \\
\mathbf{0} & \mathbf{0} & \mathbf{L}\ped{b}(z) & \mathbf{0} \\
 \mathbf{0} & \mathbf{0} & \mathbf{0} & \mathbf{L}\ped{s}\end{bmatrix} \begin{bmatrix}1 & 0 & 0 & 0 & 0 & 0 & 0 & 0 \\ 0 & 0 & 0 & 0 & 1 & 0 & 0 & 0 \\ 0 & 1 & 0 & 0 & 0 & 0 & 0 & 0 \\ 0 & 0 & 0 & 0 & 0 & 1 & 0 & 0 \\ 0 & 0 & 1 & 0 & 0 & 0 & 0 & 0 \\ 0 & 0 & 0 & 0 & 0 & 0 & 1 & 0 \\ 0 & 0 & 0 & 1 & 0 & 0 & 0 & 0 \\ 0 & 0 & 0 & 0 & 0 & 0 & 0 & 1 \end{bmatrix} = \mathbf{R}^{-1}(z), 
\end{align}
\end{subequations}
so that, collecting the Riemann variables into the vector $\mathbb{R}^8\ni \bm{w}(z,t) = [\bm{w}^{+\mathrm{T}}(z,t)\;  \bm{w}^{-\mathrm{T}}(z,t)]^{\mathrm{T}}$, with $\mathbb{R}^4 \ni \bm{w}^\pm(z,t) = [w_1^\pm(z,t)\; w_2^\pm(z,t)\; w_3^\pm(z,t) \; w_4^\pm(z,t)]^{\mathrm{T}}\triangleq [w\ped{p}^\pm(z,t)\; w\ped{a}^\pm(z,t)\; w\ped{b}^\pm(z,t)\; w\ped{s}^\pm(z,t)]^{\mathrm{T}}$, it is possible to write $\bm{w}(z,t) = \mathbf{R}(z)\bm{u}(z,t)$ and $\bm{u}(z,t)  = \mathbf{L}(z)\bm{w}(z,t)$.

Consequently, utilising the transformation~\eqref{eq:RiemannS}, combining the ODE~\eqref{eq:ODEi} with the PDEs~\eqref{eq:PDE1},~\eqref{eq:PDE2}, and~\eqref{eq:PDE3}, and setting $\mathbb{R}^{n_U}\ni \bm{U}(t) \triangleq [p\ped{ext}(t)\; v\ped{rig}(t)\; F\ped{bit}(t)]$, provides the following semilinear hyperbolic ODE--PDE interconnection:
\begin{subequations}\label{eq:ODE--PDEsystem}
\begin{align}
\dpd{\bm{X}(z,t)}{t} & = \bm{F}\bigl(\bm{X}(z,t),\bm{w}(z,t),z\bigr), \label{eq:ODE} \\
\dpd{\bm{w}(z,t)}{t} + \mathbf{\Lambda}\dpd{\bm{w}(z,t)}{z} & = \bm{G}\bigl(\bm{X}(z,t),\bm{w}(z,t),z\bigr), && (z,t) \in (0,1)\times(0,T), \label{eq:PDEsubst}\\
\begin{bmatrix} \bm{w}^+(0,t) \\ \bm{w}^-(1,t) \end{bmatrix}  & = \bm{H}\bigl(\bm{w}^+(1,t), \bm{w}^-(0,t), \bm{U}(t)\bigr), && t \in (0,T), \label{eq:BCsH} 
\end{align}
\end{subequations}
where the matrices $\mathbf{GL}_8(\mathbb{R})\cap\mathbf{Sym}_8(\mathbb{R}) \ni \mathbf{\Lambda} \succ \mathbf{0}$ are given by
\begin{align}
\mathbf{\Lambda} & \triangleq \begin{bmatrix} \mathbf{\Lambda}^+ & \mathbf{0} \\ \mathbf{0} & \mathbf{\Lambda}^{-}\end{bmatrix} = \begin{bmatrix}\mathbf{\Lambda}_0 & \mathbf{0} \\ \mathbf{0} & -\mathbf{\Lambda}_0\end{bmatrix}, && \textnormal{with} && 
\mathbf{\Lambda}_0 =\begin{bmatrix} \lambda_1 & 0 & 0 & 0 \\ 0 & \lambda_2 & 0 & 0 \\ 0 & 0 & \lambda_3 & 0 \\ 0 & 0 & 0 & \lambda_4\end{bmatrix}\triangleq \begin{bmatrix} c\ped{p} & 0 & 0 & 0 \\
0 & c\ped{a} & 0 & 0 \\
0 & 0 & c\ped{b} & 0 \\
0 & 0 & 0 & c\ped{s} 
\end{bmatrix},
\end{align} 
and $c_k \in \mathbb{R}_{>0}$, $k = \textnormal{p},\textnormal{a},\textnormal{b}, \textnormal{s}$, read
\begin{align}
c\ped{p} & = c\ped{a} \triangleq \dfrac{1}{L\ped{s}}\sqrt{\dfrac{\beta}{\rho}}, && c\ped{b}  \triangleq \dfrac{1}{L\ped{b}}\sqrt{\dfrac{\beta}{\rho}}, && c\ped{s}  \triangleq \dfrac{1}{L\ped{s}}\sqrt{\dfrac{E}{\rho\ped{s}}}.
\end{align}
In Eq.~\eqref{eq:ODE}, the ODE source function $\bm{F} : \mathbb{R}^{n_{X}+8}\times[0,1] \to \mathbb{R}^{n_{X}}$, given by $\in \mathbb{R}^{n_X}\ni \bm{F}(\bm{X},\bm{w},z) = [F_0(\bm{X},\bm{w},z)\; F_1(\bm{X},\bm{w},z) \; \dots \; F_{n_1}(\bm{X},\bm{w},z) \; F_{n_1+1}(\bm{X},\bm{w},z)\; \dots \; F_{n_1+n_2}(\bm{X},\bm{w},z)]^{\mathrm{T}}$ reads, in components
\begin{subequations}
\begin{align}
\begin{split}
F_0(\bm{X},\bm{w},z) & \triangleq -\bar{k}_0(z)g\Biggl( \dfrac{w_4^+ +w_4^-}{2},z\Biggr)f -\bar{k}_0(z)\dfrac{w_4^+ +w_4^-}{2} \\
& \quad -\bar{k}_0(z)\sum_{i=1}^{n_1}F_i(\bm{X},\bm{w},z) + \bar{k}_0(z)\sum_{i=1}^{n_2}F_{n_1+i}(\bm{X},\bm{w},z), 
\end{split}\\
F_i(\bm{X},\bm{w},z) & \triangleq -\dfrac{x_{i,1}}{\tau_{i,1}(z)} + \dfrac{f}{\bar{c}_{i,1}(z)}, \quad i \in \{1,\dots,n_1\}, \\
F_{n_1+i}(\bm{X},\bm{w},z) & \triangleq -\dfrac{x_{i,2}}{\tau_{i,2}(z)} - \dfrac{f}{\bar{c}_{i,2}(z)}, \quad i \in \{1,\dots,n_2\}, 
\end{align}
\end{subequations}
whereas the PDE source function $\bm{G} : \mathbb{R}^{n_{X}+8}\times[0,1] \to \mathbb{R}^8$ appearing in Eq.~\eqref{eq:PDEsubst}, and decomposed in the following as $\mathbb{R}^8 \ni \bm{G}(\bm{X},\bm{w},z) = [\bm{G}^{+\mathrm{T}}(\bm{X},\bm{w},z)\; \bm{G}^{-\mathrm{T}}(\bm{X},\bm{w},z)]^{\mathrm{T}}$, with $\mathbb{R}^4 \ni \bm{G}^{\pm}(\bm{X},\bm{w},z) = [G_1^\pm(\bm{X},\bm{w},z)\; G_2^\pm(\bm{X},\bm{w},z)\;G_3^\pm(\bm{X},\bm{w},z)\;G_4^\pm(\bm{X},\bm{w},z)]^{\mathrm{T}}$ is defined in components as
\begin{subequations}
\begin{align}
G_1^+(\bm{X},\bm{w},z) & = G_1^{-}(\bm{X},\bm{w},z) \triangleq f\ped{p}\bigl(\bm{X},\mathbf{L}(z)\bm{w},z\bigr)+\dfrac{w_1^+-w_1^{-}}{2A\ped{p}(z) L\ped{s}}\sqrt{\dfrac{\beta}{\rho}}\dod{A\ped{p}(z)}{z}, \\
G_2^+(\bm{X},\bm{w},z) & = G_2^{-}(\bm{X},\bm{w},z) \triangleq f\ped{a}\bigl(\bm{X},\mathbf{L}(z)\bm{w},z\bigr)+\dfrac{w_2^+-w_2^{-}}{2A\ped{a}(z) L\ped{s}}\sqrt{\dfrac{\beta}{\rho}}\dod{A\ped{a}(z)}{z}, \\
G_3^+(\bm{X},\bm{w},z) & = G_3^{-}(\bm{X},\bm{w},z) \triangleq f\ped{b}\bigl(\bm{X},\mathbf{L}(z)\bm{w},z\bigr)+\dfrac{w_3^+-w_3}{2A\ped{b}(z) L\ped{b}}\sqrt{\dfrac{\beta}{\rho}}\dod{A\ped{b}(z)}{z}, \\
G_4^+(\bm{X},\bm{w},z) & = G_4^{-}(\bm{X},\bm{w},z) \triangleq f\ped{s}\bigl(\bm{X},\mathbf{L}(z)\bm{w},z\bigr).
\end{align}
\end{subequations}
Finally, the boundary map $\bm{H} : \mathbb{R}^{8+n_U} \to \mathbb{R}^8$ in Eq.~\eqref{eq:BCs} prescribes the relationships between the ingoing and outgoing characteristics, $\bm{w}^+(z,t)$ and $\bm{w}^-(z,t)$ respectively, and the input $\bm{U}(t)$. The specific structure of $\bm{H}(\bm{w}^+(1),\bm{w}(0),\bm{U})$ must be deduced from the BCs~\eqref{eq:BCs},~\eqref{BC:open}, and~\eqref{BC:closed}, which might not yield a closed-form solution in terms of Riemann variables (as is the case for the bit configuration). However, the BCs corresponding to the open and closed pipe configurations lead to the linear expression
\begin{align}\label{eq:HLinear}
\bm{H}\bigl(\bm{w}^+(1),\bm{w}^-(0),\bm{U}(t)\bigr) = \mathbf{H}\begin{bmatrix}\bm{w}^+(1)\\ \bm{w}^-(0) \end{bmatrix} + \mathbf{B}\bm{U},
\end{align}
with $\mathbf{H}\in \mathbf{M}_8(\mathbb{R})$ and $\mathbf{B} \in \mathbf{M}_{8\times n_U}(\mathbb{R})$.
The complete expressions for the matrices $\mathbf{H}$ and $\mathbf{B}$ are omitted here, but reported in Appendix~\ref{app:closedMAP} for completeness.

\subsection{Theoretical analysis}\label{sect:HeoNal}
The present section focuses on the mathematical analysis of the semilinear hyperbolic ODE--PDE interconnection~\eqref{eq:ODE--PDEsystem}. Specifically, Sect.~\ref{sect:wellPos} delivers both local and global well-posedness results under appropriate assumptions. Subsequently, Sect.~\ref{sect:smallGain} investigates the role of friction on the stability of the system, within the framework offered by the small-gain theory for infinite-dimensional equations \cite{Krstic2}.

\subsubsection{Well-posedness}\label{sect:wellPos}
The hyperbolic ODE--PDE system~\eqref{eq:ODE--PDEsystem} is highly nonlinear. In particular, the distributed friction and viscous drag terms appearing in Eqs.~\eqref{eqs:fpfa},~\eqref{eq:fsPD}, and~\eqref{eq:fb} introduce sources of non-Lipschitzianity in the standard $L^2$-spaces typically used when proving well-posedness within the semigroup framework (see, for instance, \cite[Theorem 11.1.5]{Zwart} or \cite[Theorem 6.1.4]{Pazy}). To overcome this difficulty, the present paper considers the notion of \emph{generalised solutions} for PDEs, which are essentially solutions of~\eqref{eq:ODE--PDEsystem} obtained by integrating along the characteristic lines \cite{Strecker1,Strecker2}.

The first well-posedness result in this sense is enunciated in Theorem~\ref{thm:Local} below, which focuses on local solutions.
 
\begin{theorem}[Local existence and uniqueness of generalised solutions]\label{thm:Local} 
Assume that $\bm{F}\in C^0(\mathbb{R}^{n_{X}+8}\times[0,1];\mathbb{R}^{n_{X}})$ and $\bm{G} \in C^0(\mathbb{R}^{n_{X}+8}\times[0,1];\mathbb{R}^8)$ are locally Lipschitz with respect to $(\bm{X},\bm{w})$, uniformly with respect to $z\in[0,1]$, and that the boundary map $\bm{H}\in C^0(\mathbb{R}^{8+n_U};\mathbb{R}^8)$ satisfies the global Lipschitz condition
\begin{align}\label{eq:globalHLips}
\begin{split}
& \norm{\bm{H}(\bm{a}_1,\bm{b}_1,\bm{U})-\bm{H}(\bm{a}_2,\bm{b}_2,\bm{U})}_{\ell^1(\mathbb{R}^8)} \leq L_H\norm{(\bm{a}_1-\bm{a}_2,\bm{b_1}-\bm{b}_2)}_{\ell^1(\mathbb{R}^8)},\quad  (\bm{a}_1,\bm{a}_2,\bm{b_1},\bm{b_2},\bm{U}) \in \mathbb{R}^{16+n_U},
\end{split}
\end{align}
for some $L_H \in \mathbb{R}_{\geq 0}$, and 
\begin{align}\label{eq:h_0bound}
\sup_{t\in\mathbb{R}_{\geq 0}}\norm{\bm{h}_0(t)}_{\ell^1(\mathbb{R}^8)} \triangleq \sup_{t\in\mathbb{R}_{\geq 0}}\norm{\bm{H}(\bm{0},\bm{0},\bm{U}(t))}_{\ell^1(\mathbb{R}^8)}= M_H < \infty.
\end{align}
Then, for every input $\bm{U}\in C^0(\mathbb{R}_{\geq 0};\mathbb{R}^{n_U})$ and \emph{initial condition} (IC) $ (\bm{X}_0,\bm{w}_0^+,\bm{w}_0^-)  = (\bm{X}_0,\bm{w}_0)\triangleq (\bm{X}(\cdot,0),\bm{w}(\cdot,0)) \in C^0([0,1];\mathbb{R}^{n_{X}+8})$ satisfying the zeroth-order compatibility condition
\begin{align}
\begin{bmatrix} \bm{w}_0^+(0)\\ \bm{w}_0^-(1) \end{bmatrix} = \bm{H}\bigl(\bm{w}_0^+(1),\bm{w}_0^-(0),\bm{U}(0)\bigr),
\end{align}
there exists $t\ped{max} \in (0,\infty]$ such that the ODE--PDE system~\eqref{eq:ODE--PDEsystem} admits a unique \emph{generalised solution} $(\bm{X},\bm{w}) \in C^0\bigl([0,1]\times [0,t\ped{max});\mathbb{R}^{n_{X}+8})$. Moreover, if $t\ped{max} < \infty$, then
\begin{align}\label{eq:explosion}
 \lim_{t\uparrow t\ped{max}}  \norm{(\bm{X}(\cdot,t), \bm{w}(\cdot,t))}_{C^0([0,1];\mathbb{R}^{n_{X}+8})} = \infty. 
\end{align}
\begin{proof}
The proof is given in Appendix~\ref{sect:App1}.
\end{proof}
\end{theorem} 
The character of the existence and uniqueness result delivered by Theorem~\ref{thm:Local} is merely local, as a consequence of the nonlinear terms appearing in the PDE subsystem~\eqref{eq:PDEsubst}. However, under more restrictive assumptions on the functions $\bm{G}$ and $\bm{H}$, it is possible to infer global well-posedness, as formalised next in Theorem~\ref{thm:global}. It is also worth noting that, for $k\ped{bit} = 1$ in~Eq.~\eqref{BC:closed}, the boundary map $\bm{H}$ does not satisfy the local Lipschitz assumptions of Theorem~\ref{thm:Local} (to obviate this drawback, the nonlinear functions in Eq.~\eqref{BC:closed} may conveniently be replaced by a smooth regularisation, which would render $\bm{H}$ globally Lipschitz). 

\begin{theorem}[Global existence and uniqueness of generalised solutions]\label{thm:global} 
Under the same assumptions as Theorem~\ref{thm:Local}, suppose additionally that there exist constants $L_G,M_G\in \mathbb{R}_{\geq 0}$ such that
\begin{align}\label{eq:globalEstimateG}
\norm{\bm{G}(\bm{X},\bm{w},z)}_{\ell^1(\mathbb{R}^8)} \leq M_G+L_G\norm{(\bm{X},\bm{w})}_{\ell^1(\mathbb{R}^{n_{X}+8})}, \quad (\bm{X},\bm{w},z) \in \mathbb{R}^{n_{X}+8}\times[0,1].
\end{align}
Then, the unique generalised solution extends globally in time, that is, $t\ped{max} = \infty$ in Theorem~\ref{thm:Local}, and $(\bm{X},\bm{w})\in C^0([0,1]\times\mathbb{R}_{\geq 0};\mathbb{R}^{n_{X}+8})$. 
\begin{proof}
The proof is given in Appendix~\ref{sect:App2}.
\end{proof}
\end{theorem} 
Theorem~\ref{thm:global} promotes the local generalised solution obtained by invoking Theorem~\ref{thm:Local} to a global one, by excluding the possibility of blow-up in finite time. Essentially, this is achieved by imposing linear growth conditions on the function $\bm{G}$. In this context, it is worth observing that, for sufficiently vertical wells ($\od{\theta\ped{p}(z)}{z} \approx 0$ in Eq.~\eqref{eq:nomN}), the requirements of Theorem~\ref{thm:global} are satisfied in laminar flow regimes, where the quadratic terms relating to the viscous drag forces are negligible, and for BCs corresponding to the open, closed, and (regularised) bit pipe configurations (Eqs.~\eqref{BC:open} and~\eqref{BC:closed} with $k\ped{bit} = 0$, respectively). Finally, an important consideration about the role of friction on (global) well-posedness is formalised below.

\begin{remark}[Role of drillstring-wellbore friction on well-posedness]
It should be emphasised that, in the statement of Theorem~\ref{thm:global}, no additional assumption is postulated on $\bm{F}$; the interpretation is that drillstring-wellbore friction, being an intrinsically dissipative phenomenon, is itself non-malign for the well-posedness of the system. Aside from the viscous drag terms, the dominant source of nonlinearity in the PDEs arises indeed from the dependence of the normal force on the drillstring stress, pipe pressure, and annulus pressure, all of which evolve in closed loop with the frictional forces. Consequently, the principal obstacle to well-posedness is not friction itself, but the nonlinear feedback coupling between the generated tangential forces and the state-dependent normal load.
\end{remark}

The global existence and uniqueness results asserted by Theorem~\ref{thm:global} are propaedeutic to the stability analysis of the ODE--PDE interconenction~\eqref{eq:ODE--PDEsystem}, which is addressed next in Sect.~\ref{sect:smallGain} by relying on a small-gain strategy \cite{Krstic2}, as well as to the numerical experiments carried out in Sect.~\ref{sect:Numerics}.

\subsubsection{Stability analysis}\label{sect:smallGain}
The theoretical results advocated previously laid the basis for investigating the stability of the hydraulic-structural model~\eqref{eq:ODE--PDEsystem} in the presence of distributed friction. More specifically, the approach utilised in this paper builds on recent \emph{input-to-state stability} (ISS) results for infinite-dimensional systems \cite{Krstic2}, and considers the full nonlinear nature of the interconnection. The main motivation for relying on the proposed small-gain approach resides in the fact that, as revealed by the analytical steady-state solutions reported in Appendix~\ref{app:equilibria} for the open and closed pipe configurations, linearisation of Eqs.~\eqref{eq:ODE--PDEsystem} around a stationary input $\mathbb{R}^{n_U} \ni \bm{U}^\star = [p\ped{ext}^\star\; v\ped{rig}^\star\; F\ped{bit}^\star]^{\mathrm{T}} = [p\ped{atm}\; 0\; 0]^{\mathrm{T}}$ would correspond to zero nominal friction forces, making the analysis essentially irrelevant for the purpose of this study. 

In this context, the first result presented in the sequel is concerned with the behaviour of the friction model in isolation, as described by the spatially varying ODEs~\eqref{eq:ODEi}.

\begin{lemma}[Stability of the friction model~\eqref{eq:ODE}]\label{lemma:friction}
Under the same assumptions as Theorem~\ref{thm:Local}, consider the ODE subsystem~\eqref{eq:ODE} with dynamics~\eqref{eq:ODEi}, and suppose that the function $g \in C^0(\mathbb{R}\times[0,1];\mathbb{R}_{\geq 0})$ reads as in Eq.~\eqref{eq:g_mu}, with $\abs{v}_\varepsilon = \sqrt{v^2 + \varepsilon}$ for $\varepsilon \in \mathbb{R}_{\geq 0}$, and the friction coefficient $\mu \in C^0(\mathbb{R}\times[0,1];[\mu\ped{min},\infty))$ satisfying the linear growth condition
\begin{align}\label{eq:boundMu}
 \mu(v,z)\leq \mu_1(z)+\mu_2(z)\abs{v}_\varepsilon, \quad (v,z)\in\mathbb R\times[0,1].
\end{align}
Then, there exist constants $C_{f1}(\bm{X}_0(\cdot)),C_{X1}(\bm{X}_0(\cdot)),C_{f2},C_{X2} \in \mathbb{R}_{\geq 0}$ such that
\begin{subequations}
\begin{align}
\norm{f(\cdot,t)}_{C^0([0,1];\mathbb{R})} & \leq C_{f1}\bigl(\bm{X}_0(\cdot)\bigr) +C_{f2}\sup_{t^\prime \in [0,t]}\norm{v(\cdot,t)}_{C^0([0,1];\mathbb{R})},  \label{eq:boundfDOUS}\\
\norm{\bm{X}(\cdot,t)}_{C^0([0,1];\mathbb{R}^{n_{X}})} & \leq C_{X1}\bigl(\bm{X}_0(\cdot)\bigr) +C_{X2}\sup_{t^\prime \in [0,t]}\norm{v(\cdot,t)}_{C^0([0,1];\mathbb{R})}, \quad t\in [0,T],\label{eq:boundfDOUSX}
\end{align}
\end{subequations} 
for all $T \in (0, t\ped{max})$.
In particular,
\begin{align}\label{eq:constCf2}
C_{f2} & \triangleq \sup_{z\in[0,1]} \dfrac{2\bar{k}_0(z)}{q_f(z)} \max\bigl\{\mu_2(z),1\bigr\},
\end{align}
where $\mathbb{R}_{>0} \ni q_f(z) \triangleq 2\lambda\ped{min}(\mathbf{Q}(z))$, with $\lambda\ped{min}(\mathbf{Q}(z))$ denoting the smallest eigenvalue of the matrix $\mathbf{Sym}_{n_X}(\mathbb{R}) \ni \mathbf{Q}(z) \succ \mathbf{0}$, with entries
\begin{subequations}
\begin{align}
Q_{0,0}(z) & \triangleq \bar{k}_0(z)+ \sum_{i=1}^{n_1}\dfrac{\bar{k}_0(z)}{\bar{c}_{i,1}(z)}+ \sum_{i=1}^{n_2}\dfrac{\bar{k}_0(z)}{\bar{c}_{i,2}(z)}, \\
Q_{0,i}(z)  = Q_{i,0}(z) & \triangleq - \sqrt{\dfrac{\bar{k}_0(z)}{\bar{c}_{i,1}(z)\tau_{i,1}(z)}}, && i \in \{1,\dots,n_1\}, \\
Q_{0,n_1+i}(z) = Q_{n_1+i,0}(z) & \triangleq \sqrt{\dfrac{\bar{k}_0(z)}{\bar{c}_{i,2}(z)\tau_{i,2}(z)}}, && i \in \{1,\dots,n_2\}, \\
Q_{i,i}(z) & \triangleq \dfrac{1}{\tau_{i,1}(z)}, && i \in \{1,\dots,n_1\}, \\
Q_{n_1+i,n_1+i}(z) & \triangleq \dfrac{1}{\tau_{i,2}(z)}, && i \in \{1,\dots,n_2\}. 
\end{align}
\end{subequations}
and all remaining off-diagonal elements equal to zero.
\begin{proof}
The proof is given in Appendix~\ref{app:proofLemmaFric}.
\end{proof}
\end{lemma} 
Lemma~\ref{lemma:friction} asserts that, due to its intrinsically dissipative nature, friction cannot generate forces that grow unboundedly with the input velocity.
This aspect becomes clear from the proof of Lemma~\ref{lemma:friction}, which is based on Lyapunov-like arguments and, albeit similar to those reported in \cite{FrBD1,FrBD2}, is provided in Appendix~\ref{app:proofLemmaFric} for completeness. 

For what follows, it is worth observing that, in conjunction with the transformation $\bm{u}\ped{s}(z,t) = \mathbf{L}\ped{s}\bm{w}_4(z,t)$, with $\mathbf{L}\ped{s}(z)$ as in Eq.~\eqref{eq:invSigmaV}, the estimate~\eqref{eq:boundfDOUS} implies
\begin{align}\label{eq:boundfFric}
\begin{split}
\norm{f(\cdot,t)}_{C^0([0,1];\mathbb{R})}  & \leq C_{f1}\bigl(\bm{X}_0(\cdot)\bigr) + C_{f2}\sup_{t^\prime\in[0,t]}\norm{v(\cdot,t^\prime)}_{C^0([0,1];\mathbb{R})}  \\
& \leq \sup_{t^\prime\in[0,t]}\Bigl(C_{f1}\bigl(\bm{X}_0(\cdot)\bigr) + C_{f2}C_v\norm{\bm{w}_4(\cdot,t^\prime)}_{C^0([0,1];\ell^2(\mathbb{R}^2))}\Bigr), \quad t \in [0,T],
\end{split}
\end{align}
with $C_v = 1/\sqrt{2}$. The above bound will be instrumental for the derivation of a small-gain-like criterion ensuring the stability of the whole ODE--PDE interconnection in the presence of distributed friction dynamics. To the same end, it is also useful to define, for $\mathbb{R}_{\geq 0}^4 \ni \overline{\bm{w}} \triangleq [\overline{w}_1\; \overline{w}_2\; \overline{w}_3\; \overline{w}_4]^{\mathrm{T}}$, 
\begin{align}
\overline{\mathcal{B}}_w \triangleq \Bigl\{ \bm{w} \in C^0([0,1];\mathbb{R}^8) \mathrel{\Big|} \norm{\bm{w}_i(\cdot)}_{C^0([0,1];\ell^2(\mathbb{R}^2))} \leq \overline{w}_i, \; i \in \{1,2,3,4\}\Bigr\},
\end{align}
along with the source functions $\bm{g}_i : \mathbb{R}^{n_{X}+8}\times[0,1]\to \mathbb{R}^2$, with $\mathbb{R}^2 \ni \bm{g}_i(\bm{X},\bm{w},z) = [g_i^{+}(\bm{X},\bm{w},z)\; g_i^{-}(\bm{X},\bm{w},z)]^{\mathrm{T}}$, $i \in \{1,2,3,4\}$, as
\begin{subequations}
\begin{align}
g_1^{+}(\bm{X},\bm{w},z) &= g_1^{-}(\bm{X},\bm{w},z) \triangleq f\ped{p}\bigl(\bm{X},\mathbf{L}(z)\bm{w},z\bigr), \\
g_2^{+}(\bm{X},\bm{w},z) &= g_2^{-}(\bm{X},\bm{w},z) \triangleq f\ped{a}\bigl(\bm{X},\mathbf{L}(z)\bm{w},z\bigr), \\
g_3^{+}(\bm{X},\bm{w},z) &= g_3^{-}(\bm{X},\bm{w},z) \triangleq f\ped{b}\bigl(\bm{X},\mathbf{L}(z)\bm{w},z\bigr), \\
g_4^{+}(\bm{X},\bm{w},z) &= g_4^{-}(\bm{X},\bm{w},z) \triangleq f\ped{s}\bigl(\bm{X},\mathbf{L}(z)\bm{w},z\bigr).
\end{align}
\end{subequations}
Accordingly, under the same assumptions as Theorem~\ref{thm:Local}, and noting that the lumped states enter the PDE subsystem only via the normalised friction force $f(z,t)$, it is easy to infer the existence of constants $\phi_{i0}(\overline{\bm{w}}),\phi_{i2}(\overline{\bm{w}}),\phi_{i3}(\overline{\bm{w}}), \phi_{i4}(\overline{\bm{w}}), \phi_{if}(\overline{\bm{w}}) \in \mathbb{R}_{\geq 0}$ and $\psi_{i0},\psi_{i2},\psi_{i3}, \psi_{i4} \in \mathbb{R}_{\geq 0}$, $i \in \{1,2,3,4\}$, such that
\begin{subequations}\label{eq:boundsg_ihi}
\begin{align}\label{eq:boundsg_i}
\begin{split}
\norm{\bm{g}_i(\bm{X}(z,t),\bm{w}(z,t),z)}_{\ell^2(\mathbb{R}^2)} & \leq \phi_{i0}(\overline{\bm{w}})\norm{\bm{g}_0(\cdot)}_{C^0([0,1];\ell^2(\mathbb{R}^8))} +\sum_{j=1}^4 \phi_{ij}(\overline{\bm{w}})\norm{\bm{w}_j(\cdot,t)}_{C^0([0,1];\ell^2(\mathbb{R}^2))} \\
& \quad  +\phi_{if}(\overline{\bm{w}})\norm{f(\cdot,t)}_{C^0([0,1];\mathbb{R})} ,\\
& \qquad \qquad  (\bm{X}(\cdot,t),\bm{w}(\cdot,t),z) \in C^0([0,1];R^{n_X})\times\overline{\mathcal{B}}_w\times[0,1], 
\end{split}\\
\begin{split}
\norm{\bm{H}_i(\bm{w}^+(1,t),\bm{w}^-(0,t),\bm{U}(t))}_{\ell^2(\mathbb{R}^2)} & \leq \psi_{i0}\norm{\bm{h}_0(t)}_{\ell^2(\mathbb{R}^8)} + \sum_{j=1}^4 \psi_{ij}\norm{\bm{w}_j(\cdot,t)}_{C^0([0,1];\ell^2(\mathbb{R}^2))}, \\
& \qquad \qquad (\bm{w}(\cdot,t),\bm{U}(t)) \in \overline{\mathcal{B}}_w\times \mathbb{R}^{n_U}, \; i\in \{1,2,3,4\}.
\end{split}
\end{align}
\end{subequations}
Owing to these premises, the next Theorem~\ref{thm:Stability} delivers a conservative but analytical criterion to deliberate about the ISS of the ODE--PDE interconnection~\eqref{eq:ODE--PDEsystem}. Before enounciating Theorem~\ref{thm:Stability}, it is convenient to define the vector $\mathbb{R}^4 \ni \bm{\phi}_f(\overline{\bm{w}}) \triangleq [\phi_{1f}(\overline{\bm{w}})\; \phi_{2f}(\overline{\bm{w}})\; \phi_{3f}(\overline{\bm{w}})\; \phi_{4f}(\overline{\bm{w}})]^{\mathrm{T}}$. In this way, it is possible to formulate Assumption~\ref{ass;matrix} as done below.

\begin{assumption}\label{ass;matrix}
For $\bm{w}(\cdot,t) \in \overline{\mathcal{B}}_w$, the bounds in Eq.~\eqref{eq:boundsg_ihi} hold, and there exists a vector $\mathbb{R}^{12}\ni \bm{\eta} = [\bm{\eta}_1^{\mathrm{T}}\; \bm{\eta}_2^{\mathrm{T}}\; \bm{\eta}_3^{\mathrm{T}}\; \bm{\eta}_4^{\mathrm{T}}]^{\mathrm{T}}$, with $\mathbb{R}^3 \ni \bm{\eta}_i = [\eta_{i1}\; \eta_{i2}\; \eta_{i3}]^{\mathrm{T}}$ satisfying $\sum_{j=1}^3 \eta_{ij}^{-1} = 1$, $i \in \{1,2,3,4\}$, such that
\begin{align}\label{eq:SpectralRadius}
\rho\Bigl(\mathbf{\Theta}\bigl(\bm{\phi}_f(\overline{\bm{w}}),\bm{\eta},\overline{\bm{w}}\bigr)\Bigr) < 1,
\end{align}
 where $\rho(\mathbf{\Theta}(\bm{\phi}_f(\overline{\bm{w}}),\bm{\eta},\overline{\bm{w}}))$ denotes the spectral radius of the matrix-valued function $\mathbf{\Theta}(\cdot,\cdot,\overline{\bm{w}}) \in C^0(\mathbb{R}_{> 0}^{16};\mathbf{M}_4(\mathbb{R}_{\geq 0}))$, reading
\begin{align}\label{eq:MathTheta}
\mathbf{\Theta}(\bm{\phi}_f,\bm{\eta},\overline{\bm{w}}) \triangleq \begin{bmatrix} \theta_{11}(\bm{\eta},\overline{\bm{w}}) & \theta_{12}(\bm{\eta},\overline{\bm{w}})& \theta_{13}(\bm{\eta},\overline{\bm{w}}) & \theta_{14}(\bm{\phi}_f,\bm{\eta},\overline{\bm{w}}) \\
\theta_{21}(\bm{\eta},\overline{\bm{w}}) & \theta_{22}(\bm{\eta},\overline{\bm{w}}) & \theta_{23}(\bm{\eta},\overline{\bm{w}}) & \theta_{24}(\bm{\phi}_f,\bm{\eta},\overline{\bm{w}})\\
\theta_{31}(\bm{\eta},\overline{\bm{w}}) & \theta_{32}(\bm{\eta},\overline{\bm{w}}) & \theta_{33}(\bm{\eta},\overline{\bm{w}}) & \theta_{34}(\bm{\phi}_f,\bm{\eta},\overline{\bm{w}})\\
\theta_{41}(\bm{\eta},\overline{\bm{w}}) & \theta_{42}(\bm{\eta},\overline{\bm{w}}) & \theta_{43}(\bm{\eta},\overline{\bm{w}}) & \theta_{44}(\bm{\phi}_f,\bm{\eta},\overline{\bm{w}})\end{bmatrix},
\end{align}
with
\begin{subequations}\label{coefficientsTheta}
\begin{align}
\theta_{ij}(\bm{\eta},\overline{\bm{w}}) &\triangleq \eta_{i1}K_{i2}\psi_{ij} + \eta_{i2}K_{i3}\phi_{ij}(\overline{\bm{w}}), && i \in\{1,2,3,4\},\;j \in \{1,2,3\}, \\
\theta_{i4}(\bm{\phi}_f,\bm{\eta},\overline{\bm{w}}) & \triangleq \eta_{i1}K_{i2}\psi_{i4} + \eta_{i2}K_{i3}\phi_{i4}(\overline{\bm{w}}) +\eta_{i3}K_{i3}\phi_{if}C_{f2}C_v, && i \in \{1,2,3,4\}.
\end{align}
\end{subequations}
where the coefficients $K_{ij}\in \mathbb{R}_{>0}$, $i \in \{1,2,3,4\}$, $j \in \{1,2,3\}$ are defined as
\begin{align}\label{coeffs:K}
K_{11} & \triangleq \sup_{0\leq z^\prime \leq z \leq 1}\sqrt{1 + \alpha_1^2(z,z^\prime)} + \abs{\alpha_1(z,z^\prime)}, \quad K_{12} \triangleq \sup_{z\in[0,1]}\sqrt{1 + \alpha_1^2(z,0)} + \abs{\alpha_1(z,0)}, \nonumber \\
K_{13} & \triangleq \dfrac{1}{\lambda_1} \sup_{z \in [0,1]} \int_0^z\biggl(\sqrt{1 + \alpha_1^2(z,z^\prime)} + \abs{\alpha_1(z,z^\prime)}\biggr) \dif z^\prime, \nonumber \\
K_{21} & \triangleq \sup_{0\leq z^\prime \leq z \leq 1}\sqrt{1 + \alpha_2^2(z,z^\prime)} + \abs{\alpha_2(z,z^\prime)}, \quad K_{22} \triangleq \sup_{z\in[0,1]}\sqrt{1 + \alpha_2^2(z,0)} + \abs{\alpha_2(z,0)}, \nonumber \\
K_{23} & \triangleq \dfrac{1}{\lambda_2} \sup_{z \in [0,1]} \int_0^z\biggl(\sqrt{1 + \alpha_2^2(z,z^\prime)} + \abs{\alpha_2(z,z^\prime)}\biggr) \dif z^\prime,  \nonumber \\
K_{31} & \triangleq \sup_{0\leq z^\prime \leq z \leq 1}\sqrt{1 + \alpha_3^2(z,z^\prime)} + \abs{\alpha_3(z,z^\prime)}, \quad K_{32} \triangleq \sup_{z\in[0,1]}\sqrt{1 + \alpha_3^2(z,0)} + \abs{\alpha_3(z,0)}, \nonumber \\
K_{33} & \triangleq \dfrac{1}{\lambda_3} \sup_{z \in [0,1]} \int_0^z\biggl(\sqrt{1 + \alpha_3^2(z,z^\prime)} + \abs{\alpha_3(z,z^\prime)}\biggr) \dif z^\prime,  \nonumber \\
K_{42} & \triangleq 1,  \quad K_{41} \triangleq 1, \quad K_{43} \triangleq \dfrac{1}{\lambda_4},
\end{align}
and the functions $\alpha_i \in C^0([0,1]^2;\mathbb{R}_{\geq 0})$, $i \in \{1,2,3\}$, are given by
\begin{align}\label{eq:functionsAlpha}
\alpha_1(z,\tilde{z}) & \triangleq\dfrac{1}{2} \ln\biggl( \dfrac{A\ped{p}(z)}{A\ped{p}(\tilde{z})}\biggr), \quad \alpha_2(z,\tilde{z}) \triangleq\dfrac{1}{2} \ln\biggl( \dfrac{A\ped{a}(z)}{A\ped{a}(\tilde{z})}\biggr), \quad \alpha_3(z,\tilde{z}) \triangleq \dfrac{1}{2}\ln\biggl( \dfrac{A\ped{b}(z)}{A\ped{b}(\tilde{z})}\biggr), \quad 0\leq \tilde{z}\leq z \leq 1.
\end{align}
\end{assumption}
Under Assumption~\ref{ass;matrix}, global boundedness and ISS properties for the ODE--PDE system~\eqref{eq:ODE--PDEsystem} are asserted by Theorem~\ref{thm:Stability}.
\begin{theorem}[Global boundedness and input-to-state stability (ISS) of the PDE variables]\label{thm:Stability}
Suppose that Assumption~\ref{ass;matrix} holds. Then, under the same conditions as Theorem~\ref{thm:Local} and Lemma~\ref{lemma:friction}, for all inputs $\bm{U}\in C^0(\mathbb{R}_{\geq 0};\mathbb{R}^{n_U})$ and ICs $(\bm{X}_0,\bm{w}_0) \in C^0([0,1];\mathbb{R}^{n_X})\times \overline{\mathcal{B}}_w$ satisfying
\begin{align}\label{eq:condInitiDHDD}
\mathbf{A}(0)\begin{bmatrix} \norm{\bm{w}_{1,0}(\cdot)}_{C^0([0,1];\ell^2(\mathbb{R}^2))}\\ \norm{\bm{w}_{2,0}(\cdot)}_{C^0([0,1];\ell^2(\mathbb{R}^2))} \\ \norm{\bm{w}_{3,0}(\cdot)}_{C^0([0,1];\ell^2(\mathbb{R}^2))} \\ \norm{\bm{w}_{4,0}(\cdot)}_{C^0([0,1];\ell^2(\mathbb{R}^2))}\end{bmatrix}+ \mathbf{B}\bigl(\bm{\phi}_f(\overline{\bm{w}}),\bm{\eta}, \overline{\bm{w}}\bigr)\begin{bmatrix}C_{f1}\bigl(\bm{X}_0(\cdot)\bigr) \\  \norm{\bm{g}_0(\cdot)}_{C^0([0,1];\ell^2(\mathbb{R}^8))} \\ \sup_{t \in \mathbb{R}_{\geq 0}}\norm{\bm{h}_0(t)}_{\ell^2(\mathbb{R}^8)} \end{bmatrix}< \Bigl(\mathbf{I}_4-\mathbf{\Theta}\bigl(\bm{\phi}_f(\overline{\bm{w}}),\bm{\eta},\overline{\bm{w}}\bigr)\Bigr)\overline{\bm{w}},
\end{align}
with the matrices $\mathbf{A}\in C^0(\mathbb{R}_{\geq 0};\mathbf{M}_4(\mathbb{R}_{\geq 0}))$ and $\mathbf{B}(\bm{\phi}_f,\bm{\eta},\overline{\bm{w}}) \in \mathbf{M}_{4\times3}(\mathbb{R}_{\geq 0})$ reading
\begin{subequations}\label{eq:matricesAB}
\begin{align}
\mathbf{A}(\varrho) & \triangleq \begin{bmatrix} K_{11}\eu^{\varrho \lambda_1^{-1}} & 0 & 0 & 0 \\
0 & K_{21}\eu^{\varrho \lambda_2^{-1}} & 0 & 0 \\
0 & 0 & K_{31}\eu^{\varrho \lambda_3^{-1}} & 0 \\
0 & 0 & 0 & K_{41}\eu^{\varrho \lambda_4^{-1}} \end{bmatrix}, \\
\mathbf{B}(\bm{\phi}_f,\bm{\eta}, \overline{\bm{w}}) & \triangleq \begin{bmatrix} 
b_{11}(\bm{\phi}_f,\bm{\eta}) & b_{12}(\bm{\eta},\overline{\bm{w}}) & b_{13}(\bm{\eta}) \\
b_{21}(\bm{\phi}_f,\bm{\eta}) & b_{22}(\bm{\eta},\overline{\bm{w}}) & b_{23}(\bm{\eta}) \\
b_{31}(\bm{\phi}_f,\bm{\eta})  & b_{32}(\bm{\eta},\overline{\bm{w}}) & b_{33}(\bm{\eta}) \\
b_{41} (\bm{\phi}_f,\bm{\eta})& b_{42}(\bm{\eta},\overline{\bm{w}}) & b_{43}(\bm{\eta}) \\ \end{bmatrix},
\end{align}
\end{subequations}
with
\begin{subequations}
\begin{align}
b_{i1}(\bm{\phi}_f,\bm{\eta}) & \triangleq \eta_{i3}K_{i3}\phi_{if}, \\
b_{i2}(\bm{\eta},\overline{\bm{w}}) & \triangleq (1+\eta_i^{-1})K_{i3}\phi_{i0}(\overline{\bm{w}}),  \\
b_{i3}(\bm{\eta}) & \triangleq (1+\eta_i)K_{i2}\psi_{i0}, && i \in\{1,2,3,4\},
\end{align}
\end{subequations}
the ODE--PDE system~\eqref{eq:ODE--PDEsystem} admits a unique bounded, global generalised solution $(\bm{X},\bm{w})\in C^0([0,1]\times\mathbb{R}_{\geq 0};\mathbb{R}^{n_{X}+8})$ which is \emph{input-to-state-stable} in the PDE variables.
More precisely, $\bm{w}(\cdot,t) \in \overline{\mathcal{B}}_w$ for all $t \in \mathbb{R}_{\geq 0}$, and there exist $\Theta_X,\Theta_w,\gamma,\delta \in \mathbb{R}_{>0}$ such that 
\begin{subequations}\label{eq:BoundedandISSEst}
\begin{align}
\begin{split}
& \norm{\bm{w}(\cdot,t)}_{C^0([0,1];\mathbb{R}^{8})} \leq \Theta_w\norm{\bm{w}_0(\cdot)}_{C^0([0,1];\mathbb{R}^8)}\eu^{-\delta t} \\
& \qquad + \gamma\Biggl(C_{f1}\bigl(\bm{X}_0(\cdot)\bigr) + \norm{\bm{g}_0(\cdot)}_{C^0([0,1];\mathbb{R}^8)} + \sup_{t^\prime\in[0,t]}\norm{\bm{h}_0(t^\prime)}_{\mathbb{R}^8} \Biggr), 
\end{split}\label{eq:BoundedandISSEst1} \\
\begin{split}
& \norm{(\bm{X}(\cdot,t),\bm{w}(\cdot,t))}_{C^0([0,1];\mathbb{R}^{n_{X}+8})} \leq \Theta_X\norm{\bm{w}_0(\cdot)}_{C^0([0,1];\mathbb{R}^8)} \\
& \qquad + \gamma\Biggl(C_{X1}\bigl(\bm{X}_0(\cdot)\bigr) + C_{f1}\bigl(\bm{X}_0(\cdot)\bigr) + \norm{\bm{g}_0(\cdot)}_{C^0([0,1];\mathbb{R}^8)} + \sup_{t^\prime\in[0,t]}\norm{\bm{h}_0(t^\prime)}_{\mathbb{R}^8} \Biggr), \quad t\in \mathbb{R}_{\geq 0}. \label{eq:BoundedandISSEst2}
\end{split}
\end{align}
\end{subequations}
\begin{proof}
The proof is given in Appendix~\ref{app:finalProod}.
\end{proof}
\end{theorem}
From a mathematical perspective, Theorem~\ref{thm:Stability} extends the stability results for hyperbolic ODE--PDE and PDE--PDE loops contained in \cite{Krstic2}. More specifically, compared to \cite[Theorems 8.1 and 9.2]{Krstic2}, a key feature of Theorem~\ref{thm:Stability} is that it does not require any fading-memory assumption on the ODE subsystem. Consequently, the term $C_{f1}\bigl(\bm{X}_0(\cdot)\bigr)$ in Eq.~\eqref{eq:BoundedandISSEst1}, which depends on the initial conditions of the lumped states, does not decay over time\footnote{Strictly speaking, according to the standard definition adopted in the literature, the bound~\eqref{eq:BoundedandISSEst1} cannot be regarded as a true ISS estimate for precisely this reason. Nevertheless, if the ODE states are interpreted as inputs to the PDE subsystem, then~\eqref{eq:BoundedandISSEst1} may be viewed as an ISS estimate for the PDE variables alone.}. Within the proposed modelling framework, the estimate in Eq.~\eqref{eq:BoundedandISSEst} is instead a strict consequence of the dissipative nature of friction, as established by Eqs.~\eqref{eq:boundfDOUS} and~\eqref{eq:boundfFric}. In the same context, it is important to emphasise that the existence assumption on the parameter vector $\bm{\eta}$ is extremely difficult to remove. In fact, an analogous assumption is also required in \cite[Theorem 9.2]{Krstic2}. In practice, this condition may be verified numerically by formulating and solving a suitable optimisation problem, provided that the values of the physical parameters are known. 

Theorem~\ref{thm:Stability} concludes the theoretical developments of the paper. Before moving to the numerical simulations, some final remarks on its physical interpretation are collected below.

\begin{remark}[Global boundedness and ISS under the assumptions of Theorem~\ref{thm:global}]\label{remark:GDD}
Under the additional assumptions of Theorem~\ref{thm:global}, the coefficients $\phi_{i0}(\overline{\bm{w}}),\phi_{i2}(\overline{\bm{w}}),\phi_{i3}(\overline{\bm{w}}), \phi_{i4}(\overline{\bm{w}}), \phi_{if}(\overline{\bm{w}}) \in \mathbb{R}_{\geq 0}$ become indepndent of $\overline{\bm{w}}$, and the matrix $\mathbf{\Theta}(\bm{\phi}_f(\overline{\bm{w}}),\bm{\eta},\overline{\bm{w}})$ in Eq.~\eqref{eq:MathTheta} reduces to a matrix $\mathbf{\Theta}(\bm{\phi}_f,\bm{\eta})$, with $\mathbf{\Theta} \in C^0(\mathbb{R}_{> 0}^{16};\mathbf{M}_4(\mathbb{R}_{\geq 0}))$. In particular, since $\rho(\mathbf{\Theta}(\bm{\phi}_f,\bm{\eta}))<1$, the inverse $(\mathbf{I}_4-\mathbf{\Theta}(\bm{\phi}_f,\bm{\eta}))^{-1}$ exists and is componentwise nonnegative. Therefore, selecting $\mathbb{R}_{\geq 0}^4 \ni \overline{\bm{w}} = c_w (\mathbf{I}_4-\mathbf{\Theta}(\bm{\phi}_f,\bm{\eta}))^{-1} \bm{1}_4 $ for some sufficiently large $c_w\in\mathbb{R}_{>0}$ gives $(\mathbf{I}_4-\mathbf{\Theta}(\bm{\phi}_f,\bm{\eta})) \overline{\bm{w}} = c_w\bm{1}_4$, ensuring that inequality~\eqref{eq:condInitiDHDD} holds trivially. In this way, the estimate~\eqref{eq:BoundedandISSEst} holds for all inputs $\bm{U}\in C^0(\mathbb{R}_{\geq 0};\mathbb{R}^{n_U})$ and ICs $(\bm{X}_0,\bm{w}_0) \in C^0([0,1];\mathbb{R}^{n_X+8})$ satisfying the assumptions of Theorem~\ref{thm:global} and Lemma~\ref{lemma:friction}.
\end{remark}

\begin{remark}[Role of drillstring-wellbore friction on stability]
In contrast to the well-posedness analysis of Sect.~\ref{sect:wellPos}, which demonstrated that -- for $\od{\theta\ped{p}(z)}{z} \approx 0$ in Eq.~\eqref{eq:nomN} -- drillstring-wellbore friction dynamics plays little to no role in determining the well-posedness of the ODE--PDE interconnection~\eqref{eq:ODE--PDEsystem}, Theorem~\ref{thm:Stability} suggests that it might alter its stability, even in the absence of oscillating normal loads. To elucidate this aspect, it is possible to assume $\od{\theta\ped{p}(z)}{z} \approx 0$ in Eq.~\eqref{eq:nomN} and consider the matrix $\mathbf{\Theta}(\bm{\phi}_f,\bm{\eta})$ of Remark~\ref{remark:GDD}, along with the more conservative sufficient condition $\norm{\mathbf{\Theta}(\bm{\phi}_f,\bm{\eta})} < 1$ in place of $\rho(\mathbf{\Theta}(\bm{\phi}_f,\bm{\eta})) < 1$. Denoting by $\mathbf{M}_{4}(\mathbb{R}) \ni \mathbf{\Theta}_0(\bm{\eta}) \triangleq \mathbf{\Theta}(\bm{0},\bm{\eta})$ the matrix $\mathbf{\Theta}(\bm{\phi}_f,\bm{\eta})$ in the absence of friction, suppose that $\norm{\mathbf{\Theta}_0(\bm{\eta})} < 1$. Then, inspection of Eqs.~\eqref{eq:fsPD} and~\eqref{eq.FS} indicates that the ISS estimate~\eqref{eq:BoundedandISSEst1} is still guaranteed provided that
\begin{align}
\dfrac{C_{f2}\eta_{43}}{\sqrt{2}\rho\ped{s}}(\rho\ped{s}-\rho)\mathrm{g} \sup_{z \in [0,1]}\abs{\sin\theta\ped{p}(z)} < 1-\norm{\mathbf{\Theta}_0(\bm{\eta})}.
\end{align}
Therefore, friction may reduce the available stability margin associated with this sufficient condition. Of course, since the requirements of Theorem~\ref{thm:Stability} are only sufficient, their violation does not imply loss of stability; rather, it merely indicates that this particular criterion is no longer conclusive.
\end{remark}

\begin{remark}[Geometrical design]
The sufficient conditions of Theorem~\ref{thm:Stability} also provide useful guidelines for reducing the risk of instability and correctly designing the geometrical configuration of the pipes. Indeed, the coefficients appearing in Eq.~\eqref{coeffs:K} depend on the physical parameters governing the transport velocities, as well as on the pipe geometry through the cross-sectional areas $A\ped{p}(z)$, $A\ped{a}(z)$, and $A\ped{b}(z)$. In particular, Eqs.~\eqref{coeffs:K} and~\eqref{eq:functionsAlpha} show that the interconnection gains can be reduced by avoiding abrupt variations in the cross-sectional areas. 
\end{remark}

\section{Numerical Experiments}\label{sect:Numerics}
The analytical results derived above are primarily qualitative in nature and may yield conservative predictions in practical applications.
The experiments reported below were designed to complement the theoretical findings of Sect.~\ref{sect:HeoNal}, and intended to quantify the impact of physically relevant excitations and specific boundary configurations on the system's response, with emphasis on the gains from the rig velocity to the annulus and string structural variables for the open and closed pipe configurations. Specifically, Sect.~\ref{sect:nominalExp} describes the numerical implementation of the hydraulic-structural~\eqref{eq:ODE--PDEsystem} in MATLAB\textsuperscript{\textregistered} environment. Then, an exhaustive harmonic analysis is presented in Sect.~\ref{sect:Harmonic}, whereas Sect.~\ref{sect:MechIso} discusses the results of mechanism-isolation studies and parameter sensitivity analysis. Finally, the response of the system subjected to a realistic offshore excitation scenario is investigated in Sect.~\ref{sect:offshore}.

\subsection{Numerical implementation and nominal system's response}\label{sect:nominalExp}
The present section provides details about the numerical implementation of the hyperbolic ODE--PDE system~\eqref{eq:ODE--PDEsystem} in Riemann variables, illustrating its nominal behaviour in the presence of small excitations. 

\subsubsection{Numerical implementation}
The proposed hydraulic-structural model was discretised using the formulation based on the Riemann-invariant representation introduced in Sect.~\ref{sect:riemann}, employing the method of lines. Specifically, exploiting the fact that the drillstring and wellbore are made of similar material, and that the drillstring is much stiffer than the bit rock, a distributed FrBD model with a single auxiliary state $\mathbb{R} \ni x(z,t) \triangleq x_1(x,t)-x_2(x,t)$ was considered, and the dimensionless spatial coordinate was discretised adopting a uniform mesh with $N_z=100$ grid points, yielding a semi-discrete system comprising $10N_z$ coupled ordinary differential equations. 
Inspired by \cite{Landet2013}, Fanning friction factors for the laminar regime were modelled as
\begin{align}
f_k(v_{\textnormal{rel},k},z) = \dfrac{A_k(z)k\ped{fric}}{2\rho d_{\textnormal{hyd},k}(z)v_{\textnormal{rel},k}}, \quad k = \textnormal{p},\textnormal{a},\textnormal{b}, 
\end{align}
where $k\ped{fric} \in \mathbb{R}_{\geq 0}$ is a hydrodynamic friction factor. In the turbulent regime, the corresponding formulae from \cite{Landet2013} were adopted. 

The inclination angles $\theta\ped{p}(z)$ and $\theta\ped{b}(z)$ in Eqs.~\eqref{eqs:fpfa} and~\eqref{eq:fb} were specified respectively as
\begin{align}
\theta\ped{p}(z)= \begin{cases} 0, & L\ped{s} z < 1500, \\ \dfrac{L\ped{s} z-1500}{500}, & 1500 \leq L\ped{s} z < 1500+\dfrac{\pi}{2}\,500, \\ \dfrac{\pi}{2}, & L\ped{s} z \geq 1500+\dfrac{\pi}{2}500, \end{cases} 
\end{align}
and
\begin{align}
\theta\ped{b}(z)=\frac{\pi}{2}, \quad z\in[0,1]. 
\end{align}
Finally, the friction coefficient in Eq.~\eqref{eq:g_mu} was postulated as velocity-independent, but varying with the spatial coordinate as
\begin{align}\label{eq:mu}
\mu(z) = \mu\ped{min} + \dfrac{\mu\ped{max}-\mu\ped{min}}{2} \Biggl( 1- \tanh\biggl( \frac{z-z_0}{\delta} \biggr) \Biggr), \quad z \in [0,1],
\end{align}
with the parameters $z_0 \in [0,1]$ and $\delta \in \mathbb{R}_{>0}$ capturing the transition between steel-steel friction and steel-rock friction.

The resulting stiff nonlinear system was integrated using the MATLAB\textsuperscript{\textregistered} solver \texttt{ode15s} with relative and absolute tolerances equal to $10^{-6}$ and $10^{-8}$, respectively, and the open/closed pipe BCs were imposed analytically using the exact solution reported in Appendix~\ref{app:closedMAP}. The parameter values employed in simulation are listed in Table~\ref{tab:param}.

Unless otherwise specified, all simulations were initialised from the equilibrium solution generated by the steady-state formulation of the governing equations obtained from the stationary solution of the open/closed pipe settings (Appendix~\ref {app:equilibria}). The primary observation location was chosen as $z^\ast =0.8$, which corresponds to a point near the lower portion of the inclined section where the friction coefficient exhibits a rapid variation. Additional observation points $z^\ast \in \{0.1,0.25,0.5,0.75,1.0\}$ were employed when constructing spatial amplification profiles. Throughout the numerical study, particular attention was devoted to the hydraulic variables $p\ped{a}(z,t), q\ped{a}(z,t)$ and structural variables $v(z,t), \sigma(z,t)$. The imposed rig motion was represented by the boundary excitation $v\ped{rig}(t)$, acting as the input of the hydraulic-structural system, whereas the external pressure was assumed to be constant, with $p\ped{ext}(t) = p\ped{atm}$. All gain calculations therefore quantify the transmission of rig-motion disturbances into hydraulic and structural responses. 

\begin{table}[t] \centering 
\caption{Parameters of the hydraulic-structural drilling system model} \label{tab:parameters} 
\renewcommand{\arraystretch}{1.15} 
\begin{tabular}{llll} \toprule 
\textbf{Symbol} & \textbf{Description} & \textbf{Unit} & \textbf{Value}\\ 
\midrule 
$L$ & Total well length & m & $4000$\\ 
$L\ped{s}$ & Drillstring length & m & $2500$\\ 
$L\ped{b}$ & Bottom-hole length & m & $1500$\\ 
$\rho$ & Fluid density & kg\,m$^{-3}$ & $1200$\\ 
$\beta$ & Fluid bulk modulus & Pa & $2\times10^{9}$\\ 
$E$ & Young's modulus of the drillstring & Pa & $2\times10^{11}$\\ 
$\rho\ped{s}$ & Drillstring density & kg\,m$^{-3}$ & $7800$\\ 
$\mathrm{g}$ & Gravitational acceleration & m\,s$^{-2}$ & $9.81$\\ 
$k\ped{fric}$ & Viscous friction coefficient & Pa\,s\,m$^{-2}$ & $1144$\\ 
$d\ped{p}$ & Drillstring diameter & m & $0.110$\\ 
$d\ped{a,i}$ & Inner annulus diameter & m & $0.128$\\ 
$d\ped{a,o}$ & Outer annulus diameter & m & $0.216$\\ 
$d\ped{b}$ & Bottom-hole diameter & m & $0.216$\\ 
\midrule 
$\bar{k}_0$ & Normalised micro-stiffness coefficient & m$^{-1}$ & $10^{6}$\\ 
$\bar{c}$ & Normalised micro-damping coefficient & s\,m$^{-1}$ & $10^{2}$\\ 
$\tau$ & Relaxation time & s & $0.1$\\ 
$\varepsilon$ & Regularisation parameter & m$^2$\,s$^{-2}$ & $10^{-6}$\\ 
\midrule 
$\mu_{\min}$ & Minimum friction coefficient & -- & $0.40$\\ 
$\mu_{\max}$ & Maximum friction coefficient & -- & $0.74$\\ 
$z_0$ & Friction-transition location & -- & $0.80$\\ 
$\delta$ & Friction-transition width & -- & $0.02$\\ 
\bottomrule \end{tabular} \label{tab:param}
\end{table}

\subsubsection{Nominal hydraulic and structural response} \label{sect:Nominal}
\begin{figure}
\centering
\includegraphics[width=1\linewidth]{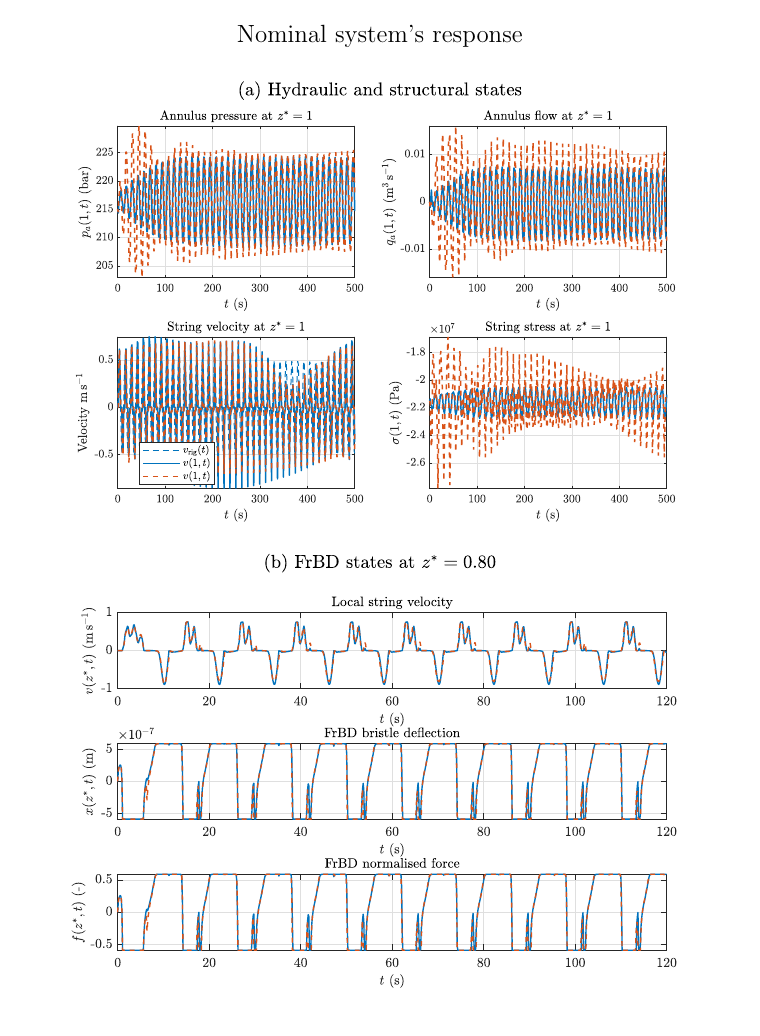} 
\caption{Nominal response of the system subjected to a harmonic excitation $v_{\rm rig}(t)=0.5\sin(0.5 t)$ $\textnormal{m}\,\textnormal{s}^{-1}$: (a) hydraulic and structural states at $z^* = 1$; (b) FrBD states at $z^* = 0.8$. Line styles: open configuration (solid blue lines), closed configuration (dashed orange lines). Model parameters as in Table~\ref{tab:param}.}
\label{fig:1}
\end{figure}

The nominal response of the nonlinear model under harmonic excitation $v_{\rm rig}(t)=0.5\sin(0.5 t)$ is visualised in Fig.~\ref{fig:1}(a) for the open (blue lines) and closed (dashed orange lines) configurations, in terms of the string velocity $v(1,t)$, stress $\sigma(1,t)$, annulus pressure $p\ped{a}(1,t)$, and annulus flow $q\ped{a}(1,t)$. The imposed rig velocity induces oscillatory behaviour in both hydraulic and structural variables, illustrating the fully coupled nature of the drilling system. The bounded character of the vibrational response indicates that pressure disturbances generated at the rig propagate through the well whilst remaining regulated by hydraulic dissipation and wave reflection mechanisms. 

Concerning the hydraulic variables, both the bottom pressure and annulus flow responses, reported in the upper panels, exhibit a transient phase followed by a nearly periodic steady-state oscillation, although with noticeably different amplitudes. 
The lower panels of Fig.~\ref{fig:1}(a) illustrate the structural response: in both the open and closed pipe configurations, the drillstring velocity remains strongly correlated with the imposed rig motion, confirming that the structural subsystem efficiently transmits boundary disturbances along the entire drillstring. 
The stress response displays significant oscillatory behaviour despite the relatively small excitation amplitude, highlighting the ability of distributed structural waves to accumulate along the drillstring. In particular, the closed pipe configuration produces wider oscillations.

These results also reveal a clear separation of dynamic time scales between the hydraulic and structural subsystems: whereas the hydraulic variables are primarily governed by transport and compressibility effects, the drillstring velocity and stress retain a stronger memory of the imposed boundary excitation and exhibit more pronounced wave-propagation dynamics. This distinction becomes even more apparent in the frequency-domain analysis presented in Sect.~\ref{sect:Harmonic}.

Figure~\ref{fig:1}(b) shows the evolution of the FrBD states for $z^* = 0.8$. The open and closed pipe systems exhibit qualitatively similar responses, with a sequence of distinct stick-slip events. In particular, friction saturation occurs for values of the nondimensional friction force slightly above $f(z^*,t) = 0.5$, which is consistent with the expression for the friction coefficient defined in Eq.~\eqref{eq:mu}.

The comparison between the open and closed pipe configurations reveals the important influence of hydraulic boundary conditions. In general, the closed configuration exhibits stronger pressure amplification, more pronounced resonant behaviour, and larger low-frequency sensitivity; these differences arise from the altered reflection properties of the hydraulic subsystem and the associated modification of wave-propagation patterns. By contrast, the structural responses display much smaller differences between the two boundary configurations. 

\subsection{Harmonic analysis}\label{sect:Harmonic} 
As a numerical complement to the stability analysis presented in Sect.~\ref{sect:smallGain}, the following experiments investigate the frequency response of the system, focusing on amplification under harmonic excitation (Sect.~\ref{sect:amplification}), nonlinear distortion effects (Sect.~\ref{section:distortion}), spatial variations (Sect.~\ref{sect:spatialAmpl}), and statistical performance indicators (Sect.~\ref{sect:RMS}).

\subsubsection{Frequency-domain amplification characteristics} \label{sect:amplification}
\begin{figure}
\centering
\includegraphics[width=1\linewidth]{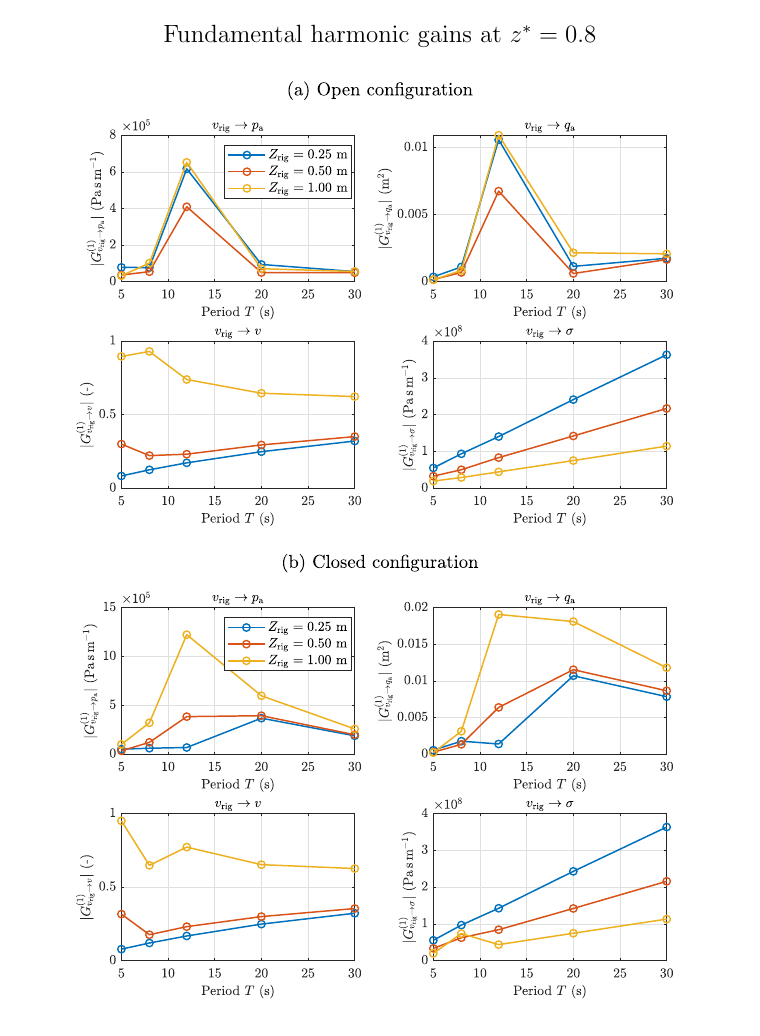} 
\caption{Fundamental harmonic gains of the system subjected to a harmonic excitation $v_{\rm rig}(t)=\frac{Z\ped{rig}}{2}\sin(0.5 t)$ $\textnormal{m}\,\textnormal{s}^{-1}$: (a) open pipe configuration; (b) closed pipe configuration. Model parameters as in Table~\ref{tab:param}.}
\label{fig:2}
\end{figure}

A harmonic gain analysis was performed to characterise disturbance propagation over a broad range of excitation frequencies. For a prescribed excitation $v_{\rm rig}(t)=\omega Z\ped{rig}\sin(\omega t)$ $\textnormal{m}\,\textnormal{s}^{-1}$, where $Z\ped{rig} \in \mathbb{R}_{>0}$ denotes the maximum rig displacement, the fundamental harmonic gain was computed as
\begin{equation}
G\ap{(1)}_{u\to y} = \frac{\widehat{y}_1}{\widehat{u}_1},
\end{equation}
where $\widehat{u}_1$ and $\widehat{y}_1$ denote the first Fourier coefficients of the input and output signals, respectively. The resulting gains are presented in Fig.~\ref{fig:2}, where both the hydraulic (upper panels) and structural (lower panels) transfer characteristics are shown. In particular, Fig.~\ref{fig:2} reveals the existence of a main resonant amplification region around $T = 12$ s for the pressure and flow gains, which may be conjectured to arise from constructive interference between hydraulic travelling waves generated at the upper boundary and reflected at the lower boundary. The magnitude of the observed resonances depend strongly on the boundary conditions, as evidenced by the differences between the open and closed pipe configurations. More specifically, the closed pipe configuration produces higher pressure peaks, whereas the resonant amplification associated with the open configuration is less severe.
In contrast, the structural gains exhibit a markedly different behaviour, with the velocity and stress gains evolving more smoothly over the investigated frequency range. This observation suggests that structural dynamics are governed primarily by distributed elastic wave propagation rather than purely hydraulic resonance phenomena. 


\subsubsection{Nonlinear harmonic distortion} \label{section:distortion}
\begin{figure}
\centering
\includegraphics[width=1\linewidth]{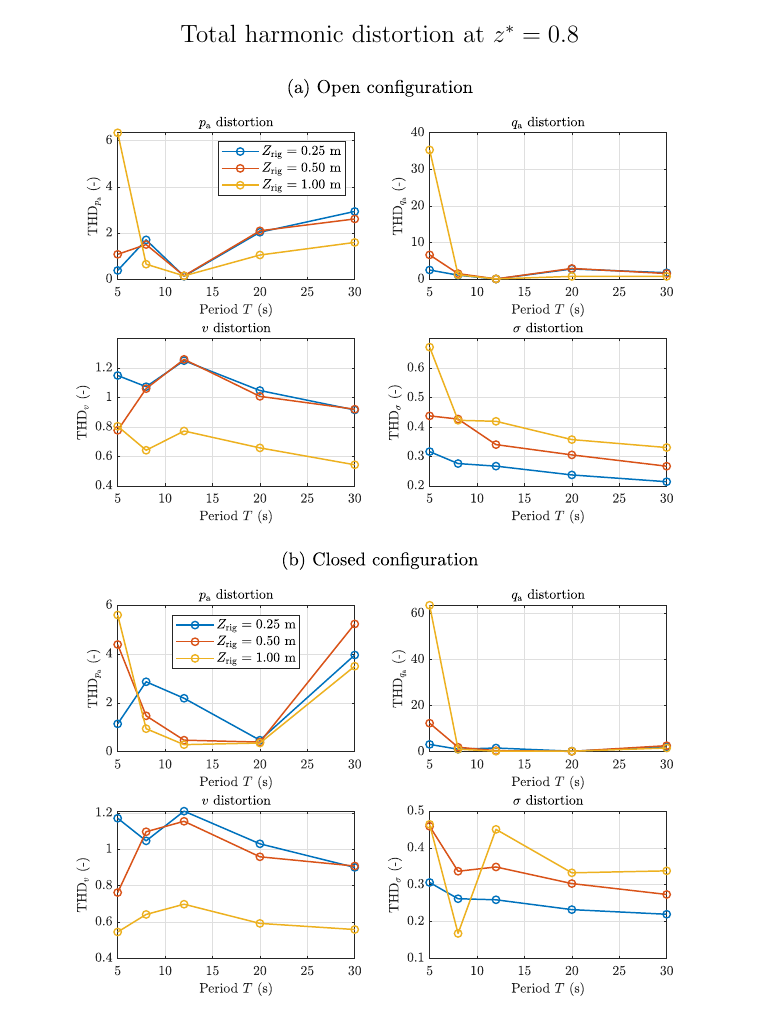} 
\caption{Total harmonic distortion of the system subjected to a harmonic excitation $v_{\rm rig}(t)=\frac{Z\ped{rig}}{2}\sin(0.5 t)$ $\textnormal{m}\,\textnormal{s}^{-1}$: (a) open pipe configuration; (b) closed pipe configuration. Model parameters as in Table~\ref{tab:param}.}
\label{fig:3}
\end{figure}

The degree of nonlinear behaviour was also quantified through the \emph{total harmonic distortion} (THD) index, defined as
\begin{align}
\textnormal{THD} = \dfrac{ \sqrt{\sum_{k=2}^{K}\abs{\widehat{y}_k}^2} }{ \abs{\widehat{y}_1} },
\end{align} 
and measuring the relative contribution of higher harmonics with respect to the fundamental response. The resulting distortion levels are shown in Fig.~\ref{fig:3}, considering the open and closed pipe configurations: overall, the distortion may be significant across the investigated operating range. Furthermore, the main conclusions that may be drawn from Fig.~\ref{fig:3} are consistent with the observations already collected in Sect.~\ref{sect:amplification}: whilst hydraulic distortion generally varies nonlinearly near resonance conditions, where pressure-wave amplification is strongest, structural distortion exhibits a nearly linear trend and remains comparatively limited, indicating that the axial drillstring dynamics is predominantly governed by harmonic response, despite nonlinear friction and coupling effects. These results support the use of harmonic gain analysis as a meaningful tool for assessing the system behaviour over a broad frequency range. 

\subsubsection{Spatial amplification of disturbances} \label{sect:spatialAmpl}
\begin{figure}
\centering
\includegraphics[width=1\linewidth]{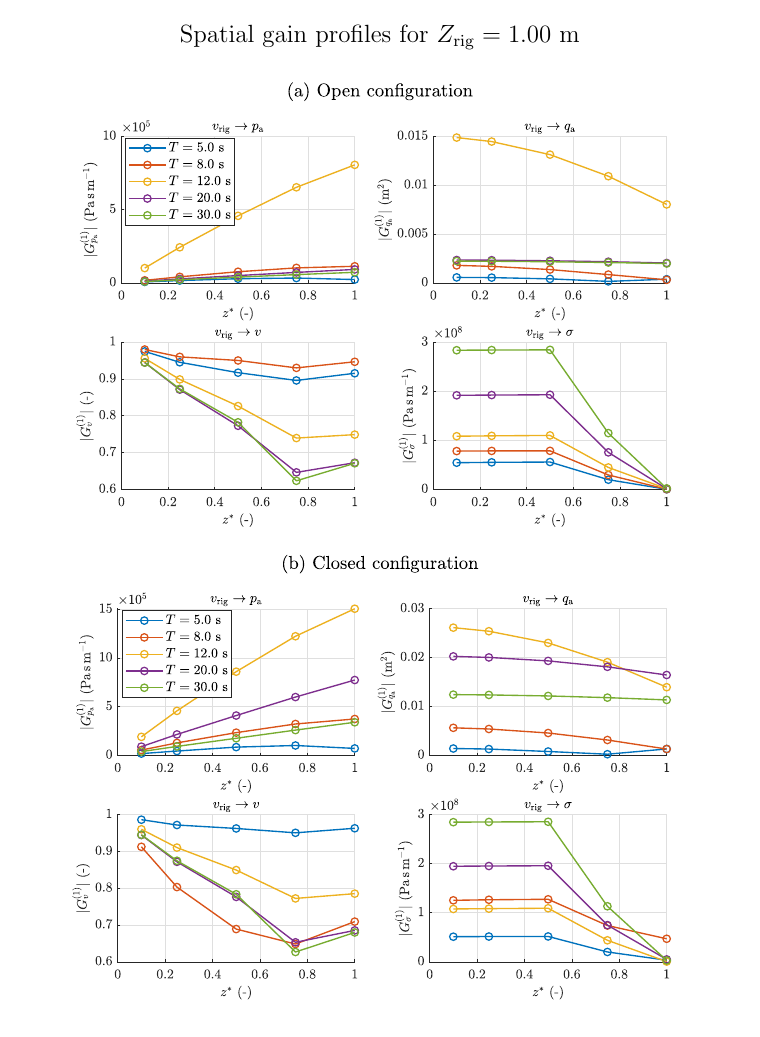} 
\caption{Spatial gain profiles for $Z\ped{rig} = 1$ m. Model parameters as in Table~\ref{tab:param}.}
\label{fig:4}
\end{figure}

Figure~\ref{fig:4} visualises the spatial distribution of amplification throughout the well. 

In particular, concerning the hydraulic variables, a pronounced increase in pressure amplification is observed with increasing depth, indicating accumulation of hydraulic-wave energy along the well and demonstrating that downhole pressure fluctuations may become substantially larger than their surface counterparts. Conversely, the flow gains exhibit an opposite trend, reflecting the influence of continuity constraints and hydraulic dissipation. Coherently with the findings of Sect.~\ref{sect:Harmonic}, both the pressure and flow gains attain their maximum at $T = 12$ s, corresponding to a resonant period.

Shifting attention to the structural variables, the velocity gains generally decrease for $z^* < 0.8$ as the distance from the rig increases due to structural attenuation mechanisms that mitigate the disturbances during propagation; the increasing trend for $z^*\geq 0.8$ may instead be attributed to the repentine friction variation occurring at $z = 0.8$. In contrast, the efficient transmission of axial loading through the drillstring causes stress gains to remain comparatively persistent throughout large sections of the well. These observations emphasise the importance of distributed modelling, and, consistently with the literature, confirm that surface measurements alone are insufficient to predict the dynamic environment experienced by downhole components. 


\subsubsection{RMS and peak amplification metrics} \label{sect:RMS}
\begin{figure}
\centering
\includegraphics[width=1\linewidth]{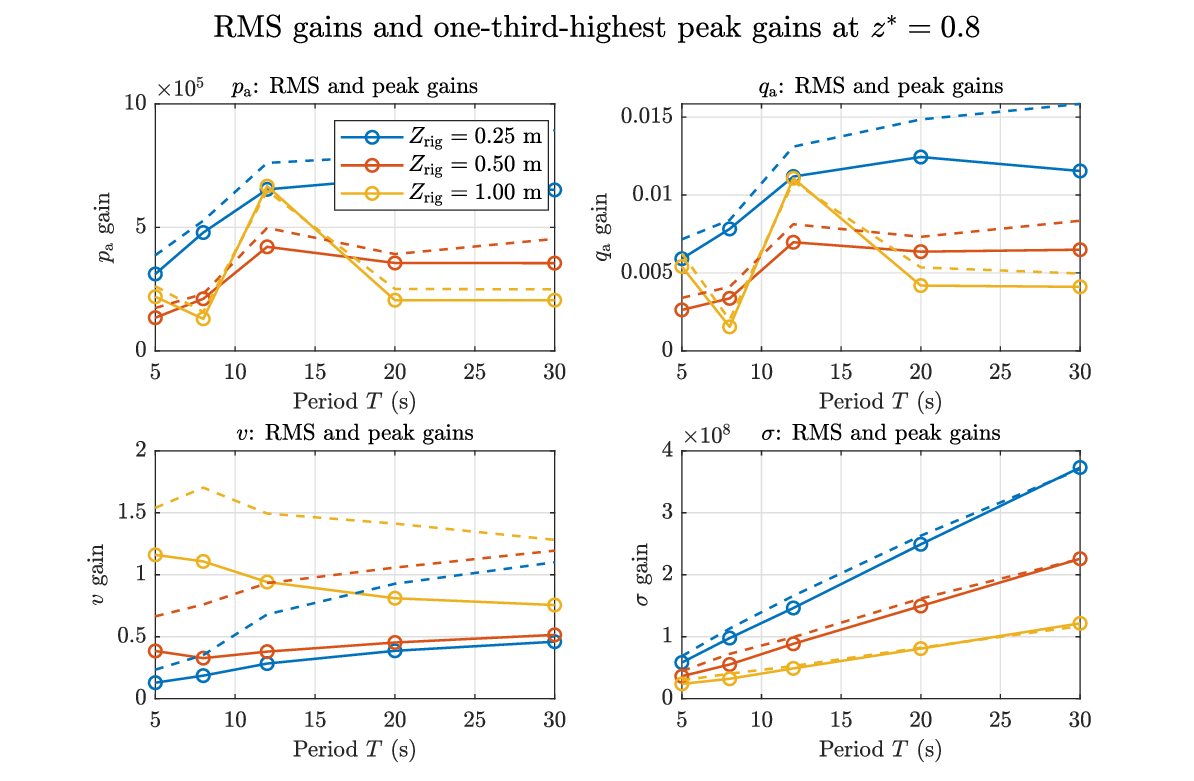} 
\caption{RMS and one-third-highest gains for the open pipe configuration. Line styles: RMS (marked lines), one-third-highest gains (dashed lines). Model parameters as in Table~\ref{tab:param}.}
\label{fig:5}
\end{figure}

Whereas harmonic gains quantify amplification at the fundamental frequency, practical drilling operations are often more sensitive to overall oscillation levels and extreme excursions. For this reason, \emph{root-mean-square} (RMS) gains and peak gains were additionally evaluated in simulation. The results, reported in Fig.~\ref{fig:5} for the open pipe configuration, demonstrate that resonant operating conditions identified through the harmonic analysis also correspond to elevated RMS and peak responses. Consequently, the resonance mechanisms revealed by Fig.~\ref{fig:2} appear to be directly relevant for fatigue loading, equipment wear, and operational risk. Particularly noteworthy is the behaviour of the structural variables, whose peak amplification can remain substantial even when the corresponding hydraulic amplification appears moderate. Overall, this finding highlights the necessity of considering structural indicators alongside pressure-based metrics when evaluating drilling-system performance.

 \subsection{Mechanism-isolation and parameter sensitivity analysis} \label{sect:MechIso}
The harmonic analysis conducted in Sect.~\ref{sect:Harmonic} provided preliminary insight into the effects of friction- and hydrodynamic-related phenomena on the structural and hydraulic variables. 

Building on these findings, the present section investigates the underlying mechanisms governing the system response (Sect.~\ref{sect:isol}), and presents a sensitivity analysis with respect to the friction parameters (Sect.~\ref{sect:sensistivity}).

 \subsubsection{Mechanism-isolation analysis} \label{sect:isol}
\begin{figure}
\centering
\includegraphics[width=1\linewidth]{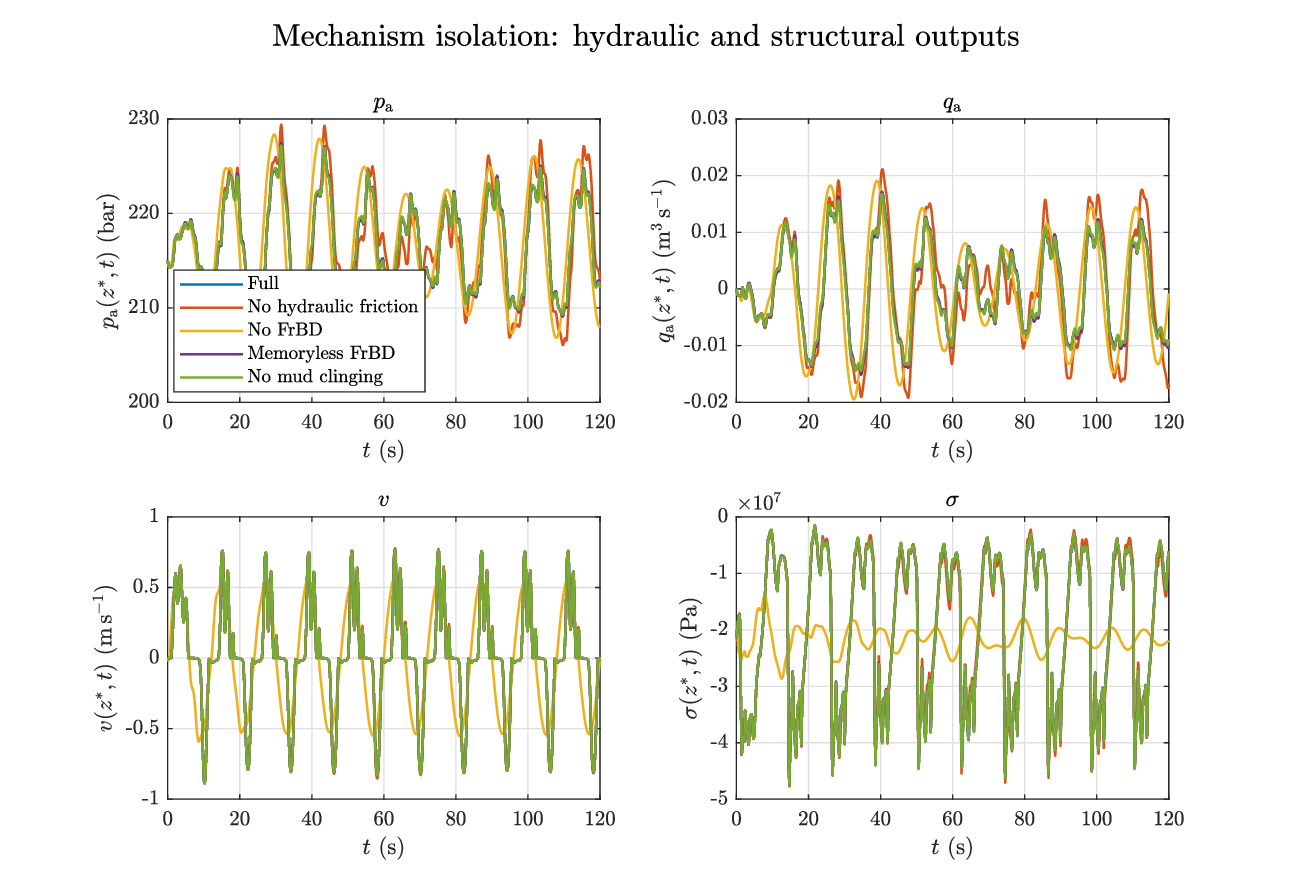} 
\caption{Mechanism isolation study of the system's response with different friction models (closed pipe configuration). Model parameters as in Table~\ref{tab:param}.}
\label{fig:6}
\end{figure}

To assess the relative importance of the individual physical phenomena, a sequence of mechanism-isolation simulations was also performed. The results are shown in Fig.~\ref{fig:6} for the closed pipe configuration, where the full model is compared with variants obtained by removing hydraulic friction, suppressing FrBD friction, replacing the dynamic FrBD formulation by a memoryless approximation based on the steady-state solution of Eq.~\eqref{eq:ODE}, and disregarding mud-clinging effects. 

As already anticipated, inspection of Fig.~\ref{fig:6} confirms that the hydraulic variables are heavily influenced by hydraulic friction and fluid-transport effects, although the inclusion of FrBD dynamics also appears to modify the pressure and flow responses. In this context, from the lower panels of Fig.~\ref{fig:6}, it may be concluded that the removal of the FrBD mechanism produces a substantial modification of both the drillstring velocity and the drillstring stress; by contrast, replacing the dynamic FrBD law with its memoryless counterpart induces comparatively small changes. This suggests that the most important factor is the presence of drillstring-wellbore friction itself, whereas memory effects associated with friction dynamics play a secondary role for the operating conditions investigated. As intuitively expected, distributed friction acts essentially as a structural mechanism influencing the interaction between the drillstring and wellbore, whereas the hydraulic variables remain fundamentally governed by fluid dynamics. 

These results suggest that the distributed motion of the string is only weakly affected by the hydraulics dynamics could therefore, in principle, be precomputed and used as an input to the hydraulic model to determine the corresponding pressures and flow rates.

\subsubsection{Parameter sensitivity analysis} \label{sect:sensistivity}

Although clarifying the qualitative role of hydrodynamic and drillstring-wellbore friction, the mechanism-isolation experiments reported in Sect.~\ref{sect:MechIso} do not quantify the system's sensitivity to variations in the corresponding model parameters. To address this point, a combinatorial sensitivity analysis was performed for the three coefficients that directly govern the hydraulic-drag and FrBD contributions, namely the hydrodynamic friction coefficient $k\ped{fric}$, the normalised micro-stiffness coefficient $\bar{k}_0$, and the relaxation time $\tau$. More specifically, the sensitivity study was conducted by varying one parameter at a time whilst retaining all other parameters at their nominal values. For $k\ped{fric}$ and $\tau$, the investigated values were defined as multiplicative perturbations of the corresponding nominal coefficients; for the micro-stiffness coefficient, a wider logarithmic range was considered, i.e., $\bar{k}_0 \in \{ 10^{5},\, 10^{6},\, 10^{7},\, 10^{8} \}$ $\textnormal{m}^{-1}$, to assess both the nominal regime and the progressively stiff limit of the FrBD dynamics. The latter sweep is especially relevant because $\bar{k}_0$ controls the deformation scale of the bristle-like contact element, and therefore determines how rapidly the dynamic friction law approaches its quasi-rigid limiting behaviour. For every combination of parameter values, the fundamental harmonic gain, RMS gain, and mean highest-third peak gain were computed. In particular, to facilitate comparisons between quantities having different physical units and substantially different magnitudes, the fundamental, RMS, and peak gains were normalised by their respective values obtained from the nominal parameter set.

\begin{figure}
\centering
\includegraphics[width=0.85\linewidth]{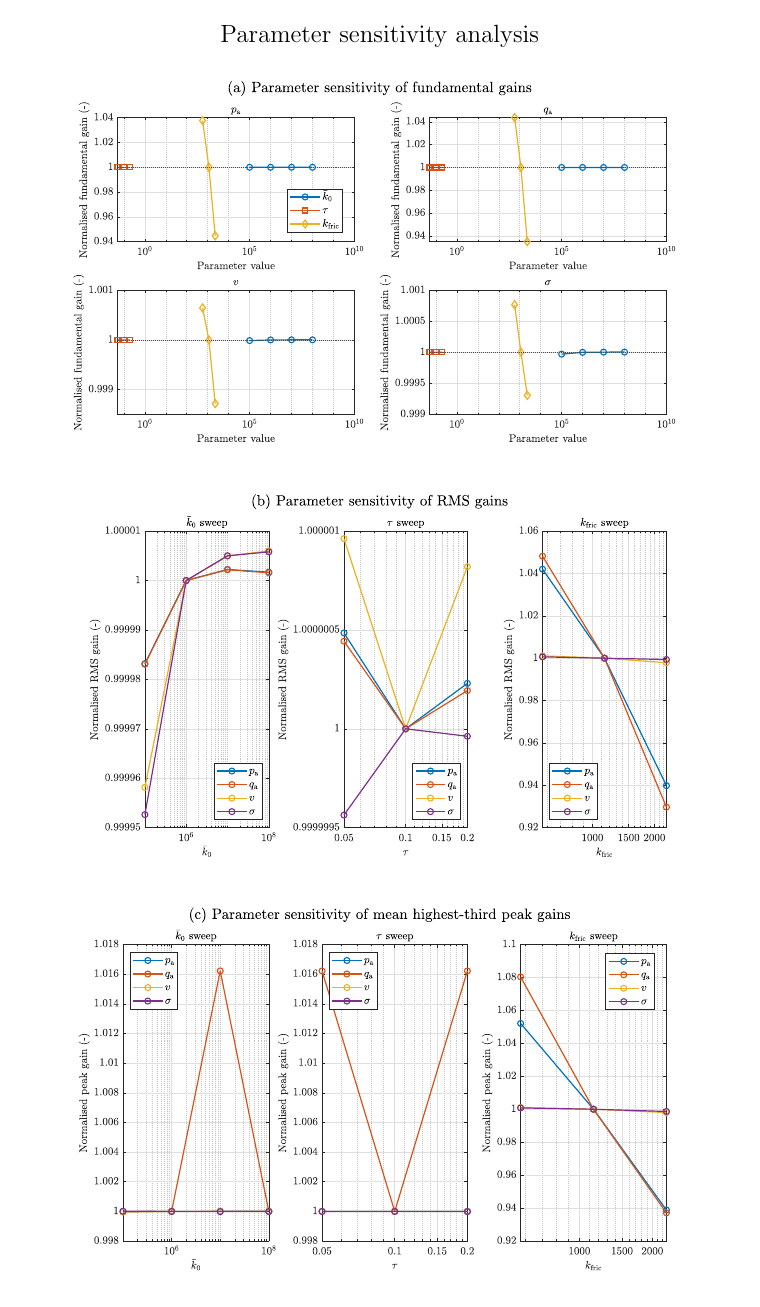} 
\caption{Sensitivity study to the friction parameters $\bar{k}_0$, $\tau$, and $k\ped{fric}$. Other model parameters as in Table~\ref{tab:param}}
\label{fig:sens}
\end{figure}

The results of the sensitivity analysis are visualised in Fig.~\ref{fig:sens} for the open pipe configuration, $z^* = 0.8$, and $T = 12$ s.
In particular, as expected from Sect.~\ref{sect:MechIso}, Fig.~\ref{fig:sens}(a) shows that $k\ped{fric}$ is the dominant parameter for the hydraulic response, with the fundamental pressure and flow gains decreasing monotonically from approximately $1.04$ to $0.94$--$0.95$ across the investigated range. Conversely, the velocity and stress gains remain within approximately $10^{-3}$ of unity. By contrast, variations in the FrBD coefficients $\bar{k}_0$ and $\tau$ produce no visible change in the fundamental gains. 
The RMS results in Fig.~\ref{fig:sens}(b) confirm this trend: bheavioural changes due to $k_0$ and $\tau$ are negligible, of the order of $10^{-5}$ and $10^{-6}$, respectively. Increasing $k\ped{fric}$, however, reduces the pressure and flow RMS gains from approximately $1.04$--$1.05$ to $0.93$--$0.94$, whilst the structural RMS gains remain nearly unchanged. The highest-third peak gains shown in Fig.~\ref{fig:sens}(c) exhibit the strongest hydraulic sensitivity, with the flow peak gain varying from approximately $1.08$ at the lowest $k\ped{fric}$ to $0.94$ at the highest value. The $\bar{k}_0$ and $\tau$ sweeps produce only a small, non-monotonic variation of approximately $1.6\%$ in the flow peak gain, with negligible changes in the other outputs. Overall, these results confirm that hydraulic amplification is controlled primarily by $k\ped{fric}$, whereas the structural outputs are robust with respect to all three parameters. The negligible dependence on $\bar{k}_0$ and $\tau$ also indicates that, under the investigated operating conditions, the response is only weakly affected by the precise stiffness and relaxation time of the FrBD model.

\subsection{Synthetic offshore excitation}\label{sect:offshore}
Albeit providing valuable insight into resonance mechanisms, simple harmonic excitations as those considered in Sects.~\ref{sect:Nominal}-\ref{sect:MechIso} may not be representative of realistic offshore operating conditions. To address this limitation, broadband rig-motion excitations were constructed using a synthetic sea-state model based on a JONSWAP spectrum with significant wave height $H\ped{s} = 4$ m and spectral peak period $T\ped{p} = 12$ s, considering a rig velocity input comprising 80 harmonic components uniformly distributed over the frequency interval $[0.025, 0.45]$ Hz, and combined with an idealised rig-heave \emph{response amplitude operator} (RAO). 

\begin{figure}
\centering
\includegraphics[width=0.9\linewidth]{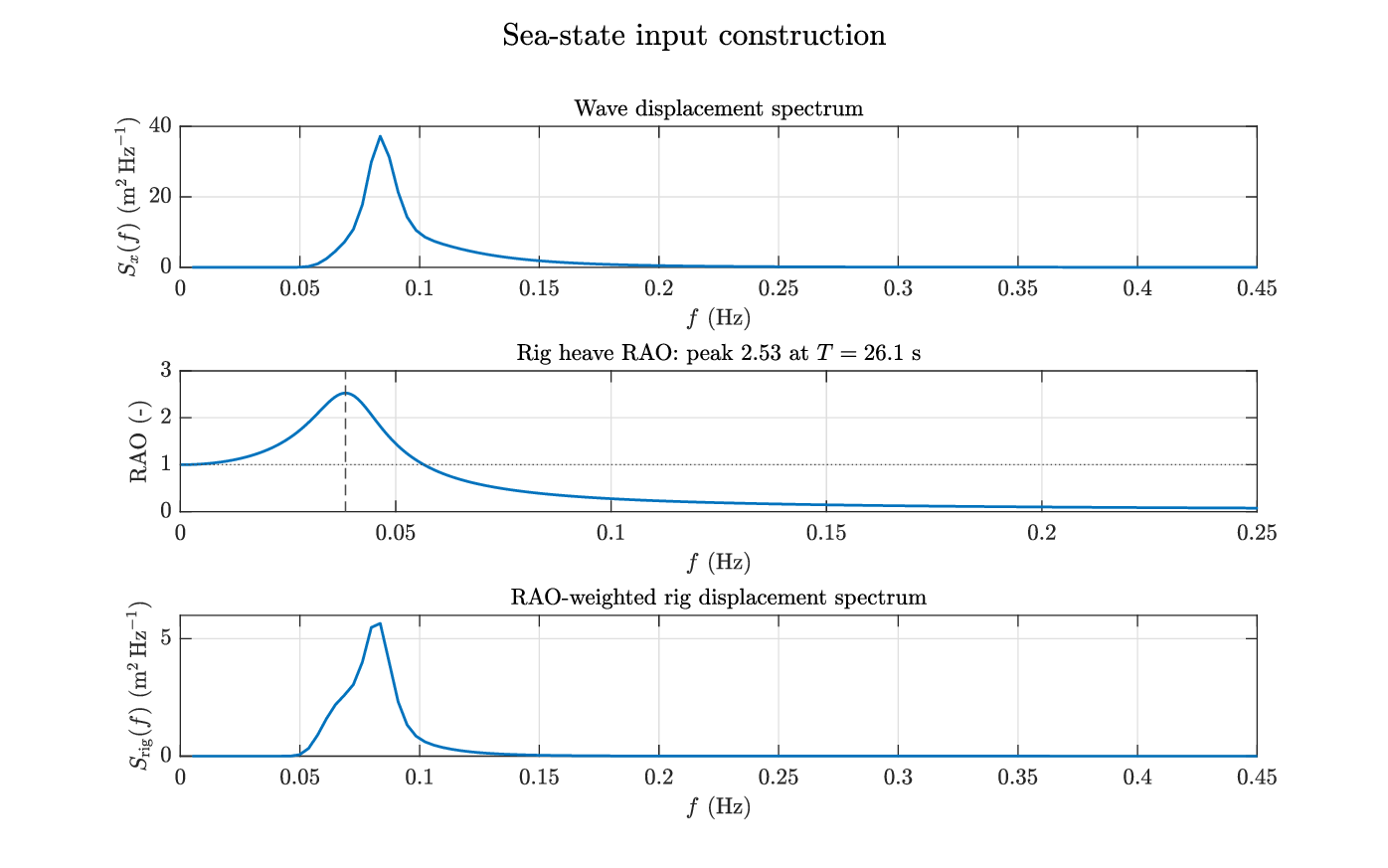} 
\caption{Synthetic offshore input.}
\label{fig:9}
\end{figure}

The corresponding spectral quantities are shown in Fig.~\ref{fig:9}, where the upper panel presents the wave-displacement spectrum $S_x(f)$, describing the distribution of wave-energy density over frequency, the middle panel displays the rig response-amplitude operator $\textnormal{RAO}(f)$, which characterises the heave dynamics of the floating platform and quantifies how efficiently wave motion is transmitted to the drilling rig, and the lower panel shows the modified spectrum $S\ped{rig}(f) \triangleq S_x(f)\textnormal{RAO}^2(f)$, which describes the displacement spectrum of the rig after accounting for the vessel dynamics. The broadband velocity excitation used in the simulations was subsequently generated from this RAO-weighted spectrum through a superposition of harmonic components with random phases. The significance of Fig.~\ref{fig:9} is that it establishes the spectral content of the imposed boundary excitation; consequently, the amplification mechanisms observed in the subsequent simulations can be interpreted relative to the underlying excitation frequencies. In particular, frequencies corresponding to peaks in both the RAO and the drilling-system transfer functions may lead to substantial dynamic amplification.

\subsubsection{Broadband sea-state simulations} 
\begin{figure}
\centering
\includegraphics[width=1\linewidth]{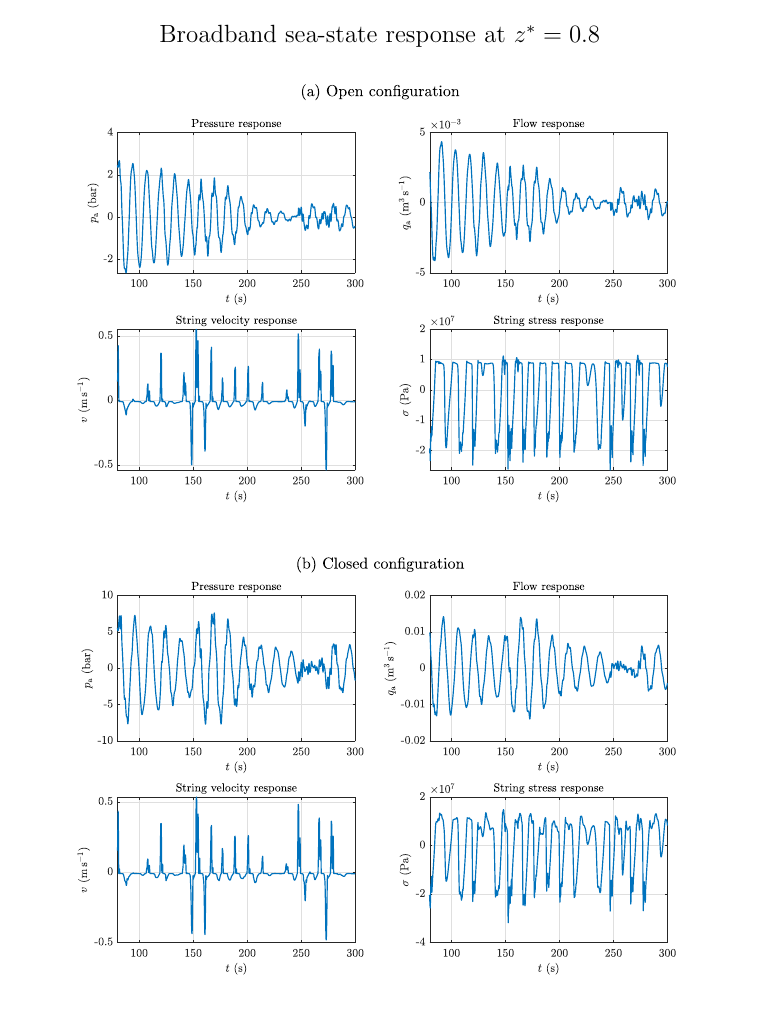} 
\caption{System's response to the synthetic offshore input: (a) open pipe configuration; (b) closed pipe configuration. Model parameters as in Table~\ref{tab:param}.}
\label{fig:7}
\end{figure}

The time-domain responses of the hydraulic and structural variables are plotted in Fig.~\ref{fig:7} for $z^* = 0.8$, concerning both the open and closed pipe configurations. The pressure, flow, velocity, and stress signals illustrated in Fig.~\ref{fig:7} all exhibit complex modulation patterns produced by the simultaneous excitation of multiple frequencies. Compared with the simple harmonic simulations of Sects.~\ref{sect:Nominal}-\ref{sect:MechIso}, broadband excitation reveals intermittent amplification events generated by the interaction of several wave components. In particular, large fluctuations may be observed in the structural variables, confirming that broadband offshore environments may significantly influence drillstring loading even when pressure fluctuations remain moderate. 

\subsubsection{Spectral amplification and coherence} 
\begin{figure}
\centering
\includegraphics[width=1\linewidth]{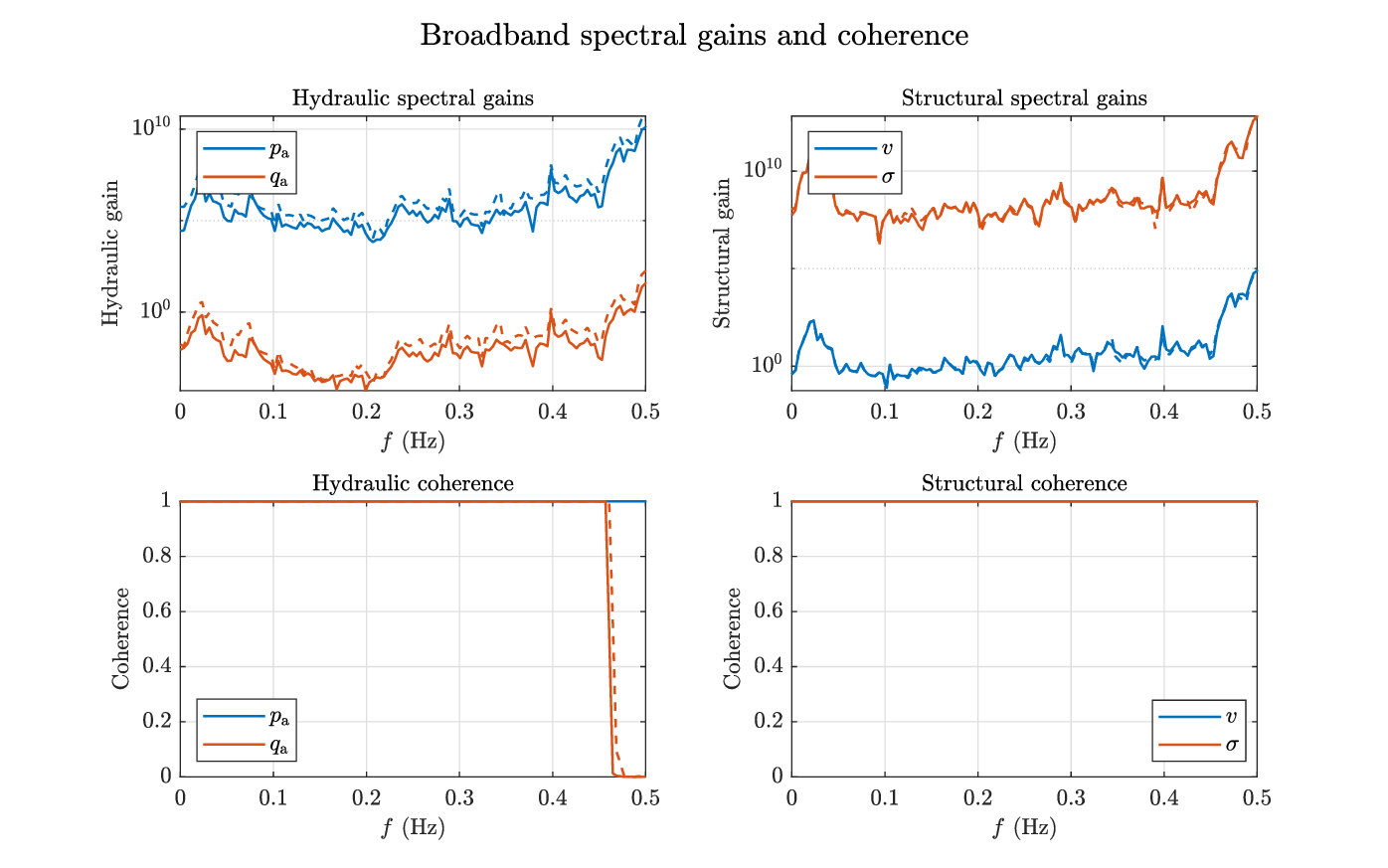} 
\caption{Hydraulic and structural system's gain and coherence indices for the synthetic offshore input. Line styles: open pipe configuration (solid lines); closed pipe configuration (dashed lines). Model parameters as in Table~\ref{tab:param}.}
\label{fig:8}
\end{figure}

The broadband simulations further enable the estimation of frequency-response characteristics directly from stochastic excitation data. Figure~\ref{fig:8} presents the resulting spectral gains and coherence functions, defined as
\begin{equation}
\gamma^2(f) = \dfrac{S_{yu}^2(f)}{S_{yy}(f)S_{uu}(f)},
\end{equation}
where $S_{uu}(f)$ and $S_{yy}(f)$ denote the power spectral densities of the input and output, respectively, and $S_{yu}(f)$ their cross-power spectral density. A value of $\gamma(f) \approx 1$ indicates that the output at frequency $f$ is almost completely explained by the input excitation; conversely, $\gamma(f) \approx0$ means that the output contains little information that can be explained by the input at that frequency.

Inspection of Fig.~\ref{fig:8} reveals that the hydraulic transfer functions exhibit frequency bands associated with elevated amplification, which coincide closely with the resonant regions identified in the harmonic analysis conducted in Sect.~\ref{sect:Harmonic}. This agreement confirms that the resonance phenomena observed under sinusoidal excitation remain relevant under realistic offshore conditions. In particular, the structural transfer functions reveal substantial amplification over a broader frequency range, particularly for drillstring stress. This observation is consistent with the mechanism-isolation study and further supports the conclusion that structural wave propagation plays a dominant role in the transmission of excitation energy. The coherence values remain close to unity over large portions of the frequency spectrum, demonstrating that the observed responses are strongly driven by the imposed rig-motion excitation rather than by numerical artefacts or autonomous oscillations.

\section{Conclusions}\label{sect:Conclusions}
This paper investigated the role of distributed friction on the disturbance propagation of heave-induced motion in offshore drilling systems. To this end, a hydraulic-structural model, combining compressible fluid transport, drillstring elasticity, hydraulic drag, mud-clinging effects, and drillstring-wellbore friction, was first developed, resulting into an interconnection of semilinear \emph{ordinary} and hyperbolic \emph{partial differential equations} (ODE--PDEs). The ODE--PDE system was then analysed theoretically and validated through an extensive numerical campaign encompassing harmonic excitation studies, nonlinear distortion analysis, spatial amplification investigations, mechanism-isolation experiments, parameter sensitivity analysis, and broadband offshore simulations. 

From a theoretical perspective, the well-posedness and stability results established in the paper revealed a clear distinction between the effects of hydraulic and frictional mechanisms on the mathematical properties of the system. In particular, distributed friction dynamics was shown to play a comparatively limited role in determining existence and uniqueness properties, but capable of affecting the stability margins of the coupled hydraulic-structural dynamics. Indeed, a conservative input-to-state stability criterion was derived using small-gain arguments, suggesting that uncontrolled friction may excite undesirable instabilities in offshore drilling systems. 

From a physical standpoint, simulation results demonstrated that offshore drilling systems behave as distributed waveguides capable of transmitting and amplifying disturbances over broad frequency ranges. Significant selectivity was observed, with resonance peaks emerging from constructive interactions between travelling and reflected waves. These amplification mechanisms were consistently identified through frequency-response computations, spectral analysis, and broadband sea-state simulations, indicating that they are not artefacts of idealised harmonic excitation but rather intrinsic features of the distributed drilling dynamics. The numerical experiments also revealed fundamentally different amplification mechanisms for hydraulic and structural responses, with the former depending primarily on fluid-wave propagation, compressibility effects, and hydraulic boundary conditions, and the latter exhibiting substantially greater sensitivity to distributed friction. 
The performed amplification analyses also demonstrated that pressure fluctuations may increase significantly as they propagate towards the lower sections of the well, indicating that surface measurements might not provide reliable information about the dynamic environment experienced by downhole components. 
Finally, broadband simulations based on synthetic offshore sea states confirmed that the amplification mechanisms identified through harmonic analysis remain relevant under realistic operating conditions. 

Overall, the results suggest that the dynamic behaviour of offshore drilling systems cannot be accurately described using purely hydraulic models. Whilst hydraulic transport and boundary conditions govern pressure amplification, distributed friction dynamics and structural wave propagation play a fundamental role in determining drillstring stability, velocity transmission, and stress loading. Future research will focus on developing refined models incorporating experimentally identified friction laws and more general boundary configurations, uncertainty quantification, as well as reduced-order control-oriented formulations suitable for real-time drilling optimisation and disturbance-rejection applications.

\section*{Funding declaration}
This research was financially supported by the project FASTEST (Reg. no. 2023-06511), funded by the Swedish Research Council. 


\section*{Compliance with Ethical Standards}

The authors declare that there are no conflicts of interest directly or indirectly related to this work.

\section*{Data availability}
The datasets used and/or analysed during the current study are available from the corresponding author on reasonable request.

\section*{Author Contribution declaration}
L.R. contributed to developing the models, carried out the mathematical analysis, implemented the code, performed all simulations, generated the figures, and wrote the manuscript. O.M.A. contributed to developing the models and provided valuable feedback. All authors reviewed and approved the final version of the manuscript.

\appendix

\section{Technical results}\label{app:techn}
This appendix contains the proofs of the technical results advocated in the manuscript. Specifically, Appendices~\ref{sect:App1} and~\ref{sect:App2} derive local and global well-posedness results for the ODE--PDE system~\eqref{eq:ODE--PDEsystem}, whereas Appendix~\ref{app;stabl} is dedicated to the stability analysis.

\subsection{Local well-posedness of the ODE--PDE system~\eqref{eq:ODE--PDEsystem}} \label{sect:App1}

The present appendix addresses the local well-posedness of the ODE--PDE interconnection~\eqref{eq:ODE--PDEsystem}. The proof invokes arguments similar to those presented in \cite[Appendix B]{Coron}, using an application of Banach's fixed point Theorem. As a first step, a transformation of the distributed states is introduced so that, when expressed in terms of the resulting Riemann variables, the exponential map $\bm{H}$ becomes a strict contraction.

\subsubsection{Exponential rescaling} 
Let $\gamma\in \mathbb{R}_{>0}$ and introduce the transformed Riemann variables $\mathbb{R}^4 \ni \widetilde{\bm{w}}^+(z,t)\triangleq \eu^{\gamma z}\bm{w}^+(z,t)$, $\mathbb{R}^4 \ni \widetilde{\bm{w}}^-(z,t)\triangleq \eu^{\gamma(1-z)}\bm{w}^-(z,t)$, so that $\mathbb{R}^8 \ni \widetilde{\bm{w}}(z,t) = [ \widetilde{\bm{w}}^{+\mathrm{T}}(z,t)\; \widetilde{\bm{w}}^{-\mathrm{T}}(z,t) ]^{\mathrm{T}}$, with $\mathbb{R}^4 \ni \widetilde{\bm{w}}^{\pm}(z,t) = [\widetilde{w}_1^{\pm}(z,t)\; \widetilde{w}_2^{\pm}(z,t)\;\widetilde{w}_3^{\pm}(z,t)\;\widetilde{w}_4^{\pm}(z,t)]^{\mathrm{T}}$.

For fixed $\gamma\in \mathbb{R}_{>0}$, this transformation is an isomorphism of $C^0([0,1];\mathbb{R}^8)$. Indeed,
\begin{align}
 \norm{\bm{w}(\cdot)}_{C^0([0,1];\mathbb{R}^8)} \leq \norm{\widetilde{\bm{w}}(\cdot)}_{C^0([0,1];\mathbb{R}^8)} \leq \eu^\gamma \norm{\bm{w}(\cdot)}_{C^0([0,1];\mathbb{R}^8)}. 
\end{align}
Therefore, local well-posedness in the transformed variables $(\bm{X}(z,t),\widetilde{\bm{w}}(z,t))$ is equivalent to local well-posedness in the original variables $(\bm{X}(z,t),\bm{w}(z,t))$. In particular, the transformed system reads
\begin{subequations} \label{eq:wp_transformed_system} 
\begin{align} \dpd{\bm{X}(z,t)}{t} &= \widetilde{\bm{F}}\bigl(\bm{X}(z,t),\widetilde{\bm{w}}(z,t),z\bigr), \label{eq:wp_transformed_system_X} \\ 
\dpd{\widetilde{\bm{w}}(z,t)}{t} + \mathbf{\Lambda} \dpd{\widetilde{\bm{w}}(z,t)}{z} &= \widetilde{\bm{G}} \bigl(\bm{X}(z,t),\widetilde{\bm{w}}(z,t),z\bigr), \quad (z,t) \in (0,1)\times(0,T), \label{eq:wp_transformed_system_w} \\ 
\begin{bmatrix} \widetilde{\bm{w}}^+(0,t)\\ \widetilde{\bm{w}}^-(1,t) \end{bmatrix} &= \widetilde{\bm{H}} \bigl( \widetilde{\bm{w}}^+(1,t), \widetilde{\bm{w}}^-(0,t), \bm{U}(t) \bigr), \quad t \in (0,T), \label{eq:wp_transformed_system_bc} 
\end{align} 
\end{subequations} where $\widetilde{\bm{F}}\in C^0(\mathbb{R}^{n_{X}+9};\mathbb{R}^{n_{X}})$ and $\widetilde{\bm{G}} \in C^0(\mathbb{R}^{n_{X}+9};\mathbb{R}^8)$ are given by
\begin{subequations}
\begin{align}
\widetilde{\bm{F}}(\bm{X},\widetilde{\bm{w}},z) & \triangleq \bm{F}\bigl(\bm{X},\mathcal{E}_\gamma(z)\widetilde{\bm{w}},z\bigr), \\ 
\widetilde{\bm{G}}(\bm{X},\widetilde{\bm{w}},z) & \triangleq \begin{bmatrix} \gamma\mathbf{\Lambda}_0\widetilde{\bm{w}}^+ + \eu^{\gamma z} \bm{G}^+\bigl(\bm{X},\mathcal{E}_\gamma(z)\widetilde{\bm{w}},z\bigr) \\ \gamma\mathbf{\Lambda}_0\widetilde{\bm{w}}^- + \eu^{\gamma(1-z)} \bm{G}^-\bigl(\bm{X},\mathcal{E}_\gamma(z)\widetilde{\bm{w}},z\bigr) \end{bmatrix},
\end{align}
\end{subequations} 
with
\begin{align}\label{eq:esponentialTransf}
\mathcal{E}_\gamma(z)\widetilde{\bm{w}} \triangleq \begin{bmatrix} \eu^{-\gamma z}\widetilde{\bm{w}}^+\\ \eu^{-\gamma(1-z)}\widetilde{\bm{w}}^- \end{bmatrix},
\end{align}
and the transformed boundary map is
\begin{align}
\widetilde{\bm{H}}(\widetilde{\bm{w}}^+,\widetilde{\bm{w}}^-,\bm{U}) \triangleq \bm{H}(\eu^{-\gamma}\widetilde{\bm{w}}^+,\eu^{-\gamma}\widetilde{\bm{w}}^-,\bm{U}). 
\end{align}
Finally, the transformed ICs are given by $\mathbb{R}^4 \ni \widetilde{\bm{w}}_0^+(z)=\eu^{\gamma z}\bm{w}_0^+(z)$, $\mathbb{R}^4 \ni \widetilde{\bm{ w}}_0^-(z)=\eu^{\gamma(1-z)}\bm{w}_0^-(z)$, with $\mathbb{R}^4 \ni \widetilde{\bm{w}}_0^{\pm}(z,t) = [\widetilde{w}_{0,1}^{\pm}(z,t)\; \widetilde{w}_{0,2}^{\pm}(z,t)\;\widetilde{w}_{0,3}^{\pm}(z,t)\;\widetilde{w}_{0,4}^{\pm}(z,t)]^{\mathrm{T}}$, and the corresponding compatibility condition becomes
\begin{align}\label{eq:compatTrans}
\begin{bmatrix} \widetilde{\bm{w}}_0^+(0)\\ \widetilde{\bm{w}}_0^-(1) \end{bmatrix} = \widetilde{\bm{H}} \bigl( \widetilde{\bm{w}}_0^+(1), \widetilde{\bm{w}}_0^-(0), \bm{U}(0) \bigr). 
\end{align}
For the sequel, it is also desirable to define the components of the functions $\widetilde{\bm{G}}$ and $\widetilde{\bm{H}}$ as $\widetilde{\bm{G}} \triangleq [\widetilde{\bm{G}}^{+\mathrm{T}}\; \widetilde{\bm{G}}^{-\mathrm{T}}]^{\mathrm{T}}$, with $\widetilde{\bm{G}}^{\pm} \triangleq [\widetilde{G}_1^{\pm}\; \widetilde{G}_2^{\pm}\; \widetilde{G}_3^{\pm}\; \widetilde{G}_4^{\pm}]^{\mathrm{T}}$, and $\widetilde{\bm{H}} \triangleq [\widetilde{\bm{H}}^{+\mathrm{T}}\; \widetilde{\bm{H}}^{-\mathrm{T}}]^{\mathrm{T}}$, with $\widetilde{\bm{H}}^{\pm} \triangleq [\widetilde{H}_1^{\pm}\; \widetilde{H}_2^{\pm}\; \widetilde{H}_3^{\pm}\;\widetilde{H}_4^{\pm}]^{\mathrm{T}}$.
 
\paragraph{Verification that $\widetilde{\bm{H}}$ is a contraction for large $\gamma$.}
For the transformed boundary map, it follows from Eq.~\eqref{eq:globalHLips} that 
\begin{align} \label{H_tildeContr}
\norm{\widetilde{\bm{H}}(\bm{a}_1,\bm{b}_1,\bm{U})-\widetilde{\bm{H}}(\bm{a}_2,\bm{b}_2,\bm{U})}_{\ell^1(\mathbb{R}^8)} \leq L_\gamma \norm{(\bm{a}_1-\bm{a}_2,\bm{b}_1-\bm{b}_2)}_{\ell^1(\mathbb{R}^8)}, \quad (\bm{a}_1,\bm{a}_2,\bm{b}_1,\bm{b}_2,\bm{U})\in \mathbb{R}^{16+n_U},
\end{align}
with $\mathbb{R}_{\geq0}\ni L_\gamma \triangleq L_H\eu^{-\gamma}$. Hence, choosing $\gamma\in \mathbb{R}_{>0}$ sufficiently large yields an arbitrarily small $ L_\gamma\in (0,1)$, which implies that the exponential rescaling renders the boundary map a strict contraction. Moreover, from Eq.~\eqref{eq:h_0bound}, it may be concluded that
\begin{align}\label{eq:supHtilde}
\sup_{t\in \mathbb{R}_{\geq 0}} \norm{\widetilde{\bm{H}}(\bm{0},\bm{0},\bm{U}(t))}_{\ell^1(\mathbb{R}^8)} = M_H.
\end{align}

\subsubsection{Preliminary bounds}
Next, some preliminary bounds are established, which are propaedeutic to the analysis conducted in Sect.~\ref{sect:analysisFixed}.

\paragraph{Norms and closed ball.} 
For notational convenience, it is beneficial to introduce $\mathbb{R}^{n_{Y}}\ni \bm{Y}(z,t) \triangleq [ \bm{X}^{\mathrm{T}}(z,t)\; \widetilde{\bm{w}}^{+\mathrm{T}}(z,t)\; \widetilde{\bm{w}}^{-\mathrm{T}}(z,t)]^{\mathrm{T}}$, with $n_{Y} \triangleq n_{X} + 8$, along with the vector of initial conditions $\mathbb{R}^{n_{Y}} \ni \bm{Y}_0(z) \triangleq \bm{Y}(z,0) = [\bm{X}_0^{\mathrm{T}}(z)\; \widetilde{\bm{w}}_0^{+\mathrm{T}}(z)\; \widetilde{\bm{w}}_0^{-\mathrm{T}}(z)]^{\mathrm{T}}$. Accordingly, the following space is considered for $\delta\in \mathbb{R}_{>0}$:
\begin{align} 
\mathcal X_\delta \triangleq C^0\bigl([0,\delta];C^0([0,1];\mathbb{R}^{n_{Y}})\bigr), 
\end{align}
equipped with the norm
\begin{align}
\norm{\bm{Y}(\cdot,\cdot)}_{\mathcal{X}_\delta} \triangleq \sup_{t\in[0,\delta]} \sup_{z\in[0,1]} \norm{\bm{Y}(z,t)}_{\ell^1(\mathbb{R}^{n_{Y}})} = \sup_{t\in[0,\delta]} \norm{\bm{Y}(\cdot,t)}_{C^0([0,1];\ell^1(\mathbb{R}^{n_{Y}}))},  
\end{align}
For $\rho(\bm{Y}_0) \in \mathbb{R}_{>0}$, the closed ball $\mathcal{B}_{\rho}(\bm{Y}_0)$ is defined as
\begin{align}\label{eq:Ball}
\mathcal{B}_{\rho}(\bm{Y}_0) \triangleq \biggl\{ \bm{Y}\in\mathcal X_\delta \mathrel{\bigg|} \bm{Y}(\cdot,0)=\bm{Y}_0, \; \norm{\bm{Y}(\cdot,\cdot)}_{\mathcal X_\delta}\leq \rho(\bm{Y}_0) \biggr\}.  
\end{align}
The set $\mathcal{B}_{\rho}(\bm{Y}_0)$ in Eq.~\eqref{eq:Ball} is closed in the complete space $\mathcal{X}_\delta$, and is therefore complete.

\paragraph{Bounds.} 
Defining $\bm{Y}_1(z,t) = (\bm{X}_1(z,t),\widetilde{\bm{w}}_1(z,t))$ and $\bm{Y}_2(z,t) = (\bm{X}_2(z,t),\widetilde{\bm{w}}_2(z,t))$ yields
\begin{align}\label{eq:LipschitzFG}
\begin{split}
& \norm{\widetilde{\bm{F}}(\bm{Y}_1(z,t),z)-\widetilde{\bm{F}}(\bm{Y}_2(z,t),z)}_{\ell^1(\mathbb{R}^{n_{X}})}, \norm{\widetilde{\bm{G}}(\bm{Y}_1(z,t),z)-\widetilde{\bm{G}}(\bm{Y}_2(z,t),z)}_{\ell^1(\mathbb{R}^{n_{8}})} \\
&\qquad \leq L_\rho\bigl(\rho(\bm{Y}_0)\bigr)\norm{\bm{Y}_1(z,t)-\bm{Y}_2(z,t)}_{\ell^1(\mathbb{R}^{n_Y})}, \quad \mathcal{B}_{\rho}(\bm{Y}_0)\times\mathcal{B}_{\rho}(\bm{Y}_0)\times[0,1],
\end{split}
\end{align}
for some $L_\rho(\rho(\bm{Y}_0))\in \mathbb{R}_{\geq 0}$.
Moreover, from the condition $\bm{F}(\bm{0},\cdot) \in C^0([0,1];\mathbb{R}^{n_X})$ and $\bm{G}(\bm{0},\cdot) \in C^0([0,1];\mathbb{R}^{8})$, it follows the existence of $M_{FG} \in \mathbb{R}_{\geq 0}$ such that
\begin{align}\label{eq:boundFG}
\norm{\widetilde{\bm{F}}(\bm{0},z)}_{\ell^1(\mathbb{R}^{n_{X}})}, \norm{\widetilde{\bm{G}}(\bm{0},z)}_{\ell^1(\mathbb{R}^{n_{X}})}& \leq M_{FG}, \quad z \in [0,1].
\end{align}

\subsubsection{Fixed-point map}\label{sect:analysisFixed}
The adopted strategy relies on invoking Banach's fixed-point theorem to prove local existence and uniqueness in time.
To this end, for $\bm{Y}\in\mathcal B_{\rho}(\bm{Y}_0)$, the operator $\bm{\Phi} : B_{\rho}(\bm{Y}_0) \to \mathcal{X}_\delta$, with $\bm{\Phi}(\bm{Y})=[\bm{\Phi}^{\bm{X}\mathrm{T}}(\bm{Y})\; \bm{\Phi}^{+\mathrm{T}}(\bm{Y})\; \bm{\Phi}^{-\mathrm{T}}(\bm{Y})]^{\mathrm{T}}$ is introduced, with $\bm{\Phi}^{\pm}(\bm{Y}) = [\Phi_1^{\pm}(\bm{Y})\; \Phi_2^{\pm}(\bm{Y})\; \Phi_3^{\pm}(\bm{Y}) \; \Phi_4^{\pm}(\bm{Y})]^{\mathrm{T}}$ obtained from the characteristic formulae associated with the transformed system. 
In particular, the ODE component is 
\begin{align}\label{eq:ODE_Phi}
\bigl(\bm{\Phi}^{\bm{X}}(\bm{Y})\bigr)(z,t) \triangleq \bm{X}_0(z) + \int_{0}^{t} \widetilde{\bm{F}} \bigl(\bm{Y}(z,t^\prime),z\bigr)\dif t^\prime. 
\end{align}
Concerning the PDE variables, it is first beneficial to define the positive characteristics through $(z,t)$ and the corresponding hitting times at $z=0$: 
\begin{subequations}\label{eq:tauZ+}
\begin{align}
 z_i^+(z,t,\tilde{t}) & \triangleq z-\lambda_i(t-\tilde{t}),  \\
\tau_i^+(z,t) & \triangleq t-\dfrac{z}{\lambda_i}, && i \in \{1,2,3,4\}.
\end{align}
\end{subequations}
Similarly, the negative characteristics through $(z,t)$ and their hitting times at $z=1$ are defined as 
\begin{subequations}
\begin{align}
z_i^-(z,t,\tilde{t}) & \triangleq z+\lambda_i(t-\tilde{t}), \\ 
\tau_i^-(z,t) & \triangleq t-\frac{1-z}{\lambda_i}, && i \in \{1,2,3,4\},
\end{align}
\end{subequations}
Consequently, the operators $\bm{\Phi}^+(\bm{Y})$ and $\bm{\Phi}^-(\bm{Y})$ may be deduced in components as
\begin{subequations}\label{eq:PhiPDEs}
\begin{align}
\bigl(\Phi_i^{\pm}(\bm{Y})\bigr)(z,t) & = \widetilde{w}_{0,i}^{\pm} \bigl(z_i^{\pm}(z,t,0)\bigr) + \int_{0}^{t} \widetilde{G}_i^{\pm} \bigl(\bm{Y}(z_i^{\pm}(z,t,t^\prime),t^\prime),z_i^{\pm}(z,t,t^\prime)\bigr)\dif t^\prime, \quad \tau_i^{\pm}(z,t)\leq 0, \label{eq:PHitah<} \\
\begin{split}
\bigl(\Phi_i^{\pm}(\bm{Y})\bigr)(z,t) &= \widetilde{H}_i^{\pm} \bigl( \widetilde{\bm{w}}^+(1,\tau_i^{\pm}(z,t)), \widetilde{\bm{w}}^-(0,\tau_i^{\pm}(z,t)), \bm{U}(\tau_i^{\pm}(z,t)) \bigr) \\ &\quad+ \int_{\tau_i^{\pm}(z,t)}^{t} \widetilde{G}_i^{\pm} \bigl(\bm{Y}(z_i^{\pm}(z,t,t^\prime),t^\prime),z_i^{\pm}(z,t,t^\prime)\bigr)\dif t^\prime, \quad \tau_i^{\pm}(z,t)>0, \; i \in \{1,2,3,4\}.\label{eq:PHitah>} 
\end{split}
\end{align}
\end{subequations}
Next, it must be proved that, for $\delta\in \mathbb{R}_{>0}$ sufficiently small, $\bm{\Phi}(\mathcal{B}_{\rho}(\bm{Y}_0))\subset \mathcal{B}_{\rho}(\bm{Y}_0)$.

\paragraph{Ball mapping.} Starting from the ODE variables, it follows immediately from Eqs.~\eqref{eq:ODE_Phi},~\eqref{eq:LipschitzFG}, and~\eqref{eq:boundFG} that
\begin{align} \label{eq:ODeestimate}
\begin{split}
\norm{\bm{\Phi}^{\bm{X}}(\bm{Y})(z,t)}_{\ell^1(\mathbb{R}^{n_{X}})} & \leq \norm{\bm{X}_0(z)}_{\ell^1(\mathbb{R}^{n_{\bm{X}}})} + \int_{0}^{t} \norm{\widetilde{\bm{F}} \bigl(\bm{Y}(z,t^\prime),z\bigr)}_{\ell^1(\mathbb{R}^{n_{X}})}\dif t^\prime \\
& \leq \norm{\bm{Y}_0(\cdot)}_{C^0([0,1];\ell^1(\mathbb{R}^{n_Y}))} + \Bigl( L_\rho\bigl(\rho(\bm{Y}_0)\bigr)\rho(\bm{Y}_0) + M_{FG}\Bigr)\delta.
\end{split}
\end{align}
Regarding the PDE states, the cases $\tau_i^\pm(z,t)\leq 0$, $i \in \{1,2,3,4\}$, are first considered. 
For both positive and negative characteristics reaching the initial line, Eqs.~\eqref{eq:PHitah<},~\eqref{eq:LipschitzFG}, and~\eqref{eq:boundFG} give
\begin{align} \label{eq:estInterPhi+-tau<0}
\begin{split}
\abs{(\Phi_i^{\pm}(\bm{Y}))(z,t)} &\leq \abs{\widetilde w_{0,i}^{\pm} \bigl(z_i^{\pm}(z,t,0)\bigr)} + \int_{0}^{t} \abs{\widetilde{G}_i^{\pm} \bigl(\bm{Y}(z_i^{\pm}(z,t,t^\prime),t^\prime),z_i^{\pm}(z,t,t^\prime)\bigr)}\dif t^\prime \\
& \leq \norm{\bm{Y}_0(\cdot)}_{C^0([0,1];\ell^1(\mathbb{R}^{n_Y}))} + \Bigl( L_\rho\bigl(\rho(\bm{Y}_0)\bigr)\rho(\bm{Y}_0) + M_{FG}\Bigr)\delta.
\end{split}
\end{align}
Considering instead the cases $\tau_i^{\pm}(z,t)>0$, $i \in \{1,2,3,4\}$, Eqs.~\eqref{H_tildeContr},~\eqref{eq:supHtilde},~\eqref{eq:LipschitzFG}, and~\eqref{eq:boundFG} provide the following estimate:  
\begin{align} \label{eq:estInterPhi+}
\begin{split}
\abs{(\Phi_i^{\pm}(\bm{Y}))(z,t)} & \leq \abs{ \widetilde{H}_i^\pm \bigl( \widetilde{\bm{w}}^+(1,\tau_i^\pm(z,t)), \widetilde{\bm{w}}^-(0,\tau_i^\pm(z,t)), \bm{U}(\tau_i^\pm(z,t)) \bigr)}  \\
& \quad + \int_{\tau_i^\pm(z,t)}^{t} \abs{\widetilde{G}_i^{\pm} \bigl(\bm{Y}(z_i^{\pm}(z,t,t^\prime),t^\prime),z_i^{\pm}(z,t,t^\prime)\bigr)}\dif t^\prime \\
& \leq L_\gamma \rho(\bm{Y}_0) + \Bigl( L_\rho\bigl(\rho(\bm{Y}_0)\bigr)\rho(\bm{Y}_0) + M_{FG}\Bigr)\delta.
\end{split}
\end{align} 
Thus, combining Eq.~\eqref{eq:estInterPhi+-tau<0} with~\eqref{eq:estInterPhi+} yields
\begin{align}
\abs{(\Phi_i^{\pm}(\bm{Y}))(z,t)} &\leq  \norm{\bm{Y}_0(\cdot)}_{C^0([0,1];\ell^1(\mathbb{R}^{n_Y}))} +  L_\gamma \rho(\bm{Y}_0) + \Bigl(L_\rho\bigl(\rho(\bm{Y}_0)\bigr)\rho(\bm{Y}_0) + M_{FG}\Bigr)\delta,
\end{align}
for $i \in \{1,2,3,4\}$. Summing over the different components and adding Eq.~\eqref{eq:ODeestimate} leads finally to
\begin{align}
\norm{(\bm{\Phi}(\bm{Y}))(\cdot,\cdot)}_{\mathcal{X}_\delta} \leq 9\norm{\bm{Y}_0(\cdot)}_{C^0([0,1];\ell^1(\mathbb{R}^{n_Y}))} + 8L_\gamma \rho(\bm{Y}_0)+ 8M_H + 9\Bigl( L_\rho\bigl(\rho(\bm{Y}_0)\bigr) \rho(\bm{Y}_0)+ M_{FG}\Bigr)\delta.
\end{align}
Hence, selecting $\delta \in \mathbb{R}_{>0}$ as
\begin{align}
\delta \leq \dfrac{9\norm{\bm{Y}_0(\cdot)}_{C^0([0,1];\ell^1(\mathbb{R}^{n_Y}))} + 8L_\gamma \rho(\bm{Y}_0)+ 8M_H}{9 L_\rho\bigl(\rho(\bm{Y}_0)\bigr)\rho(\bm{Y}_0) + 9M_{FG}},
\end{align}
recalling that $L_\gamma \in (0,1)$ may be made arbitrarily small, and specifying $\rho(\bm{Y}_0) \in \mathbb{R}_{>0}$ as
\begin{align}
\rho(\bm{Y}_0) & \triangleq \dfrac{18\norm{\bm{Y}_0(\cdot)}_{C^0([0,1];\ell^1(\mathbb{R}^{n_Y}))}+16M_H}{1-16L_\gamma},
\end{align}
gives
\begin{align}
\norm{(\bm{\Phi}(\bm{Y}))(\cdot,\cdot)}_{\mathcal{X}_\delta} \leq \rho(\bm{Y}_0),
\end{align}
implying that $\bm{\Phi}(\bm{Y})\in\mathcal{B}_{\rho}(\bm{Y}_0)$. Since $\bm{Y}\in\mathcal{B}_{\rho}(\bm{Y}_0)$ was arbitrary, it may be concluded that $\bm{\Phi}(\mathcal{B}_{\rho}(\bm{Y}_0)) \subset \mathcal{B}_{\rho}(\bm{Y}_0)$. In the following, it is proved that $\bm{\Phi}$ is a strict contraction on $\mathcal{B}_{\rho}(\bm{Y}_0)$.

 \paragraph{Contraction estimate.} 
Let $\bm{Y}_1,\bm{Y}_2\in\mathcal{B}_{\rho}(\bm{Y}_0)$. For the ODE component, Eq.~\eqref{eq:LipschitzFG} immediately provides
\begin{align}
\norm{(\bm{\Phi}^{\bm{X}}(\bm{Y}_1))(z,t)-(\bm{\Phi}^{\bm{X}}(\bm{Y}_2))(z,t)}_{\ell^1(\mathbb{R}^{n_X})} \leq L_\rho\bigl(\rho(\bm{Y}_0)\bigr)\delta \norm{\bm{Y}_1(\cdot,\cdot)-\bm{Y}_2(\cdot,\cdot)}_{\mathcal{X}_\delta}. 
\end{align}
Additionally, for PDE components whose backward characteristics reach the initial line, the initial data cancel, and the same estimate holds:
\begin{align}
\abs{(\Phi_i^{\pm}(\bm{Y}_1))(z,t)-(\Phi_i^{\pm}(\bm{Y}_2))(z,t)} \leq L_\rho\bigl(\rho(\bm{Y}_0)\bigr)\delta \norm{\bm{Y}_1(\cdot,\cdot)-\bm{Y}_2(\cdot,\cdot)}_{\mathcal{X}_\delta}. 
\end{align}
On the other hand, for PDE components whose backward characteristics reach the boundary, the additional boundary contribution is controlled by the contraction property of $\widetilde{\bm{H}}$:
 \begin{align}
\begin{split} 
&\abs{\widetilde{H}_i^{\pm} \bigl( \widetilde{\bm{w}}_1^+(1,\tau_i^{\pm}(z,t)), \widetilde{\bm{w}}_1^-(0,\tau_i^{\pm}(z,t)), \bm{U}(\tau_i^{\pm}(z,t)) \bigr) - \widetilde{H}_i^{\pm} \bigl( \widetilde{\bm{w}}_2^+(1,\tau_i^{\pm}(z,t)), \widetilde{\bm{w}}_2^-(0,\tau_i^{\pm}(z,t)), \bm{U}(\tau_i^{\pm}(z,t)) \bigr)} \\ &\leq  L_\gamma \norm{\bm{Y}_1(\cdot,\cdot)-\bm{Y}_2(\cdot,\cdot)}_{\mathcal{X}_\delta}. 
\end{split}
\end{align} 
Therefore, simple manipulations yield
\begin{align}\label{eq:contract}
\norm{(\bm{\Phi}(\bm{Y}_1))(\cdot,\cdot)-(\bm{\Phi}(\bm{Y}_2))(\cdot,\cdot)}_{\mathcal{X}_\delta} \leq \Bigl(8 L_\gamma+9L_\rho\bigl(\rho(\bm{Y}_0)\bigr)\delta \Bigr) \norm{\bm{Y}_1(\cdot,\cdot)-\bm{Y}_2(\cdot,\cdot)}_{\mathcal{X}_\delta}, 
\end{align}
Again, since $L_\gamma \in (0,1)$ may be chosen arbitrarily small, $\delta \in \mathbb{R}_{>0}$ may be selected to satisfy
\begin{align}\label{€q:contrConst}
\delta \leq \delta\ped{min}(\bm{Y}_0) \triangleq \min\bigggl\{\dfrac{9\norm{\bm{Y}_0(\cdot)}_{C^0([0,1];\ell^1(\mathbb{R}^{n_Y}))} + 8L_\gamma \rho(\bm{Y}_0)+ 8M_H}{9L_\rho\bigl(\rho(\bm{Y}_0)\bigr)\rho(\bm{Y}_0) + 9M_{FG}}, \dfrac{1-8L_\gamma}{9L_\rho\bigl(\rho(\bm{Y}_0)\bigr)}\bigggr\},
\end{align}
rendering~\eqref{eq:contract} a strict contraction.

\subsubsection{Proof of Theorem~\ref{thm:Local}}

The proof of Theorem~\ref{thm:Local} is finally given below.
\begin{proof}[Proof of Theorem~\ref{thm:Local}]
The estimate~\eqref{eq:contract} together with~\eqref{€q:contrConst} implies that, for sufficiently small $\delta \in \mathbb{R}_{>0}$, the mapping $\bm{\Phi} : \mathcal{B}_{\rho}(\bm{Y}_0) \to \mathcal{X}_\delta$ defined as in Eqs.~\eqref{eq:ODE_Phi} and~\eqref{eq:PhiPDEs} is a strict contraction on $\mathcal{B}_{\rho}(\bm{Y}_0)$. By Banach's fixed-point theorem, $\bm{\Phi}(\bm{Y})$ admits a unique fixed point $\bm{Y}=(\bm{X},\widetilde{\bm{w}}) \in C^0([0,\delta];C^0([0,1];\mathbb{R}^{n_{Y}}))$, which is the unique generalised solution of the transformed system~\eqref{eq:wp_transformed_system}. Since the exponential change of variables~\eqref{eq:esponentialTransf} is an isomorphism of $C^0([0,1];\mathbb{R}^8)$, the original system~\eqref{eq:ODE--PDEsystem} is also locally well-posed in $C^0([0,\delta];C^0([0,1];\mathbb{R}^{n_{X}+8})) \cong C^0([0,1]\times[0,\delta];\mathbb{R}^{n_{X}+8})$. Moreover, repeating the procedure above with ICs $\bm{Y}(\cdot,\delta) = (\bm{X}(\cdot,\delta),\widetilde{\bm{w}}(\cdot,\delta)) \in C^0([0,1];\mathbb{R}^{n_{Y}})$ satisfying the zeroth-order compatibility condition at the initial time $\delta$, a new solution may be obtained on an interval $[\delta,\delta+\eta]$, where $\eta \in \mathbb{R}_{>0}$ depends only on the IC $\bm{Y}(z,\delta)$. 

\paragraph{Blow-up alternative.}
Let now $t\ped{max}\in (0,\infty]$ be the maximal existence time of the solution. If $t\ped{max}<\infty$, then $\lim_{t\uparrow t\ped{max}} \norm{\bm{Y}(\cdot,t)}_{C^0([0,1];\mathbb{R}^{n_Y})}= \infty$, since otherwise there would exist a sequence $t_n\uparrow t\ped{max}$ such that $\norm{\bm{Y}(\cdot,t)}_{C^0([0,1];\mathbb{R}^{n_Y})} \leq R$ for all $n$ and some $R \in \mathbb{R}_{\geq 0}$. From the arguments above, this would imply that, for each $t_n$ close enough to $t\ped{max}$, the solution $\bm{Y}(z,t)$ obtained on $[0,t_n]$ can be extended on $[0,t_n + \eta]$ with $\eta \in \mathbb{R}_{>0}$ independent of $t_n$, and $t_n + \eta > t\ped{max}$, contradicting its maximality.
Finally, recalling the exponential transformation~\eqref{eq:esponentialTransf}, it may be easily concluded that $\lim_{t\uparrow t\ped{max}} \norm{\bm{Y}(\cdot,t)}_{C^0([0,1];\mathbb{R}^{n_Y})}= \infty$ is equipollent to Eq.~\eqref{eq:explosion}.
\end{proof}

\subsection{Global well-posedness for the ODE--PDE system~\eqref{eq:ODE--PDEsystem}}\label{sect:App2}
This appendix is devoted to establishing global existence and uniqueness results for the generalised solution of the ODE--PDE interconnection, under the additional assumptions of Theorem~\ref{thm:global}. The proposed strategy employs the system's formulation in terms of the transformed variables $(\bm{X}(z,t),\widetilde{\bm{w}}(z,t))$, and relies on an application of Grönwall-Bellman's inequality to exclude the blow-up alternative of Theorem~\ref{thm:Local}. To this end, it is first necessary to provide preliminary bounds on the functions involved.

\paragraph{Preliminary bounds} 
First, it may be observed that the assumptions of Theorem~\ref{thm:global} ensure the existence of constants $L_G,M_G\in \mathbb{R}_{\geq 0}$ such that
\begin{align}\label{eq:globalEstimateTildeG}
\norm{\widetilde{\bm{G}}(\bm{X},\widetilde{\bm{w}},z)}_{\ell^1(\mathbb{R}^8)} \leq M_G+\widetilde{L}_G\norm{(\bm{X},\widetilde{\bm{w}})}_{\ell^1(\mathbb{R}^{n_{X}+8})}, \quad (\bm{X},\widetilde{\bm{w}},z) \in \mathbb{R}^{n_{X}+8}\times[0,1],
\end{align}
for some $\widetilde{L}_G\in \mathbb{R}_{\geq 0}$. 

Additionally, the same global bounds as in Eqs.~\eqref{H_tildeContr}~\eqref{eq:supHtilde} hold for the transformed boundary map $\tilde{\bm{H}}$.

\paragraph{Estimate for the ODE subsystem.} 

Substituting~\eqref{eq:ODEx1} and~\eqref{eq:ODEx2} into~\eqref{eq:ODEf} yields 
\begin{align} \label{eq:ODEfrewritten}
\begin{split}
\dpd{f(z,t)}{t} &= -\alpha\bigl(v(z,t),z\bigr)f(z,t) -\bar k_0(z)v(z,t) \\
& \quad + \bar k_0(z) \sum_{i=1}^{n_1} \frac{x_{1,i}(z,t)}{\tau_{1,i}(z)} - \bar k_0(z) \sum_{i=1}^{n_2} \frac{x_{2,i}(z,t)}{\tau_{2,i}(z)}, \quad (z,t) \in (0,1)\times(0,T), 
\end{split}
\end{align} 
where $\alpha \in C^0(\mathbb{R}\times[0,1];\mathbb{R}_{>0})$ reads
\begin{align}
\alpha(v,z) \triangleq \bar{k}_0(z) \Biggl[ g(v,z) + \sum_{i=1}^{n_1}\frac{1}{\bar{c}_{1,i}(z)} + \sum_{i=1}^{n_2}\frac{1}{\bar{c}_{2,i}(z)} \Biggr]. 
\end{align}
For fixed $z\in [0,1]$, the explicit solution to the ODE~\eqref{eq:ODEfrewritten} may be recovered as
\begin{align} 
f(z,t) &= \eu^{-\int_{0}^{t}\alpha(v(z,t^\prime),z)\dif t^\prime}f_0(z) \\ &\quad+ \int_{0}^{t} \eu^{-\int_{t^\prime}^t \alpha(v(z,\tilde{t}),z)\dif \tilde{t}} \bigggl[ -\bar{k}_0(z)v(z,t^\prime) + \bar k_0(z)\sum_{i=1}^{n_1}\frac{x_{1,i}(z,t^\prime)}{\tau_{1,i}(z)} - \bar k_0(z)\sum_{i=1}^{n_2}\frac{x_{2,i}(z,t^\prime)}{\tau_{2,i}(z)} \bigggr]\dif t^\prime, \\
& \qquad \qquad \qquad \qquad \qquad \qquad \qquad\qquad \qquad \qquad \qquad \qquad \qquad \qquad (z,t) \in [0,1]\times[0,T],
\end{align}
implying the existence of a constant $C_f\in \mathbb{R}_{>0}$ such that
\begin{align}\label{eq:boundf}
\abs{f(z,t)} \leq \abs{f_0(z)} + C_f \int_{0}^{t} \Bigl( \abs{v(z,t^\prime)} + \norm{\bm{X}(z,t^\prime)}_{\ell^1(\mathbb{R}^{n_{X}})} \Bigr)\dif t^\prime, \quad (z,t) \in [0,1]\times[0,T]. 
\end{align}
Similarly, from Eqs.~\eqref{eq:ODEx1} and~\eqref{eq:ODEx2}, the following bound may be deduced:
\begin{align}\label{eq:boundx}
\abs{x_{j,i}(z,t)} \leq \abs{x_{j,i}(z,0)} + C_x\int_{0}^{t}\abs{f(z,t^\prime)}\dif t^\prime, \quad (z,t) \in [0,1]\times[0,T],\; i \in \{1,\dots,n_j\},\; j \in\{1,2\},
\end{align}
for some $C_x \in \mathbb{R}_{>0}$.
Combining inequalities~\eqref{eq:boundf} and~\eqref{eq:boundx}, recalling Eq.~\eqref{eq:RiemannS} together with the exponential transformation~\eqref{eq:esponentialTransf}, and taking the supremum over $z\in[0,1]$ leads to
\begin{align}
\begin{split}
\norm{\bm{X}(\cdot,t)}_{C^0([0,1];\ell^1(\mathbb{R}^{n_{X}}))}& \leq C_X\norm{\bm{X}_0(\cdot)}_{C^0([0,1];\ell^1(\mathbb{R}^{n_{X}}))} \\
& \quad+ C_X \int_{0}^{t} \Bigl( \norm{\widetilde{\bm{w}}(\cdot,t^\prime)}_{C^0([0,1];\ell^1(\mathbb{R}^8))} + \norm{\bm{X}(\cdot,t^\prime)}_{C^0([0,1];\ell^1(\mathbb{R}^{n_{X}}))} \Bigr)\dif t^\prime,\quad t \in [0,T], 
\end{split}
\end{align}
for some constant $C_X\in \mathbb{R}_{>0}$.
Thus, defining
\begin{subequations}
\begin{align}
X^\ast(t) & \triangleq \sup_{\tilde{t}\in[0,t]} \norm{X(\cdot,\tilde{t})}_{C^0([0,1];\ell^1(\mathbb{R}^{n_{X}}))}, \\
 \widetilde{W}^\ast(t) & \triangleq \sup_{\tilde{t}\in[0,t]} \norm{\widetilde{\bm{w}}(\cdot,\tilde{t})}_{C^0([0,1];\ell^1(\mathbb{R}^8))}, \\
Y^\ast(t) & \triangleq  X^\ast(t) + \widetilde{W}^\ast(t),
\end{align}
\end{subequations}
yields 
\begin{align}\label{eq:boundXast}
X^\ast(t) \leq C_X\norm{\bm{X}_0(\cdot)}_{C^0([0,1];\ell^1(\mathbb{R}^{n_{X}}))} + C_X \int_{0}^{t} Y^\ast(t^\prime) \dif t^\prime, \quad t \in [0,T]. 
\end{align}
It is worth noting that
\begin{align}
Y^\ast(t) \geq \sup_{\tilde{t}\in[0,t]}\norm{\bm{Y}(\cdot,\tilde{t})}_{C^0([0,1];\ell^1(\mathbb{R}^{n_{X}+8}))} =  \sup_{\tilde{t}\in[0,t]}\norm{(\bm{X}(\cdot,\tilde{t}),\widetilde{\bm{w}}(\cdot,\tilde{t}))}_{C^0([0,1];\ell^1(\mathbb{R}^{n_{X}+8}))}.
\end{align}

\paragraph{Estimate for the PDE subsystem.} Next, it is necessary to bound the transformed Riemann variables. If a characteristic reaches the initial line before the boundary, Eq.~\eqref{eq:PHitah<} gives
\begin{align}
\abs{\widetilde{w}_i^\pm(z,t)} \leq \norm{\widetilde{\bm{w}}_0(\cdot)}_{C^0([0,1];\ell^1(\mathbb{R}^{8}))} + \int_{0}^{t} \abs{\widetilde{G}_i^\pm(\bm{Y}(z_i^\pm(z,t,t^\prime),t^\prime),z_i^\pm(z,t,t^\prime))}\dif t^\prime, \quad \tau_i^\pm(z,t) < 0. 
\end{align}
Using the global growth estimate~\eqref{eq:globalEstimateTildeG} then yields 
\begin{align}
\abs{\widetilde{w}_i^\pm(z,t)} \leq \norm{\widetilde{\bm{w}}_0(\cdot)}_{C^0([0,1];\ell^1(\mathbb{R}^{8}))} + \widetilde{C}_G \int_{0}^{t} \bigl(1+Y^\ast(t^\prime)\bigr)\dif t^\prime, \quad \tau_i^\pm(z,t) < 0,
\end{align}
with $\mathbb{R}_{>0}\ni \widetilde{C}_G \triangleq \max\{M_G,\widetilde{L}_G\}$.

If a positive characteristic reaches the boundary $z=0$, then
 \begin{align} 
\begin{split}
\abs{\widetilde{w}_i^+(z,t)} &\leq \abs{\widetilde{H}_i^+\bigl( \widetilde{\bm{w}}^+(1,\tau_i^+(z,t)), \widetilde{\bm{w}}^-(0,\tau_i^+(z,t)), \bm{U}(\tau_i^+(z,t)) \bigr)} \\
& \quad + \int_{\tau_i^+(z,t)}^{t} \abs{\widetilde{G}_i^\pm(\bm{Y}(z_i^\pm(z,t,t^\prime),t^\prime),z_i^\pm(z,t,t^\prime))}\dif t^\prime, \quad \tau_i^+(z,t) \geq 0.
\end{split}
\end{align} 
By the growth assumption on $\widetilde{\bm{H}}$, 
\begin{align}
\abs{\widetilde{H}_i^+\bigl( \widetilde{\bm{w}}^+(1,\tau_i^+(z,t)), \widetilde{\bm{w}}^-(0,\tau_i^+(z,t)), \bm{U}(\tau_i^+(z,t)) \bigr)} \leq M_H+L_\gamma\widetilde{W}^\ast(t). 
\end{align}
Therefore 
\begin{align}
\abs{\widetilde{w}_i^+(z,t)} \leq \norm{\widetilde{\bm{w}}_0(\cdot)}_{C^0([0,1];\ell^1(\mathbb{R}^{8}))} + M_H+L_\gamma\widetilde{W}^\ast(t) + \widetilde{C}_G \int_{0}^{t} \bigl(1+Y^\ast(t^\prime)\bigr)\dif t^\prime, \quad t \in [0,T], \; i \in \{1,2,3,4\}. 
\end{align}
The same estimate holds for negative characteristics reaching the boundary $z=1$. Thus, taking the supremum over all $z\in[0,1]$ and all times $\tilde{t}\in[0,t]$, and summing over all components yields
\begin{align}\label{eq:boundWtildeStar}
\widetilde{W}^\ast(t) \leq 8C_{\widetilde{w}_0} + 8M_H + 8L_\gamma \widetilde{W}^\ast(t) + 8\widetilde{C}_G \int_{0}^{t} \bigl( 1+Y^\ast(t^\prime) \bigr)\dif t^\prime, \quad t \in [0,T], 
\end{align}
where $\mathbb{R}_{>0}\ni C_{\widetilde{w}_0} \triangleq \norm{\widetilde{\bm{w}}_0(\cdot)}_{C^0([0,1];\ell^1(\mathbb{R}^{8}))}$. Since $L_\gamma \in (0,1)$ may be made arbitrarily small, the boundary term appearing in Eq.~\eqref{eq:boundWtildeStar} may be absorbed into the left-hand side, ensuring the existence of a $C_{\widetilde{W}}$ such that 
\begin{align}\label{eq:boundWstarFInal}
\widetilde{W}^\ast(t) \leq C_{\widetilde{W}} + C_{\widetilde{W}} \int_{0}^{t} \bigl( 1+Y^\ast(t^\prime)\bigr)\dif t^\prime, \quad t\in[0,T]. 
\end{align}

\paragraph{Final estimate and proof of Theorem~\ref{thm:global}.} 

Adding the two estimates~\eqref{eq:boundXast} and~\eqref{eq:boundWstarFInal} finally gives 
\begin{align}
Y^\ast(t) \leq C_0+C_1t + C_1 \int_{0}^{t}Y^\ast(t^\prime)\dif t^\prime, \quad t \in [0,T]
\end{align}
for suitable constants $C_0,C_1\in \mathbb{R}_{>0}$. By Grönwall-Bellman's inequality, for every finite $T\in \mathbb{R}_{>0}$, 
\begin{align}\label{eq:Yaprioriestimate}
Y^\ast(t) \leq (C_0+C_1T) \eu^{C_1T} \quad t\in[0,T]\cap[0,t\ped{max}), 
\end{align}
implying that
\begin{align}
\sup_{t\in[0,T]\cap[0,t\ped{max})} \norm{\bm{Y}(\cdot,t)}_{C^0([0,1];\mathbb{R}^{n_{Y}})} <\infty, 
\end{align}
and
\begin{align}
\sup_{t\in[0,T]\cap[0,t\ped{max})}  \norm{(\bm{X}(\cdot,t), \bm{w}(\cdot,t))}_{C^0([0,1];\mathbb{R}^{n_{X}+8})} < \infty. 
\end{align}

The proof of Theorem~\ref{thm:global} is finally given below.
\begin{proof}[Proof of Theorem~\ref{thm:global}]
Consider $t\ped{max}$, the maximal existence time of the local solution. From the blow-up alternative already proved in Theorem~\ref{thm:Local} , either $t\ped{max}=\infty$, or Eq.~\eqref{eq:explosion} holds. However, under the stated assumptions, the a priori estimate~\eqref{eq:Yaprioriestimate} shows that $\norm{(\bm{X}(\cdot,t), \bm{w}(\cdot,t))}_{C^0([0,1];\mathbb{R}^{n_{X}+8})}$ remains bounded on every finite interval $[0,T]$, therefore excluding finite-time blow-up and implying necessarily $t\ped{max}=\infty$. In particular, the solution satisfies $(\bm{X},\bm{w})\in C^0(\mathbb{R}_{\geq 0};C^0([0,1];\mathbb{R}^{n_{X}+8}))  \cong C^0([0,1]\times \mathbb{R}_{\geq 0};\mathbb{R}^{n_{X}+8})$. 
\end{proof}

\subsection{Stability of the ODE--PDE interconnection}\label{app;stabl}
The proof for the stability of the entire ODE--PDE system~\eqref{eq:ODE--PDEsystem} is worked out in two steps: first, the stability of the ODE subsystem~\eqref{eq:ODE} is investigated in isolation; second, by exploiting the inherent dissipative nature of friction, the result is proved for the whole interconnection by relying on small-gain-like arguments.
  
\subsubsection{Proof of Lemma~\ref{lemma:friction}}\label{app:proofLemmaFric}
The proof of Lemma~\ref{lemma:friction} is similar to those presented in \cite{FrBD1,FrBD2}, the principal distinction being that the coefficient $C_{f2}$ must here be determined explicitly.

\begin{proof} 
First, it should be remarked that, under the same assumptions as Theorem~\ref{thm:Local}, the ODE--PDE system~\eqref{eq:ODE--PDEsystem} admits a unique generalised solution $(\bm{X},\bm{w})\in C^0([0,1]\times [0,T];\mathbb{R}^{n_{X}+8})$ on every finite interval $[0,T]$ with $T \in (0,t\ped{max})$. In particular, combined with Eq.~\eqref{eq:ODE}, this type of regularity also ensures $\bm{X}(\cdot,z) \in C^1([0,T];\mathbb{R}^{n_{X}})$.
Accordingly, the following Lyapunov function is considered:
\begin{align}\label{eq:Lyapunov}
\begin{split}
V\bigl(\bm{X}(z,t)\bigr) \triangleq & \dfrac{1}{2}\bigggl( \dfrac{f^2(z,t)}{\bar{k}_0(z,t)} + \sum_{i=1}^{n_1}\dfrac{\bar{c}_{i,1}(z,t)}{\tau_{i,1}(z)}x_{i,1}^2(z,t) + \sum_{i=1}^{n_2}\dfrac{\bar{c}_{i,2}(z,t)}{\tau_{i,2}(z)}x_{i,2}^2(z,t)\bigggr),
\end{split}
\end{align}
which is differentiable with respect to time in the classical sense.
In fact, taking the time derivative of Eq.~\eqref{eq:Lyapunov} along the dynamics~\eqref{eq:ODEi} yields
\begin{align}
\begin{split}
\dpd{V(z,t)}{t} & \leq \abs{v(z,t)}_\varepsilon\abs{f(z,t)} - \dfrac{\abs{v(z,t)}_\varepsilon}{\mu\bigl(v(z,t),z\bigr)}f^2(z,t) \\
& \quad - \sum_{i=1}^{n_1}\bar{c}_{i,1}(z)\biggl(\dpd{x_{i,1}(z,t)}{t}\biggr)^2 - \sum_{i=1}^{n_2}\bar{c}_{i,2}(z)\biggl(\dpd{x_{i,2}(z,t)}{t}\biggr)^2,  \\
& \leq \abs{v(z,t)}_\varepsilon\abs{f(z,t)} - \abs{v(z,t)}_\varepsilon \min\Biggl\{\dfrac{1}{\mu\bigl(v(z,t),z\bigr)},\dfrac{1}{\abs{v(z,t)}_\varepsilon}\Biggr\}\\
& \quad  \times \bigggl[ f^2(z,t) + \sum_{i=1}^{n_1}\bar{c}_{i,1}(z)\biggl(\dpd{x_{i,1}(z,t)}{t}\biggr)^2 + \sum_{i=1}^{n_2}\bar{c}_{i,2}(z)\biggl(\dpd{x_{i,2}(z,t)}{t}\biggr)^2 \bigggr], \quad (z,t) \in [0,1]\times [0,T]. 
\end{split}
\end{align}
Recalling Eq.~\eqref{eq:Lyapunov}, lengthy but straightforward manipulations lead in turn to
\begin{align}
\begin{split}
\dpd{V(z,t)}{t} & \leq \abs{v(z,t)}_\varepsilon\sqrt{2\bar{k}_0(z)} \sqrt{V(z,t)} \\
& \quad - \abs{v(z,t)}_\varepsilon q_f(z)\min\Biggl\{\dfrac{1}{\mu\bigl(v(z,t),z\bigr)},\dfrac{1}{\abs{v(z,t)}_\varepsilon}\Biggr\}V(z,t), \quad (z,t) \in [0,1]\times[0,T],
\end{split}
\end{align}
where $\mathbb{R}_{>0} \ni q_f(z) \triangleq 2\lambda\ped{min}(\mathbf{Q}(z))$.
Furthermore, defining $W(\bm{X}(z,t)) \triangleq \sqrt{V(\bm{X}(z,t))}$ with some abuse of notation yields, for $W(z,t) > 0$,
\begin{align}
\begin{split}
\dpd{W(z,t)}{t} &\leq -\dfrac{q_f(z)}{2} \min\Biggl\{\dfrac{\abs{v(z,t)}_\varepsilon}{\mu\bigl(v(z,t),z\bigr)},1\Biggr\} \\
& \quad \times \biggl[ W(z,t) - \dfrac{\sqrt{2\bar{k}_0(z)}}{q_f(z)} \max\Bigl\{\mu\bigl(v(z,t),z\bigr),\abs{v(z,t)}_\varepsilon\Bigr\} \biggr], \quad (z,t) \in [0,1]\times [0,T].
\end{split}
\end{align}
Thus, by comparison
\begin{align}\label{eq:Wcomp}
W(z,t) \leq \max\Biggl\{ W(z,0), \dfrac{\sqrt{2\bar{k}_0(z)}}{q_f(z)} \sup_{t^\prime\in[0,t]} \max\Bigl\{\mu\bigl(v(z,t^\prime),z\bigr),\abs{v(z,t^\prime)}_\varepsilon\Bigr\} \Biggr\}, \quad (z,t) \in [0,1]\times[0,T].
\end{align} 
Multiplying Eq.~\eqref{eq:Wcomp} by $\sqrt{2\bar{k}_0(z)}$, and using $\abs{f(z,t)}\leq \sqrt{2\bar{k}_0(z)}W(z,t)$ also gives
\begin{align}\label{eq:fcomp}
\abs{f(z,t)} \leq \max\Biggl\{ \sqrt{2\bar{k}_0(z)V(z,0)}, \frac{2\bar{k}_0(z)}{q_f(z)} \sup_{t^\prime\in[0,t]} \max\Bigl\{\mu\bigl(v(z,t^\prime),z\bigr),\abs{v(z,t^\prime)}_\varepsilon\Bigr\} \Biggr\}, \quad (z,t) \in [0,1]\times[0,T].
\end{align} 
Moreover, utilising the bound in Eq.~\eqref{eq:boundMu} and observing that $\abs{v}_\varepsilon=\sqrt{v^2+\varepsilon} \leq \abs{v}+\sqrt{\varepsilon}$ for all $\varepsilon \in \mathbb{R}_{\geq 0}$, the following inequality is deduced:
\begin{align}
\max\bigl\{\mu(v,z),\abs{v}_\varepsilon\bigr\} \leq \mu_1(z) + \max\bigl\{\mu_2(z),1\bigr\}\sqrt{\varepsilon} + \max\bigl\{\mu_2(z),1\bigr\}\abs{v}, \quad (v,z) \in \mathbb{R}\times[0,1],
\end{align}
which implies
\begin{align}\label{eq:boundsupMu}
\sup_{t^\prime\in[0,t]} \max\Bigl\{\mu\bigl(v(z,t^\prime\bigr),z),\abs{v}_\varepsilon\Bigr\} \leq \mu_1(z) + \max\bigl\{\mu_2(z),1\bigr\}\sqrt{\varepsilon} & + \max\bigl\{\mu_2(z),1\bigr\}\sup_{t^\prime\in[0,t]} \abs{v(z,t^\prime)}, \\
& \quad (v,z) \in C^0([0,1]\times[0,T];\mathbb{R})\times[0,1].
\end{align}
From Eqs.~\eqref{eq:Wcomp},~\eqref{eq:fcomp},~\eqref{eq:boundsupMu}, it follows that
\begin{subequations}
\begin{align}
W(z,t) & \leq \max\Biggl\{ W(z,0), \frac{\sqrt{2\bar{k}_0(z)}}{q_f(z)} \bigl[ \mu_1(z) + \max\{\mu_2(z),1\}\sqrt{\varepsilon} \bigr] + \frac{\sqrt{2k_0(z)}}{q_f(z)} \max\{\mu_2(z),1\} \sup_{t^\prime\in[0,t]}\abs{v(z,t^\prime)} \Biggr\}, \\
\abs{f(z,t)} & \leq \max\Biggl\{ \sqrt{2\bar{k}_0(z)V(z,0)}, \dfrac{2\bar{k}_0(z)}{q_f(z)} \bigl[ \mu_1(z) + \max\{\mu_2(z),1\}\sqrt{\varepsilon} \bigr] + \dfrac{2\bar{k}_0(z)}{q_f(z)} \max\{\mu_2(z),1\} \sup_{t^\prime\in[0,t]}\abs{v(z,t^\prime)} \Biggr\}.
\end{align}
\end{subequations}
Using the elementary inequality $\max\{a,c+dM\}\leq\max\{a,c\}+dM$ that holds for all $a,c,d,M \in \mathbb{R}_{\geq 0}$, it may be concluded that
\begin{subequations}
\begin{align}
\norm{W(\cdot,z)}_{C^0([0,1];\mathbb{R})}& \leq C_{W1} + C_{W2}\sup_{t^\prime\in[0,t]}\norm{v(\cdot,t^\prime)}_{C^0([0,1];\mathbb{R})}, \label{eq:finalBoundsfWW}\\
\norm{f(\cdot,z)}_{C^0([0,1];\mathbb{R})} & \leq C_{f1}\bigl(\bm{X}_0(\cdot)\bigr) + C_{f2}\sup_{t^\prime\in[0,t]}\norm{v(\cdot,t^\prime)}_{C^0([0,1];\mathbb{R})}, && t \in [0,T],\label{eq:finalBoundsfWf}
\end{align}
\end{subequations} 
with $C_{W1}(\bm{X}_0(\cdot)), C_{f1}(\bm{X}_0(\cdot)), C_{W2}, C_{f2} \in \mathbb{R}_{\geq 0}$ given by
\begin{subequations}
\begin{align}
C_{W1}\bigl(\bm{X}_0(\cdot)\bigr) & \triangleq \sup_{z\in[0,1]} \max\Biggl\{ W\bigl(\bm{X}_0(z)\bigr), \dfrac{\sqrt{2\bar{k}_0(z)}}{q_f(z)} \bigl[ \mu_1(z) + \max\{\mu_2(z),1\}\sqrt{\varepsilon} \bigr] \Biggr\},\label{eq:constCW1fb} \\
C_{W2} & \triangleq \sup_{z\in[0,1]} \dfrac{\sqrt{2\bar{k}_0(z)}}{q_f(z)} \max\bigl\{\mu_2(z),1\bigr\}, \label{eq:constC21fb}\\
C_{f1}\bigl(\bm{X}_0(\cdot)\bigr)  & \triangleq \sup_{z\in[0,1]} \max\Biggl\{ \sqrt{2\bar{k}_0(z)V\bigl(\bm{X}_0(z)\bigr)}, \dfrac{2\bar{k}_0(z)}{q_f(z)} \bigl[ \mu_1(z) + \max\{\mu_2(z),1\}\sqrt{\varepsilon} \bigr] \Biggr\}, \label{eq:constCf1fb}\\
C_{f2} & \triangleq \sup_{z\in[0,1]} \dfrac{2\bar{k}_0(z)}{q_f(z)} \max\bigl\{\mu_2(z),1\bigr\}. \label{eq:constCf2fb}
\end{align}
\end{subequations}
Equations~\eqref{eq:finalBoundsfWf},~\eqref{eq:constCf1fb}, and~\eqref{eq:constCf2fb} prove~\eqref{eq:boundfDOUS} and~\eqref{eq:constCf2}. The estimate~\eqref{eq:boundfDOUSX} is instead a consequence of Eqs.~\eqref{eq:finalBoundsfWW},~\eqref{eq:constCW1fb}, and~\eqref{eq:constCW1fb} by noting that $W(\bm{X}(z,t))$ is equivalent to a norm on $\mathbb{R}^{n_{X}}$.
\end{proof}

\subsubsection{Proof of Theorem~\ref{thm:Stability}}\label{app:finalProod}
The proof of Theorem~\ref{thm:Stability} is presented below. The main idea underlying the proposed approach is to transform the original PDE subsystem~\eqref{eq:PDEsubst} into a homodirectional one, thereby enabling the derivation of a closed-form solution for the Riemann variables propagating with the same transport velocity. This reformulation allows the PDEs~\eqref{eq:PDEsubst} to be reinterpreted as an interconnection of four sets of Riemann variables.
\begin{proof}[Proof of Theorem~\ref{thm:Stability}]
The proof starts by noting that, under the stated assumptions, Theorem~\ref{thm:Local} ensures the existence of a unique, maximal generalised solution $(\bm{X},\bm{w}) \in C^0([0,t\ped{max});C^0([0,1];\mathbb{R}^{n_X+8}))$ for some $t\ped{max} \in (0,\infty]$.
To prove that the solution is global and bounded, the variables $\mathbb{R}^2 \ni \bm{\xi}_i(z,t) = [\xi_i^+(z,t)\; \xi_i^-(z,t)]^{\mathrm{T}}$, $i \in \{1,2,3,4\}$, are first introduced, with
\begin{align}\label{eq:xiRdos}
\xi_i^{+}(z,t) \triangleq w_i^{+}(z,t), && \xi_i^{-}(z,t) \triangleq w_i^{-}(1-z,t). 
\end{align}
The above transformation is an isomorphism of $C^0([0,1];\mathbb{R}^2)$. Indeed,
\begin{align}
 \norm{\bm{\xi}_i(\cdot)}_{C^0([0,1];\mathbb{R}^2)} = \norm{\bm{w}_i(\cdot)}_{C^0([0,1];\mathbb{R}^2)}, \quad i \in \{1,2,3,4\}.
\end{align}
Moreover, inserting~\eqref{eq:xiRdos} into the PDE dynamics~\eqref{eq:PDEsubst} and BC~\eqref{eq:BCs} provides
\begin{subequations}\label{eq:ODE--PDEsystemphi}
\begin{align}
\dpd{\bm{\xi}_i(z,t)}{t} + \lambda_i\dpd{\bm{\xi}_i(z,t)}{z} & = \mathbf{A}_i(z)\bm{\xi}_i(z,t) + \bm{g}_i\bigl(\bm{X}(z,t),\bm{\xi}(z,t),z\bigr), && (z,t) \in (0,1)\times(0,t\ped{max}), \; i \in \{1,2,3,4\},\label{eq:PDEsubstphi}\\
\bm{\xi}(0,t)  & = \bm{H}\bigl(\bm{\xi}(0,t), \bm{U}(t)\bigr), && t \in (0,t\ped{max}), \label{eq:BCxi}
\end{align}
\end{subequations}
where the matrix-valued functions $\mathbf{A}_i \in C^0([0,1];\mathbf{M}_2(\mathbb{R}))$, $i \in \{1,2,3,4\}$, read
\begin{align}
\mathbf{A}_1(z) & \triangleq \dfrac{1}{2A\ped{p}(z)L\ped{s}}\sqrt{\dfrac{\beta}{\rho}}\dod{A\ped{p}(z)}{z}\begin{bmatrix} 1& -1 \\ 1 & -1\end{bmatrix},  && \mathbf{A}_2(z)  \triangleq \dfrac{1}{2A\ped{a}(z)L\ped{s}}\sqrt{\dfrac{\beta}{\rho}}\dod{A\ped{a}(z)}{z}\begin{bmatrix} 1& -1 \\ 1 & -1\end{bmatrix}, \nonumber \\
\mathbf{A}_3(z) & \triangleq \dfrac{1}{2A\ped{b}(z)L\ped{b}}\sqrt{\dfrac{\beta}{\rho}}\dod{A\ped{b}(z)}{z}\begin{bmatrix} 1& -1 \\ 1 & -1\end{bmatrix}, && \mathbf{A}_4(z) \triangleq \mathbf{0}.
\end{align}
Accordingly, defining $z_i(z,t,\tilde{t}) \triangleq z_i^+(z,t,\tilde{t})$ and $\tau_i(z,t) \triangleq \tau_i^+(z,t)$, $i \in \{1,2,3,4\}$, as in~\eqref{eq:tauZ+}, integrating Eq.~\eqref{eq:PDEsubstphi} along the characteristic lines and imposing the BC~\eqref{eq:BCxi} yields
\begin{subequations}\label{eq:solutionZI}
\begin{align}
\begin{split}
\bm{\xi}_i(z,t) & = \mathbf{\Phi}_i\bigl(z,z_i(z,t,0)\bigr)\bm{\xi}_{i,0}\bigl(z_i(z,t,0)\bigr)  \\
& \quad + \int_0^t \mathbf{\Phi}_i\bigl(z,z_i(z,t,t^\prime)\bigr) \widetilde{\bm{g}}_i\bigl(z_i(z,t,t^\prime),t^\prime\bigr) \dif t^\prime, \quad \tau_i(z,t) \leq 0, 
\end{split} \\
\begin{split}
\bm{\xi}_i(z,t) & =\mathbf{\Phi}_i(z,0) \widetilde{\bm{h}}_i\bigl(\tau_i(z,t)\bigr) \\
& \quad + \int_{\tau_i(z,t)}^t \mathbf{\Phi}_i\bigl(z,z_i(z,t,t^\prime)\bigr) \widetilde{\bm{g}}_i\bigl(z_i(z,t,t^\prime),t^\prime\bigr) \dif t^\prime, \quad \tau_i(z,t) > 0, \; i \in \{1,2,3,4\},
\end{split}
\end{align}
\end{subequations}
where the spatial state-transition matrices $\mathbf{\Phi}_i(z,\tilde{z}) \in \mathbf{M}_2(\mathbb{R})$ generated by $\mathbf{A}_i(z)$, $i \in \{1,2,3,4\}$, write
\begin{align}
\mathbf{\Phi}_1(z,\tilde{z}) & \triangleq \mathbf{I}_2 + \dfrac{1}{2}\ln\biggl( \dfrac{A\ped{p}(z)}{A\ped{p}(\tilde{z})}\biggr) \begin{bmatrix}1 & -1\\ 1 & -1\end{bmatrix}, 
&& \mathbf{\Phi}_2(z,\tilde{z}) \triangleq \mathbf{I}_2 + \dfrac{1}{2}\ln\biggl( \dfrac{A\ped{a}(z)}{A\ped{a}(\tilde{z})}\biggr) \begin{bmatrix}1 & -1\\ 1 & -1\end{bmatrix}, \nonumber \\
\mathbf{\Phi}_3(z,\tilde{z}) & \triangleq \mathbf{I}_2 + \dfrac{1}{2}\ln\biggl( \dfrac{A\ped{b}(z)}{A\ped{b}(\tilde{z})}\biggr) \begin{bmatrix}1 & -1\\ 1 & -1\end{bmatrix}, &&
\mathbf{\Phi}_4(z,\tilde{z}) \triangleq \mathbf{I}_2.
\end{align}
The next step consists of estimating upper bounds for the different components of the solution~\eqref{eq:solutionZI}. To this end, the vectors $\mathbb{R}_{\geq 0}^4 \ni \bm{W}(t) \triangleq [W_{1}\; W_{2}\; W_{3}\; W_{4}]^{\mathrm{T}}$, $\mathbb{R}_{\geq 0}^4 \ni \bm{W}_0(t) =  [W_{0,1}\; W_{0,2}\; W_{0,3}\; W_{0,4}]^{\mathrm{T}} \triangleq \bm{W}(0)$, and $\bm{\beta}\in \mathbb{R}_{\geq 0}^3$ are first defined as
\begin{subequations}\label{eq:W0Betas}
\begin{align}
W_{i}(t) & \triangleq  \norm{\bm{w}_{i}(\cdot, t)}_{C^0([0,1];\ell^2(\mathbb{R}^2))}, \\
W_{0,i} & \triangleq  \norm{\bm{w}_{i,0}(\cdot)}_{C^0([0,1];\ell^2(\mathbb{R}^2))}, && i \in \{1,2,3,4\}, \\
\bm{\beta}(t) & \triangleq \begin{bmatrix} C_{f1}\bigl(\bm{X}_0(\cdot)\bigr)\\ \norm{\bm{g}_0(\cdot)}_{C^0([0,1];\ell^2(\mathbb{R}^8))}, \\ \sup_{t \in \mathbb{R}_{\geq 0}} \norm{\bm{h}_0(t)}_{\ell^2(\mathbb{R}^8)} \end{bmatrix}.
\end{align}
\end{subequations}
Utilising Eq.~\eqref{eq:W0Betas}, it may be observed that the condition~\eqref{eq:condInitiDHDD} implies that 
\begin{align}\label{eq:condInitiDHDD2}
\mathbf{A}(0)\bm{W}_0+ \mathbf{B}\bigl(\bm{\phi}_f(\overline{\bm{w}}),\bm{\eta}, \overline{\bm{w}}\bigr)\bm{\beta}(t)< \Bigl(\mathbf{I}_4-\mathbf{\Theta}\bigl(\bm{\phi}_f(\overline{\bm{w}}),\bm{\eta},\overline{\bm{w}}\bigr)\Bigr)\overline{\bm{w}},
\end{align}
and, since $\mathbf{B}(\bm{\phi}_f(\overline{\bm{w}}),\bm{\eta},\overline{\bm{w}}) \in \mathbf{M}_{4\times 3}(\mathbb{R}_{\geq 0})$, also
\begin{align} 
\mathbf{A}(0)\bm{W}_0 < \Bigl( \mathbf{I}_4- \mathbf{\Theta} \bigl( \bm{\phi}_f(\overline{\bm{w}}), \bm{\eta}, \overline{\bm{w}} \bigr) \Bigr) \overline{\bm{w}} \leq \overline{\bm{w}}. 
\end{align} 
Moreover, by Eq.~\eqref{coeffs:K}, $\mathbf{A}(0)\succeq \mathbf{I}_4$, yielding $\bm{W}_0<\overline{\bm{w}}$ and thus $\bm{w}_0 \in \interior\overline{\mathcal{B}}_w$. Consequently, by continuity of $t\mapsto\bm{w}(\cdot,t)$ in $C^0([0,1];\mathbb{R}^8)$, the first-exit time 
\begin{align}\label{eq:firstExitTime} 
t_{w} \triangleq \inf\Bigl\{ t\in[0,t_{\max}) \mathrel{\Big|} \bm{w}(\cdot,t) \notin \interior\overline{\mathcal{B}}_w \Bigr\}, 
\end{align} 
with the convention $\inf\varnothing=t_{\max}$, is strictly positive, i.e., $t_w \in \mathbb{R}_{>0}$. By the definition of $t_w$, $\bm{w}(\cdot,t) \in \interior\overline{\mathcal{B}}_w$ on every interval $t\in[0,T]$ with $T \in (0,t_w)$; therefore, all the bounds in Eq.~\eqref{eq:boundsg_i}, and in particular the coefficients $\phi_{i0}(\overline{\bm{w}})$, $\phi_{ij}(\overline{\bm{w}})$, and $\phi_{if}(\overline{\bm{w}})$, are valid throughout $[0,T]$.

Then, defining the functions $\widetilde{\bm{g}}_i, \widetilde{\bm{h}}_i \in C^0([0,1]\times[0,T];\mathbb{R}^2) $ as $\widetilde{\bm{g}}_i(z,t) \triangleq \bm{g}_i(\bm{X}(z,t),\bm{\xi}(z,t),z)$ and $\widetilde{\bm{h}}_i(z,t) \triangleq \bm{H}_i(\bm{\xi}^-(0,t),\bm{U}(t))$, $i \in \{1,2,3,4\}$, and introducing
\begin{align}
\bar{g}_i(t) & \triangleq \phi_{i0}(\overline{\bm{w}}) \norm{\bm{g}_0(\cdot)}_{C^0([0,1];\ell^2(\mathbb{R}^8))} +\sum_{j=1}^4 \phi_{ij}(\overline{\bm{w}})\norm{\bm{w}_j(\cdot,t)}_{C^0([0,1];\ell^2(\mathbb{R}^2))}, \quad i \in \{1,2,3,4\},
\end{align}
provides, in conjunction with Eq.~\eqref{eq:boundsg_i},
\begin{align}\label{eq:gbarBound}
\norm{\widetilde{\bm{g}}(\cdot,t)}_{C^0([0,1];\ell^2(\mathbb{R}^2))} \leq \bar{g}_i(t) +\phi_{if}(\overline{\bm{w}})\norm{f(\cdot,t)}_{C^0([0,1];\mathbb{R})}, \quad i \in \{1,2,3,4\}.
\end{align}
Moreover,
\begin{subequations}
\begin{align}
\norm{\mathbf{\Phi}_1(z,\tilde{z})} & = \sqrt{1 + \dfrac{1}{4}\ln^2\biggl( \dfrac{A\ped{p}(z)}{A\ped{p}(\tilde{z})}\biggr)} +\dfrac{1}{2} \abs{\ln\biggl( \dfrac{A\ped{p}(z)}{A\ped{p}(\tilde{z})}\biggr)} = \sqrt{1+\alpha_1^2(z,\tilde{z})} + \abs{\alpha_1(z,\tilde{z})},\\
 \norm{\mathbf{\Phi}_2(z,\tilde{z})} & = \sqrt{1 + \dfrac{1}{4}\ln^2\biggl( \dfrac{A\ped{a}(z)}{A\ped{a}(\tilde{z})}\biggr)} +\dfrac{1}{2} \abs{\ln\biggl( \dfrac{A\ped{a}(z)}{A\ped{a}(\tilde{z})}\biggr)}= \sqrt{1+\alpha_2^2(z,\tilde{z})} + \abs{\alpha_2(z,\tilde{z})}, \\
\norm{\mathbf{\Phi}_3(z,\tilde{z})} & = \sqrt{1 +\dfrac{1}{4} \ln^2\biggl( \dfrac{A\ped{b}(z)}{A\ped{b}(\tilde{z})}\biggr)} + \dfrac{1}{2}\abs{\ln\biggl( \dfrac{A\ped{b}(z)}{A\ped{b}(\tilde{z})}\biggr)}= \sqrt{1+\alpha_3^2(z,\tilde{z})} + \abs{\alpha_3(z,\tilde{z})}, \\
\norm{\mathbf{\Phi}_4(z,\tilde{z})} & = 1,
\end{align}
\end{subequations}
with $\alpha_i(z,\tilde{z})$, $i \in \{1,2,3\}$ as in Eq.~\eqref{eq:functionsAlpha}.

Hence, combining Eq.~\eqref{eq:solutionZI} with~\eqref{eq:boundfFric},~\eqref{coeffs:K}, and~\eqref{eq:gbarBound}, and exploiting the fact that the ICs vanish for $t\geq \lambda_i^{-1}$, it is possible to infer the existence of an arbitraty constant $\varrho \in \mathbb{R}_{>0}$ verifying
\begin{align}\label{eq:bound1Xi}
\begin{split}
\norm{\bm{\xi}_i(z,t)}_{\ell^2(\mathbb{R}^2)} & \leq K_{i1}\eu^{-\varrho(t-\lambda_i^{-1})}\norm{\bm{\xi}_{i,0}(\cdot)}_{C^0([0,1];\ell^2(\mathbb{R}^2))} + K_{i2} \sup_{t^\prime\in[0,t]}\norm{\bm{\widetilde{h}}_i(t^\prime)}_{\ell^2(\mathbb{R}^2)} +K_{i3}\sup_{t^\prime\in[0,t]}\bar{g}_i(t^\prime) \\
& \quad + K_{if}(\overline{\bm{w}})\sup_{t^\prime\in[0,t]}\Bigl(C_{f1}\bigl(\bm{X}_0(\cdot)\bigr)+ C_{f2}C_v\norm{\bm{w}_4(\cdot,t^\prime)}_{C^0([0,1];\ell^2(\mathbb{R}^2))}\Bigr), \\
& \qquad \qquad \qquad \qquad \qquad \qquad \qquad \qquad \qquad \qquad t \in [0,T],\; i\in \{1,2,3,4\},
\end{split}
\end{align}
with $\mathbb{R}_{\geq 0} \ni K_{if} (\overline{\bm{w}})\triangleq K_{i3}\phi_{if}(\overline{\bm{w}})$, $i \in \{1,2,3,4\}$.
Furthermore, using the fact that the inequality
\begin{align}
\begin{split}
& K_{i2} \sup_{t^\prime\in[0,t]}\norm{\bm{\widetilde{h}}_i(t^\prime)}_{\ell^2(\mathbb{R}^2)} + \quad K_{i3}\sup_{t^\prime\in[0,t]}\bar{g}_i(t^\prime) \\
& \quad + K_{if}(\overline{\bm{w}})\sup_{t^\prime\in[0,t]}\Bigl(C_{f1}\bigl(\bm{X}_0(\cdot)\bigr)+ C_{f2}C_v\norm{\bm{w}_4(\cdot,t^\prime)}_{C^0([0,1];\ell^2(\mathbb{R}^2))}\Bigr)\\
& \quad \leq \sup_{t^\prime \in [0,t]}\Biggl[ \eta_{i1}K_{i2}\norm{\bm{\widetilde{h}}_i(t^\prime)}_{\ell^2(\mathbb{R}^2)} +\eta_{i2} K_{i3}\bar{g}(t^\prime) + \eta_{i3}K_{if}(\overline{\bm{w}})\sup_{t^\prime\in[0,t]}\Bigl(C_{f1} + C_{f2}C_v\norm{\bm{w}_4(\cdot,t^\prime)}_{C^0([0,1];\ell^2(\mathbb{R}^2))}\Bigr) \Biggr]
\end{split}
\end{align}
holds for every $\eta_{i1},\eta_{i2},\eta_{i3} \in \mathbb{R}_{>0}$ satisfying $\sum_{j=1}^3 \eta_{ij}^{-1}=1$, $i \in \{1,2,3,4\}$, the following ISS estimate is obtained from Eq.~\eqref{eq:bound1Xi}:
\begin{align}
\begin{split}
&  \norm{\bm{w}_i(z,t)}_{\ell^2(\mathbb{R}^2)} \leq \sup_{t^\prime \in [0,t]}\Biggl[ \eta_{i1}K_{i2}\norm{\bm{\widetilde{h}}_i(t^\prime)}_{\ell^2(\mathbb{R}^2)} +\eta_{i2} K_{i3}\bar{g}(t^\prime) \\
& \quad + \eta_{i3}K_{if}(\overline{\bm{w}})\sup_{t^\prime\in[0,t]}\Bigl(C_{f1}\bigl(\bm{X}_0(\cdot)\bigr) + C_{f2}C_v\norm{\bm{w}_4(\cdot,t^\prime)}_{C^0([0,1];\ell^2(\mathbb{R}^2))}\Bigr) \Biggr]\\
& \qquad+  K_{i1}\eu^{-\varrho(t-\lambda_i^{-1})}\norm{\bm{w}_{i,0}(\cdot)}_{C^0([0,1];\ell^2(\mathbb{R}^2))}, \quad t \in [0,T],\; i\in \{1,2,3,4\}.
\end{split}
\end{align}
Since the semigroup property holds for the solution~\eqref{eq:solutionZI}, $i \in\{1,2,3,4\}$, it is possible to invoke the fading memory inequality \cite[Lemma~7.1]{Krstic2} to conclude the existence of an arbitrary small constant $\epsilon \in \mathbb{R}_{>0}$ such that
\begin{align}\label{eq:fadingEstimate}
\begin{split}
&  \norm{\bm{w}_i(z,t)}_{\ell^2(\mathbb{R}^2)}\eu^{\delta t} \\
& \quad \leq (1+\epsilon)\sup_{t^\prime \in [0,t]}\eu^{\delta t^\prime}\biggl( \eta_{i1}K_{i2}\norm{\bm{\widetilde{h}}_i(t^\prime)}_{\ell^2(\mathbb{R}^2)} +\eta_{i2} K_{i3}\bar{g}(t^\prime) + \eta_{i3}K_{if}(\overline{\bm{w}}) C_{f2}C_v\norm{\bm{w}_4(\cdot,t^\prime)}_{C^0([0,1];\ell^2(\mathbb{R}^2))} \biggr)\\
& \qquad+  K_{i1}\eu^{\varrho\lambda_i^{-1}}\norm{\bm{w}_{i,0}(\cdot)}_{C^0([0,1];\ell^2(\mathbb{R}^2))} +(1+\epsilon)\eta_{i3}K_{if}(\overline{\bm{w}})C_{f1}\bigl(\bm{X}_0(\cdot)\bigr)\eu^{\delta t}, \quad t \in [0,T],\; i\in \{1,2,3,4\},
\end{split}
\end{align}
where $\delta \in (0,\varrho)$ depends only on $\varrho$ and $\lambda_i,K_{i1}$, $i \in \{1,2,3,4\}$. Consequently, recalling the bounds~\eqref{eq:boundsg_ihi} and defining the vectors $\mathbb{R}_{\geq 0}^4 \ni \bm{W}^\delta(t) \triangleq [W_1^\delta(t)\; W_2^\delta(t)\; W_3^\delta(t)\; W_4^\delta(t)]^{\mathrm{T}}$ and $\bm{\beta}^\delta (t) \in \mathbb{R}_{\geq 0}^3$ as
\begin{subequations}
\begin{align}
W_i^\delta(t) & \triangleq \sup_{t^\prime \in [0,t]}\eu^{\delta t^\prime} \norm{\bm{w}_i(\cdot,t^\prime)}_{C^0([0,1];\ell^2(\mathbb{R}^2))}, && i \in \{1,2,3,4\}, \\
\bm{\beta}^\delta(t) & \triangleq \begin{bmatrix} \eu^{\delta t}C_{f1}\bigl(\bm{X}_0(\cdot)\bigr)\\ \eu^{\delta t}\norm{\bm{g}_0(\cdot)}_{C^0([0,1];\ell^2(\mathbb{R}^8))}, \\ \sup_{t^\prime \in[0,t]} \eu^{\delta t^\prime} \norm{\bm{h}_0(t^\prime)}_{\ell^2(\mathbb{R}^8)} \end{bmatrix}.
\end{align}
\end{subequations}
using both bounds in~\eqref{eq:boundsg_ihi}, taking the supremum in~\eqref{eq:fadingEstimate}, and collecting the four resulting inequalities, provides
\begin{align}\label{eq:EqEdelta} 
\bigl( \mathbf{I}_4- (1+\epsilon) \mathbf{\Theta}_w \bigr) \bm{W}^\delta(t) \leq \mathbf{A}(\varrho)\bm{W}_0 + (1+\epsilon) \mathbf{B}_w \bm{\beta}^\delta(t), \quad t\in[0,T], \end{align}
where, to alleviate the notation, the following matrices $\mathbf{\Theta}_w \in \mathbf{M}_4(\mathbb{R}_{\geq 0})$ and $\mathbf{B}_w \in \mathbf{M}_{4\times 3}(\mathbb{R}_{\geq 0})$ have been defined:
\begin{subequations} 
\begin{align} 
\mathbf{\Theta}_{w} &\triangleq \mathbf{\Theta} \bigl( \bm{\phi}_f(\overline{\bm{w}}), \bm{\eta}, \overline{\bm{w}} \bigr), \\ 
\mathbf{B}_{w} &\triangleq \mathbf{B} \bigl( \bm{\phi}_f(\overline{\bm{w}}), \bm{\eta}, \overline{\bm{w}} \bigr). 
\end{align} 
\end{subequations}
By virtue of Eq.~\eqref{eq:SpectralRadius}, $\rho(\mathbf{\Theta}_{w})<1$, and, since the spectral radius is continuous, there exists $\epsilon_0\in \mathbb{R}_{>0}$ such that $\rho( (1+\epsilon)\mathbf{\Theta}_{w} )<1$ for all $\epsilon \in (0,\epsilon_0]$. In particular, the matrix-valued function $\mathbf{R}_w \in C^0([0,\epsilon_0];\mathbf{M}_4(\mathbb{R}_{\geq 0}))$, reading 
\begin{align} \mathbf{R}_w(\epsilon) \triangleq \bigl( \mathbf{I}_4- (1+\epsilon)\mathbf{\Theta}_{w} \bigr)^{-1} = \sum_{k=0}^\infty (1+\epsilon)^k \mathbf{\Theta}_{w}^k, 
\end{align} 
is well-defined and componentwise nonnegative. Thus, premultiplying Eq.~\eqref{eq:EqEdelta} by $\mathbf{R}_w(\epsilon)$, then multiplying by $\eu^{-\delta t}$, and recalling Eq.~\eqref{eq:W0Betas} yields 
\begin{align}\label{eq:DwWW_dina} 
\bm{W}(t) \leq \mathbf{R}_w(\epsilon) \Bigl[ \mathbf{A}(\varrho)\bm{W}_0\eu^{-\delta t} + (1+\epsilon) \mathbf{B}_{w}\bm{\beta} \Bigr], \quad t\in[0,T], 
\end{align}
since $\eu^{-\delta t}\bm{\beta}^\delta(t)\leq\bm{\beta}$ componentwise. It remains to show that $\varrho \in \mathbb{R}_{>0}$ and $\epsilon \in (0,\epsilon_0]$ may be selected so that $\bm{w}(\cdot,t) \in \overline{\mathcal{B}}_w$ for all times. To this end, the following function $\bm{\mathcal{F}} \in C^0(\mathbb{R}_{\geq 0}\times[0,\epsilon_0];\mathbb{R}^4)$ is conveniently introduced: 
\begin{align}\label{eq:strictMarginMap} 
\bm{\mathcal{F}}(\varrho,\epsilon) \triangleq \overline{\bm{w}} - \mathbf{R}_w(\epsilon)\bigl[ \mathbf{A}(\varrho)\bm{W}_0 + (1+\epsilon) \mathbf{B}_{w}\bm{\beta} \bigr]. 
\end{align} 
From Eq.~\eqref{eq:condInitiDHDD}, it follows that 
\begin{align} 
\bm{\mathcal{F}}(0,0) = \overline{\bm{w}} - \mathbf{R}_w(0) \bigl[ \mathbf{A}(0)\bm{W}_0 + \mathbf{B}_{w}\bm{\beta} \bigr] >\bm{0}, 
\end{align} 
since multiplication by the componentwise nonnegative matrix $\mathbf{R}_w(\epsilon)$ preserves the strict componentwise inequality. Furthermore, the mapping $(\varrho,\epsilon)\mapsto \bm{\mathcal{F}}(\varrho,\epsilon)$ is continuous in a neighbourhood of $(0,0)$, implying the existence of $\varrho_0 \in \mathbb{R}_{>0}$, $\epsilon_1\in(0,\epsilon_0]$, and $\mathbb{R}_{>0}^4 \ni \bm{\nu} \triangleq [\nu_1\; \nu_2\; \nu_3\; \nu_4]^{\mathrm{T}}$ such that 
\begin{align}\label{eq:uniformStrictMargin} 
\bm{\mathcal{F}}(\varrho,\epsilon) \geq\bm{\nu}>\bm{0} 
\end{align} for every $(\varrho,\epsilon)\in(0,\varrho_0]\times(0,\epsilon_1]$. Hence, choosing sufficiently small $(\varrho,\epsilon)\in(0,\varrho_0]\times(0,\epsilon_1]$, and using the fact that $\eu^{-\delta t}\leq1$, Eqs.~\eqref{eq:DwWW_dina} and~\eqref{eq:uniformStrictMargin} give 
\begin{align}\label{eq:strictBoxBound} 
\bm{W}(t) \leq \overline{\bm{w}}-\bm{\nu}, \quad t\in[0,T]. 
\end{align} 
Since $T\in (0,t_w)$ was arbitrary, Eq.~\eqref{eq:strictBoxBound} holds for every $t\in[0,t_w)$. Suppose, by contradiction, that $t_w<t\ped{max}$: passing to the limit as $t\uparrow t_w$ in Eq.~\eqref{eq:strictBoxBound}, and exploiting the continuity of $t\mapsto\bm{w}(\cdot,t)$ provides
\begin{align} 
\norm{\bm{w}_i(\cdot,t_w)}_{C^0([0,1];\ell^2(\mathbb{R}^2))} \leq \overline{w}_i-\nu_i < \overline{w}_i, \quad i\in\{1,2,3,4\}, 
\end{align} 
implying that $\bm{w}(\cdot,t_w) \in \interior\overline{\mathcal{B}}_w$. By continuity, there exists $h\in \mathbb{R}_{>0}$ such that $\bm{w}(\cdot,t)\in \interior\overline{\mathcal{B}}_w$ for all $t\in[t_w, t_w+h]$, contradicting the definition of $t_w$ in Eq.~\eqref{eq:firstExitTime}. Thus, $t_w = t\ped{max}$, and $\sup_{t\in[0,t\ped{max})} \norm{\bm{w}(\cdot,t)}_{C^0([0,1];\mathbb{R}^8)} <\infty$. Consequently, recalling Eq.~\eqref{eq:boundfDOUSX}, it may be concluded that $\sup_{t\in[0,t\ped{max})} \norm{ ( \bm{X}(\cdot,t),\bm{w}(\cdot,t) )}_{C^0([0,1];\mathbb{R}^{n_X+8})} <\infty$, which excludes the blow-up alternative of Theorem~\ref{thm:Local}. Hence, $t_{\max}=\infty$, and Eq.~\eqref{eq:strictBoxBound} proves that $\bm{w}(\cdot,t)\in\overline{\mathcal{B}}_w$ for all $t\in\mathbb{R}_{\geq0}$. Finally, since the bound~\eqref{eq:DwWW_dina} is valid globally in time, combining~\eqref{eq:DwWW_dina} with~\eqref{eq:boundfDOUSX}, and using equivalence of norms on finite-dimensional spaces, permits deducing the existence of constants $\Theta_X,\Theta_w,\gamma\in\mathbb{R}_{>0}$ such that the estimate~\eqref{eq:BoundedandISSEst} holds for all $t\in\mathbb{R}_{\geq0}$.
\end{proof}

\section{Additional details}\label{app:details}
This appendix provides additional details on the ODE--PDE system~\eqref{eq:ODE--PDEsystem}. Specifically, Appendix~\ref{app:equilibria} reports the analytical solution for the equilibria of the open and closed pipe configurations, whereas the corresponding expressions for the boundary map are given in Appendix~\ref{app:closedMAP}.

\subsection{Equilibria of the open/closed pipe configurations}\label{app:equilibria}
For a zero velocity rig input $\bm{U}^\star =[p\ped{ext}^\star\; v\ped{rig}^\star\; F\ped{bit}^\star]^{\mathrm{T}} = [p\ped{atm} \; 0\; 0]^{\mathrm{T}}$, the equilibria $(\bm{X}^\star(z),\bm{u}^\star(z))$ of the hydraulic-structural model~\eqref{eq:ODE--PDEsystem} may be recovered in closed form with $(\bm{X}^\star(z), q\ped{p}^\star(z),q\ped{a}^\star(z),q\ped{b}^\star(z),v^\star(z)) = (\bm{0},0,0,0,0)$, and
\begin{subequations}\label{eq:steadyStates}
\begin{align}
p^\star\ped{p}(z,t) & = p^\star\ped{a}(z,t) = p\ped{atm} +\rho L\ped{s}\mathrm{g} \int_0^z \cos\theta\ped{p}(z^\prime) \dif z^\prime, \\
p^\star\ped{b}(z,t) & = p\ped{a}^\star(1) + \rho L\ped{b}\mathrm{g}\int_0^z \cos\theta\ped{b}(z^\prime) \dif z^\prime, \\
\sigma^\star(z,t) & =- p^\star\ped{a}(1)\eu^{-2\alpha_4(z,1)} +\rho\ped{s}L\ped{s}\mathrm{g}\int_{z}^{1} \eu^{-2\alpha_4(z,z^\prime)}\cos\theta\ped{p}(z^\prime) \dif z^\prime,  \quad z \in [0,1],
\end{align}
\end{subequations}
where the function $\alpha_4 \in C^0([0,1];\mathbb{R}_{\geq 0})$ is given by
\begin{align}
\alpha_4(z,\tilde{z}) \triangleq \dfrac{1}{2} \ln \biggl(\dfrac{A\ped{s}(z)}{A\ped{s}(\tilde{z})}\biggr), \quad 0\leq \tilde{z}\leq z \leq 1.
\end{align}
Equations~\eqref{eq:steadyStates} may be employed directly to compute an expression for the stationary effective stress $\sigma\ped{e}^\star(z) \in \mathbb{R}$, defined according to~\eqref{eq:effstress}. A cleaner, alternative derivation exploits the fact that, in steady-state conditions, Eqs.~\eqref{eq:qp} and~\eqref{eq:qa} imply
\begin{subequations}\label{eq:ImplsPstat}
\begin{align}
\dpd{}{z}\bigl(A\ped{p}(z)p\ped{p}^\star(z) \bigr) & = \rho L\ped{s}\mathrm{g}A\ped{p}(z)\cos\theta\ped{p}(z) + \dod{A\ped{p}(z)}{z}p\ped{p}^\star(z), \\
\dpd{}{z}\bigl(A\ped{i}(z)p\ped{a}^\star(z) \bigr) & = \rho L\ped{s}\mathrm{g}A\ped{i}(z)\cos\theta\ped{p}(z)+ \dod{A\ped{i}(z)}{z}p\ped{a}^\star(z), \quad z\in(0,1).
\end{align}
\end{subequations}
Combining the above Eqs.~\eqref{eq:ImplsPstat} with~\eqref{eq:effstress} yields the following ODE for the stationary effective stress:
\begin{align}\label{eq:ODEeffectiveStress}
\dpd{}{z}\bigl(A\ped{s}(z)\sigma\ped{e}^\star(z) \bigr) = -(\rho\ped{s}-\rho)L\ped{s}\mathrm{g}A\ped{s}(z)\cos\theta\ped{p}(z), \quad z \in (0,1),
\end{align}
which coincides with the equation obtained in \cite{Sheppard} (where the coordinate $z \in [0,1]$ is defined with an opposite convention, i.e., from the bit to the rig). In particular, enforcing the BC~\eqref{eq:bCsigma}, the solution to~\eqref{eq:ODEeffectiveStress} may be recovered as
\begin{align}
A\ped{s}(z)\sigma\ped{e}^\star(z) = (\rho\ped{s}-\rho)L\ped{s}\mathrm{g}\int_{z}^{1}A\ped{s}(z^\prime)\cos\theta\ped{p}(z^\prime) \dif z^\prime, \quad z\in[0,1],
\end{align}
which is consistent with the expression reported in \cite{Aarsnes2}.

\subsection{Boundary map for the open and closed pipe configurations}\label{app:closedMAP}

The present appendix reports the complete expressions for the matrices $\mathbf{H}$ and $\mathbf{B}$ in Eq.~\eqref{eq:HLinear} for the open and closed pipe configurations. For notational convenience, the characteristic impedances $Z\ped{f},Z\ped{s}\in \mathbb{R}_{>0}$ are defined in both cases as
\begin{align}
Z\ped{f} & \triangleq \sqrt{\beta\rho}, && Z\ped{s} \triangleq \sqrt{E\rho_s}.
\end{align}

\paragraph{Open pipe configuration.}
For the open pipe configuration, the matrix $\mathbf{H}$ reads
\begin{align}
\mathbf{H}=
\begin{bmatrix}
0&0&0&0&1&0&0&0 \\
0&0&0&0&0&1&0&0 \\
\dfrac{2Z\ped{s}A\ped{b}(0)}{Z_0} &\dfrac{2Z\ped{s}A\ped{b}(0)}{Z_0} &0 &\dfrac{2Z\ped{s}A\ped{b}(0)A\ped{s}(1)}{Z_0} &0 &0 &\dfrac{Z_1}{Z_0} &0 \\
0&0&0&0&0&0&0&-1 \\
\dfrac{Z_2}{Z_0} &-\dfrac{2Z\ped{s}A\ped{p}(1)}{Z_0} &0 &-\dfrac{2Z\ped{s}A\ped{p}(1)A\ped{s}(1)}{Z_0} &0 &0
&\dfrac{2Z\ped{s}A\ped{p}(1)}{Z_0} &0 \\
-\dfrac{2Z\ped{s}A\ped{a}(1)}{Z_0}
&\dfrac{Z_3}{Z_0}
&0
&-\dfrac{2Z\ped{s}A\ped{a}(1)A\ped{s}(1)}{Z_0}
&0
&0
&\dfrac{2Z\ped{s}A\ped{a}(1)}{Z_0}
&0
\\
0&0&-1&0&0&0&0&0
\\
-\dfrac{2Z\ped{f}}{Z_0}
&-\dfrac{2Z\ped{f}}{Z_0}
&0
&\dfrac{Z_4}{Z_0}
&0
&0
&\dfrac{2Z\ped{f}}{Z_0}
&0
\end{bmatrix},
\end{align}
with $Z_0,Z_1,Z_2,Z_3,Z_4 \in \mathbb{R}$ given by
\begin{subequations}
\begin{align}
Z_0 & \triangleq Z\ped{s}\bigl(A\ped{a}(1)+A\ped{b}(0)+A\ped{p}(1)\bigr)+Z\ped{f}A\ped{s}(1), \\
Z_1 & \triangleq Z\ped{s}\bigl(A\ped{a}(1)-A\ped{b}(0)+A\ped{p}(1)\bigr)+Z\ped{f}A\ped{s}(1), \\
Z_2 & \triangleq Z\ped{s}\bigl(A\ped{a}(1)+A\ped{b}(0)-A\ped{p}(1)\bigr)+Z\ped{f}A\ped{s}(1), \\
Z_3 & \triangleq -Z\ped{s}\bigl(A\ped{a}(1)-A\ped{b}(0)-A\ped{p}(1)\bigr)+Z\ped{f}A\ped{s}(1), \\
Z_4 & \triangleq Z\ped{s}\bigl(A\ped{a}(1)+A\ped{b}(0)+A\ped{p}(1)\bigr)-Z\ped{f}A\ped{s}(1).
\end{align}
\end{subequations}
The matrix $\mathbf{B}$ is
\begin{align}\label{eq:matrixBopen}
\mathbf{B}=
\begin{bmatrix}
\dfrac{2A\ped{p}(0)}{Z\ped{f}}&0&0
\\
\dfrac{2A\ped{a}(0)}{Z\ped{f}}&0&0
\\
0&0&0
\\
0&2&0
\\
0&0&0
\\
0&0&0
\\
0&0&0
\\
0&0&0
\end{bmatrix}.
\end{align}

\paragraph{Closed pipe configuration.}
For the closed pipe configuration, the matrix $\mathbf{H}$ reads
\begin{align}
\mathbf{H}=
\begin{bmatrix}
0&0&0&0&1&0&0&0
\\
0&0&0&0&0&1&0&0
\\
\dfrac{2Z\ped{f}A\ped{b}(0)A\ped{i}(1)}{Z_0}
&\dfrac{Z_1}{Z_0}
&0
&\dfrac{Z_2}{Z_0}
&0
&0
&\dfrac{Z_3}{Z_0}
&0
\\
0&0&0&0&0&0&0&-1
\\
-\dfrac{Z_4}{Z_0}
&-\dfrac{2Z\ped{f}A\ped{i}(1)A\ped{p}(1)}{Z_0}
&0
&\dfrac{Z_5}{Z_0}
&0
&0
&\dfrac{2Z\ped{f}A\ped{i}(1)A\ped{p}(1)}{Z_0}
&0
\\
-\dfrac{2Z\ped{f}A\ped{a}(1)A\ped{i}(1)}{Z_0}
&\dfrac{Z_6}{Z_0}
&0
&-\dfrac{Z_7}{Z_0}
&0
&0
&\dfrac{Z_8}{Z_0}
&0
\\
0&0&-1&0&0&0&0&0
\\
\dfrac{Z_9}{Z_0}
&-\dfrac{2Z\ped{f}A\ped{i}(1)}{Z_0}
&0
&\dfrac{Z_{10}}{Z_0}
&0
&0
&\dfrac{2Z\ped{f}A\ped{i}(1)}{Z_0}
&0
\end{bmatrix},  
\end{align}
with $Z_0,Z_1,Z_2,Z_3,Z_4, Z_5, Z_6, Z_7, Z_8, Z_9, Z_{10}\in \mathbb{R}$ given by
\begin{subequations}
\begin{align}
Z_0 & \triangleq Z\ped{s}A\ped{s}(1)\bigl(A\ped{a}(1)+A\ped{b}(0)\bigr) +Z\ped{f}\Bigl[A\ped{p}(1)\bigl(A\ped{a}(1)+A\ped{b}(0)\bigr)+A\ped{i}^2(1)\Bigr], \\
Z_1 & \triangleq 2A\ped{b}(0)\bigl(Z\ped{s}A\ped{s}(1)+Z\ped{f}A\ped{p}(1)\bigr), \\
Z_2 & \triangleq 2Z\ped{s}A\ped{b}(0)A\ped{s}(1)A\ped{i}(1), \\
Z_3 & \triangleq Z\ped{s}A\ped{s}(1)\bigl(A\ped{a}(1)-A\ped{b}(0)\bigr)+Z\ped{f}\Bigl[A\ped{p}(1)\bigl(A\ped{a}(1)-A\ped{b}(0)\bigr)+A\ped{i}^2(1)\Bigr], \\
Z_4 & \triangleq Z\ped{s}A\ped{s}(1)\bigl(A\ped{a}(1)+A\ped{b}(0)\bigr)-Z\ped{f}\Bigl[A\ped{p}(1)\bigl(A\ped{a}(1)+A\ped{b}(0)\bigr)-A\ped{i}^2(1)\Bigr], \\
Z_5 & \triangleq 2Z\ped{s}A\ped{p}(1)A\ped{s}(1)\bigl(A\ped{a}(1)+A\ped{b}(0)\bigr), \\
Z_6 & \triangleq -Z\ped{s}A\ped{s}(1)\bigl(A\ped{a}(1)-A\ped{b}(0)\bigr)-Z\ped{f}\Bigl[A\ped{p}(1)\bigl(A\ped{a}(1)-A\ped{b}(0)\bigr)-A\ped{i}^2(1)\Bigr], \\
Z_7 & \triangleq 2Z\ped{s} A\ped{a}(1)A\ped{s}(1)A\ped{i}(1)\\
Z_8 & \triangleq 2A\ped{a}(1)\bigl(Z\ped{s}A\ped{s}(1)+Z\ped{f}A\ped{p}(1)\bigr), \\
Z_9 & \triangleq 2Z\ped{f}\bigl(A\ped{a}(1)+A\ped{b}(0)\bigr), \\
Z_{10} & \triangleq Z\ped{s}A\ped{s}(1)\bigl(A\ped{a}(1)+A\ped{b}(0)\bigr)-Z\ped{f}\Bigl[A\ped{p}(1)\bigl(A\ped{a}(1)+A\ped{b}(0)\bigr)+A\ped{i}^2(1)\Bigr].
\end{align}
\end{subequations}
The matrix $\mathbf{B}$ is the same as in Eq.~\eqref{eq:matrixBopen}.


\begin{thebibliography}{99}

\bibitem[Petersen et al.(2008)]{Petersen2008}
Petersen, J., Rommetveit, R., Bj\o rkevoll, K.S., Fr\o yen, J., Bj\o rkevoll, O., Svanes, K. and Lage, A.C.V.M. (2008).
A general dynamic model for single and multi-phase flow operations during drilling, completion, well control and intervention.
In: \emph{Proceedings of the IADC/SPE Drilling Conference and Exhibition}.
Paper IADC/SPE 114688. doi: \url{https://doi.org/10.2118/114688-MS}.

\bibitem[Mitchell(1988)]{Mitchell1988}
Mitchell, R.F. (1988).
Dynamic surge/swab pressure predictions.
\emph{SPE Drilling Engineering}, 3(3), 325-333.
doi: \url{https://doi.org/10.2118/16156-PA}.

\bibitem[Wagner et al.(1993)]{Wagner1993}
Wagner, R.R., Halal, A.S. and Goodman, M.A. (1993).
Surge field tests highlight dynamic fluid response.
In: \emph{Proceedings of the SPE/IADC Drilling Conference}, 883--892.
Paper SPE/IADC 25771. doi: \url{https://doi.org/10.2118/25771-MS}.

\bibitem[Pavlov et al.(2010)]{Pavlov2010}
Pavlov, A., Kaasa, G.-O. and Imsland, L. (2010).
Experimental disturbance rejection on a full-scale drilling rig.
In: \emph{IFAC Proceedings Volumes}, 43(14), 1338--1343.
doi: \url{https://doi.org/10.3182/20100901-3-IT-2016.00296}.

\bibitem[Kaasa et al.(2012)]{Kaasa2012}
Kaasa, G.-O., Stamnes, \O.N., Imsland, L. and Aamo, O.M. (2012).
Simplified hydraulics model used for intelligent estimation of downhole pressure for a managed-pressure-drilling control system.
\emph{SPE Drilling \& Completion}, 27(1), 127--138.
doi: \url{https://doi.org/10.2118/143097-PA}.


\bibitem[Samuel et al.(2003)]{Samuel2003}
Samuel, G.R., Sunthankar, A., McColpin, G., Bern, P. and Flynn, T. (2003).
Field validation of transient swab-surge response with real-time downhole pressure data.
\emph{SPE Drilling \& Completion}, 18(4), 280--283.
doi: \url{https://doi.org/10.2118/85109-PA}.

\bibitem[Stecki and Davis(1986a)]{Stecki1986a}
Stecki, J.S. and Davis, E.J. (1986a).
Fluid transmission lines---distributed parameter models. Part 1: A review of the state of the art.
\emph{Proceedings of the Institution of Mechanical Engineers, Part A: Power and Process Engineering}, 200(4), 215--228.

\bibitem[Stecki and Davis(1986b)]{Stecki1986b}
Stecki, J.S. and Davis, E.J. (1986b).
Fluid transmission lines---distributed parameter models. Part 2: Comparison of models.
\emph{Proceedings of the Institution of Mechanical Engineers, Part A: Power and Process Engineering}, 200(4), 229--236.

\bibitem{Aamo00}
Godhavn, J.-M., Olorunju, B., Gorski, D., Kvernland, M., Sant'Ana, M., Aamo, O.M. and Sangesland, S. (2021).
Significant surge and swab offshore Brazil induced by rig heave during drillpipe connections.
\emph{SPE Drilling \& Completion}, 36(3), 552--559.
doi: \url{https://doi.org/10.2118/200518-PA}.

\bibitem{Aamo01}
Godhavn, J.-M., Olorunju, B., Gorski, D., Kvernland, M., Sant'Ana, M., Aamo, O.M. and Sangesland, S. (2020).
Surge and swab during drill pipe connections on floaters may affect pressure windows in calm weather.
\emph{Drilling Contractor}, 76(2), 53--54.

\bibitem[Spanos et al.(2003)]{Spanos2003}
Spanos, P.D., Chevallier, A.M., Politis, N.P. and Payne, M.L. (2003).
Oil and gas well drilling: A vibrations perspective.
\emph{Shock and Vibration Digest}, 35(2), 85--103.
doi: \url{https://doi.org/10.1177/0583102403035002564}.

\bibitem[Leine et al.(2002)]{Leine2002}
Leine, R.I., van Campen, D.H. and Keultjes, W.J.G. (2002).
Stick-slip whirl interaction in drillstring dynamics.
\emph{Journal of Vibration and Acoustics}, 124(2), 209--220.
doi: \url{https://doi.org/10.1115/1.1452745}.

\bibitem[Leine and van Campen(2005)]{Leine2005}
Leine, R.I. and van Campen, D.H. (2005).
Stick-slip whirl interaction in drillstring dynamics.
In: \emph{IUTAM Symposium on Chaotic Dynamics and Control of Systems and Processes in Mechanics}, 287--296.

\bibitem[Richard et al.(2004)]{Richard2004}
Richard, T., Germay, C. and Detournay, E. (2004).
Self-excited stick-slip oscillations of drill bits.
\emph{Comptes Rendus M\'ecanique}, 332(8), 619--626.

\bibitem[Richard et al.(2007)]{Richard2007}
Richard, T., Germay, C. and Detournay, E. (2007).
A simplified model to explore the root cause of stick-slip vibrations in drilling systems with drag bits.
\emph{Journal of Sound and Vibration}, 305(3), 432--456.

\bibitem[Besselink et al.(2011)]{Besselink2011}
Besselink, B., van de Wouw, N. and Nijmeijer, H. (2011).
A semi-analytical study of stick-slip oscillations in drilling systems.
\emph{Journal of Computational and Nonlinear Dynamics}, 6(2), 021006.
doi: \url{https://doi.org/10.1115/1.4002386}.

\bibitem[Liao et al.(2011)]{Liao2011}
Liao, C.-M., Balachandran, B., Karkoub, M. and Abdel-Magid, Y.L. (2011).
Drill-string dynamics: Reduced-order models and experimental studies.
\emph{Journal of Vibration and Acoustics}, 133(4), 041008.
doi: \url{https://doi.org/10.1115/1.4003406}.

\bibitem[Kovalyshen(2014)]{Kovalyshen2014}
Kovalyshen, Y. (2014).
Understanding root cause of stick-slip vibrations in deep drilling with drag bits.
\emph{International Journal of Non-Linear Mechanics}, 67, 331--341.

\bibitem[Depouhon and Detournay(2014)]{Depouhon2014}
Depouhon, A. and Detournay, E. (2014).
Instability regimes and self-excited vibrations in deep drilling systems.
\emph{Journal of Sound and Vibration}, 333(7), 2019--2039.

\bibitem[Landet et al.(2013)]{Landet2013}
Landet, I.S., Pavlov, A. and Aamo, O.M. (2013).
Modeling and control of heave-induced pressure fluctuations in managed pressure drilling.
\emph{IEEE Transactions on Control Systems Technology}, 21(4), 1340--1351.
doi: \url{https://doi.org/10.1109/TCST.2012.2204751}.

\bibitem{Aamo03}
Aamo, O.M. (2013).
Disturbance rejection in $2\times2$ linear hyperbolic systems.
\emph{IEEE Transactions on Automatic Control}, 58(5), 1095--1106.
doi: \url{https://doi.org/10.1109/TAC.2012.2228035}.

\bibitem{Aamo04}
Aarsnes, U.J.F., Gleditsch, M.S., Aamo, O.M. and Pavlov, A. (2014).
Modelling and avoidance of heave-induced pressure resonances in offshore drilling.
\emph{SPE Drilling \& Completion}, 29(4).

\bibitem{Aamo05}
Albert, A., Aamo, O.M., Godhavn, J.-M. and Pavlov, A. (2015).
Suppressing pressure oscillations in offshore drilling: Control design and experimental results.
\emph{IEEE Transactions on Control Systems Technology}, 23(2), 813--819.

\bibitem[Strecker et al.(2017)]{Strecker2017}
Strecker, T., Aamo, O.M. and Manum, H. (2017).
Simulation of heave-induced pressure oscillations in Herschel--Bulkley muds.
\emph{SPE Journal}, 22(5), 1635--1653.
doi: \url{https://doi.org/10.2118/185947-PA}.

\bibitem{Aamo06}
Strecker, T. and Aamo, O.M. (2019).
Attenuating heave-induced pressure oscillations in offshore drilling by downhole flow control.
\emph{IFAC Journal of Systems and Control}, 8, 100058.
doi: \url{https://doi.org/10.1016/j.ifacsc.2019.100058}.

\bibitem{Sheppard}
Sheppard, M.C., Wick, C. and Burgess, T. (1987).
Designing well paths to reduce drag and torque.
\emph{SPE Drilling Engineering}, 2(4), 344--350.

\bibitem{FrBD1}
Romano, L., Aamo, O.M., \AA slund, J. and Frisk, E. (2026).
Linear viscoelastic rheological FrBD models.
\emph{arXiv preprint arXiv:2601.13799}.
doi: \url{https://doi.org/10.48550/arXiv.2601.13799}.

\bibitem{FrBD2}
Romano, L. (2026).
Hyperbolic dynamic friction models for viscoelastic sliding and rolling contact.
\emph{arXiv preprint arXiv:2606.09128}.
doi: \url{https://doi.org/10.48550/arXiv.2606.09128}.

\bibitem{FrBD}
Romano, L., Aamo, O.M., \AA slund, J. and Frisk, E. (2026).
First-order friction models with bristle dynamics: Lumped and distributed formulations.
\emph{IEEE Transactions on Control Systems Technology}, early access.
doi: \url{https://doi.org/10.1109/TCST.2026.3696416}.

\bibitem{Aarsnes2}
Aarsnes, U.J.F. and Shor, R.J. (2018).
Torsional vibrations with bit off bottom: Modeling, characterization and field data validation.
\emph{Journal of Petroleum Science and Engineering}, 163, 712--721.
doi: \url{https://doi.org/10.1016/j.petrol.2017.11.024}.

\bibitem{Zwart}
Curtain, R.F. and Zwart, H. (2020).
\emph{Introduction to Infinite-Dimensional Systems Theory: A State-Space Approach}.
Springer, New York, 2020.

\bibitem{Pazy}
Pazy, A. (1983).
\emph{Semigroups of Linear Operators and Applications to Partial Differential Equations}.
Springer, New York.
doi: \url{https://doi.org/10.1007/978-1-4612-5561-1}.

\bibitem{Strecker1}
Strecker, T. and Aamo, O.M. (2017).
Output feedback boundary control of $2\times2$ semilinear hyperbolic systems.
\emph{Automatica}, 83, 290--302.
doi: \url{https://doi.org/10.1016/j.automatica.2017.06.026}.

\bibitem{Strecker2}
Strecker, T., Aamo, O.M. and Cantoni, M. (2020).
Output feedback boundary control of heterodirectional semilinear hyperbolic systems.
\emph{Automatica}, 117, 108990.
doi: \url{https://doi.org/10.1016/j.automatica.2020.108990}.




\bibitem{Krstic2}
Karafyllis, I. and Krstić, M. (2018).
\emph{Input-to-State Stability for PDEs}.
Springer, Cham.

\bibitem{Coron}
Bastin, G. and Coron, J.-M. (2016).
\emph{Stability and Boundary Stabilization of 1-D Hyperbolic Systems}.
Birkh\"auser, Cham.
doi: \url{https://doi.org/10.1007/978-3-319-32062-5}.




\end{thebibliography}
\end{document}